\documentclass[%
  twoside,
  floatfix,
  reprint,
  amsmath,amssymb,
  aps,
  pra,
  nofootinbib,
  superscriptaddress,
  a4paper,
]{revtex4-1}

\usepackage{graphicx}%
\usepackage[dvipsnames]{xcolor}
\usepackage{siunitx}
\usepackage{mathtools}
\usepackage{bm}
\usepackage{enumitem}

\usepackage{lipsum}

\definecolor{alexis}{rgb}{0,0,0}

\usepackage[
  bookmarks=false,
  colorlinks,
  linkcolor=blue,
  urlcolor=blue,
  citecolor=blue,
  plainpages=false,
  pdfpagelabels,
  final,
  breaklinks=true
]{hyperref}
\hypersetup{
pdftitle={Circular dichroism in high-order harmonic generation: Heralding topological phase transitions in Chern insulators},
pdfauthor={Alexis Chac\'on, Dasol Kim, Wei Zhu, Shane P. Kelly,  Alexandre Dauphin, Emilio Pisanty,  Andrew S. Maxwell,  Antonio Pic\'on,   Marcelo F. Ciappina,  Dong Eon Kim,  Christopher Ticknor, Avadh Saxena, and Maciej Lewenstein},
}

\usepackage{natbib}
\makeatletter \def\NAT@def@citea{\def\@citea{\NAT@separator\,}} \makeatother

\usepackage{physics}

\renewcommand{\Im}{\operatorname{Im}}

\newcommand{\CD}{\mathrm{CD}}
\newcommand{\RCP}{\mathrm{RCP}}
\newcommand{\LCP}{\mathrm{LCP}}

\DeclareSIUnit{\au}{{a.u.}}

\begin{document}

\title{Circular dichroism in high-order harmonic generation: Heralding topological {\color{alexis}phases and transitions} in Chern insulators}

\author{Alexis Chac\'on}
\email{achacon@postech.ac.kr}
\affiliation{Center for Nonlinear Studies and Theoretical Division, Los Alamos National Laboratory, Los Alamos, New Mexico 87545, USA}
\affiliation{Department of Physics and Center for Attosecond Science and Technology, POSTECH, 7 Pohang 37673, South Korea; Max Planck POSTECH/KOREA Research Initiative, Pohang, 37673, South Korea}
\author{Dasol Kim}
\affiliation{Department of Physics and Center for Attosecond Science and Technology, POSTECH, 7 Pohang 37673, South Korea; Max Planck POSTECH/KOREA Research Initiative, Pohang, 37673, South Korea}
\author{Wei Zhu}
\affiliation{Center for Nonlinear Studies and Theoretical Division, Los Alamos National Laboratory, Los Alamos, New Mexico 87545, USA}
\author{Shane P. Kelly}
\affiliation{Center for Nonlinear Studies and Theoretical Division, Los Alamos National Laboratory, Los Alamos, New Mexico 87545, USA}
\affiliation{Physics and Astronomy Department, University of California Riverside, Riverside, California 92521, USA}
\author{\color{alexis}Alexandre Dauphin}
\affiliation{ICFO -- Institut de Ciencies Fotoniques, The Barcelona Institute of Science and Technology, 08860 Castelldefels (Barcelona), Spain}
\author{\color{alexis}Emilio Pisanty}
\affiliation{ICFO -- Institut de Ciencies Fotoniques, The Barcelona Institute of Science and Technology, 08860 Castelldefels (Barcelona), Spain}
\author{\color{alexis}Andrew S.  Maxwell}
\affiliation{ICFO -- Institut de Ciencies Fotoniques, The Barcelona Institute of Science and Technology, 08860 Castelldefels (Barcelona), Spain}
\affiliation{Department of Physics \& Astronomy, University College London, Gower Street, London WC1E 6BT, United Kingdom}
\author{\color{alexis}Antonio Pic\'on}
\affiliation{ICFO -- Institut de Ciencies Fotoniques, The Barcelona Institute of Science and Technology, 08860 Castelldefels (Barcelona), Spain}
\affiliation{Departamento de Qu\'imica, Universidad Aut\'onoma de Madrid, 28049 Madrid, Spain}
\author{\color{alexis}Marcelo F. Ciappina}
\affiliation{ICFO -- Institut de Ciencies Fotoniques, The Barcelona Institute of Science and Technology, 08860 Castelldefels (Barcelona), Spain}
\affiliation{Institute of Physics of the ASCR, ELI-Beamlines project, Na Slovance 2, 182 21 Prague, Czech Republic}
\author{\color{alexis}Dong Eon Kim}
\affiliation{Department of Physics and Center for Attosecond Science and Technology, POSTECH, 7 Pohang 37673, South Korea; Max Planck POSTECH/KOREA Research Initiative, Pohang, 37673, South Korea}
\author{\color{alexis}Christopher Ticknor}
\affiliation{Center for Nonlinear Studies and Theoretical Division, Los Alamos National Laboratory, Los Alamos, New Mexico 87545, USA}
\author{\color{alexis}Avadh Saxena}
\affiliation{Center for Nonlinear Studies and Theoretical Division, Los Alamos National Laboratory, Los Alamos, New Mexico 87545, USA}
\author{\color{alexis}Maciej Lewenstein}
\affiliation{ICFO -- Institut de Ciencies Fotoniques, The Barcelona Institute of Science and Technology, 08860 Castelldefels (Barcelona), Spain}
\affiliation{ICREA, Pg. Llu\'is Companys 23, 08010 Barcelona, Spain}

\date{\today}

\begin{abstract}
Topological materials are of interest to both fundamental science and advanced technologies, because topological states are robust with respect to perturbations and dissipation. Experimental detection of topological invariants is thus in great demand, but it remains extremely challenging. Ultrafast laser-matter interactions,  and in particular high-harmonic generation (HHG), meanwhile, were proposed several years ago as tools to explore the structural and dynamical properties of various matter targets.
Here we show that the {\color{alexis}high-harmonic} emission signal produced by {\color{alexis}a circularly-polarized laser} contains signatures of the topological phase transition in the paradigmatic Haldane model. In addition to clear shifts of the overall emissivity and harmonic cutoff, the {\color{alexis}high-harmonic} emission shows a unique circular dichroism, which exhibits clear changes in behavior at the topological phase boundary.~Our findings pave the way to understand fundamental questions about the ultrafast electron-hole pair dynamics in topological materials via {\color{alexis}non-linear high-harmonic generation spectroscopy}.
\end{abstract}
\maketitle

\section{Introduction}


The study of topological order was initiated by the discoveries of the {\color{alexis}Berezinskii-Kosterlitz-Thouless} transition in 2D as well as the integer and fractional quantum Hall effects~\cite{KosterlitzNobel2016, KlitzingNobel1985, StormerNobel1998}.
These materials constitute a new paradigm, since they are characterized by a global order parameter: this goes beyond the standard Landau theory of phase transitions, which uses \textit{local} order parameters to describe materials.~Topological order, due to its robustness and resistance to perturbations, has already found applications in standards and metrology (most notably, via the integer quantum Hall effect, a key ingredient of the recent revision to the SI system of units~\cite{Haddad2016}), and promises numerous applications from quantum spintronics and valleytronics to quantum computing.\\ 
One question of particular interest is the possible applications of topological insulators (TI) and superconductors~\cite{ShouCheng2011}.
These systems are insulating in their bulk, but have conducting surface states protected by the topological invariant of the bulk.
A considerable number of solid-state systems with these properties have been proposed in recent years, both in the context of real topological materials~\cite{Hasan2010} and in synthetic ones, employing ultracold atoms~\cite{Goldman2014}, photonic systems~\cite{Ozawa2018} and mechanical systems~\cite{Huber2016}, among others.
Nevertheless, new methods for the creation and physical characterization of topological phases are still being sought for real materials.
Here we focus on the detection of topological phase transitions and different topological phases by the highly non-linear optical responses of the medium -- specifically, high-harmonic generation (HHG) -- as a counterpart to recent proposals based on linear-optical properties~\cite{KortKamp2017, Tran2017}.

HHG is a highly non-perturbative process in which a material medium (which may be an atom, molecule or solid) is driven by an intense laser field, producing {\color{alexis} extreme ultraviolet (XUV)} photons as harmonics of the driving frequency. The XUV emission can carry information about the structure and dynamics of the medium, as first proposed for molecules by Itatani et al.~\cite{Itatani2004} (for a recent review of various processes in strong laser fields, see Ref.~\cite{Symphony2019}).~In HHG, an ultrashort (5-50 fs) intense mid-infrared (MIR) laser pulse causes partial ionization of an electron in an atom or molecule; the resulting electronic wave packet is accelerated in the laser field, returns to and recombines with the parent ion, producing high-order harmonics~\cite{Corkum1993, Lewenstein1994}. The efficiency of this process depends directly on the atomic or molecular orbital that the electron leaves and recombines with.~The electronic and molecular structure of the target can similarly be probed by the re-scattered electron (without recombination) via laser-induced electron diffraction~\cite{Zuo1996, JensScience}.

In the last few years, the subject of HHG from solid-state targets has attracted considerable attention~\cite{GhimireNatPhy2011, VampaJPB2017, EOsikaPRX2017}.
In particular, Vampa and co-workers investigated the role of intra-band and inter-band currents in ZnO experimentally, employing HHG to characterize structural information such as the energy dispersions~\cite{VampaNat2015, VampaPRB2015, VampaPRL2015} from their nontrivial coupling~\cite{GoldePRB2008}, where inter-band dynamics governed the emission. However, depending on material structure, Golde {\it et al.}~\cite{GoldePRB2008} 
also {\color{alexis}showed that the} intra-band mechanism might govern over the inter-band. Additionally, real-space approaches that, in some regimes, discard the band-structure picture altogether have recently been developed~\cite{Lakhotia2020}. The latter shows that inter- and intra-band mechanisms cannot trivially be disentangled, and their natures and relationship remain a matter of intense discussion in ultrafast science.

Generally, existing studies have dealt with standard materials, where the topology does not play a role.
Nevertheless, Berry-phase effects have been explored in some to\-po\-lo\-gi\-cally-trivial materials:
this includes experimental studies of HHG in atomically-thin semiconductors~\cite{Liu2017} and in quasi-2D models~\cite{Luu2018}, which show the sensitivity of harmonic emission to symmetry breaking (specifically, the breaking of inversion symmetry in monolayer MoS$_2$ and $\alpha$-quartz, via the appearance of even harmonics), 
as well as the effects of Berry curvature on the HHG spectrum of a time-reversal-invariant material~\cite{Avetissian2020}.
On a broader outlook, recent works have focused on HHG in 
graphene~\cite{AlNaib2014, Dimitrovski2017, Yoshikawa2017, Plaja2018}, 
monolayer transition-metal dichalcogenides~\cite{Tamaya2017, Yoshikawa2019}, 
and strained MoS$_2$~\cite{Guan2019,Jia2020},
as well as bilayer graphene~\cite{Avetissian2020bilayer}.
HHG from strongly-correlated materials has also been investigated recently, with work probing dynamical Mott-insulator transitions~\cite{SilvaNatPhoton2018} as well as strongly-correlated spin systems~\cite{Takayoshi2019}.

{\color{alexis} Very} recent work, however, has started to consider more directly the question of whether HHG is sensitive to topological order, 
starting from the contributions of edge states in a 1D-chain model of the Su-Schrieffer-Heeger (SSH) system~\cite{SSH}, which can significantly influence the harmonic emission~\cite{Bauer2018, Drueke2019,Juerss2019}.
In particular, recent work has also shown that the topological transitions in the Haldane model -- a paradigmatic Chern insulator -- leave explicit traces in the helicity of the emitted harmonics~\cite{Misha2018}.
Further work has also probed HHG in 
carbon nanotubes~\cite{DeVega2020},
few-layer~\cite{Jia2019}
and 3D~\cite{Avetissian2018} topological insulators,
and in materials with nodal topology~\cite{Lee2019}.

\begin{figure*}[t]
\begin{center}
\includegraphics[width=0.99\textwidth,height=0.33\textheight]{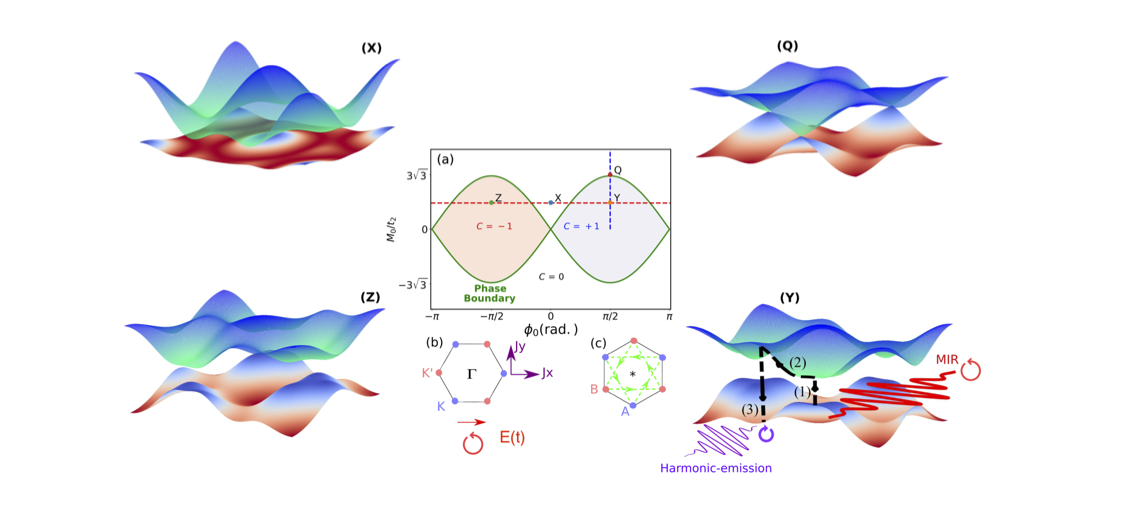}
\caption{%
Phase diagram and band structures of the Haldane model.
(a) Phase diagram in the plane ($\phi_0,\,M_0/t_2$), with the different phases labelled by their Chern number $C$; the green line shows the boundary of the topological phase transition.
The band-structure diagrams correspond to the points X, Y, Z and Q as marked in (a), with the point Q at the phase transition showing the gapless Dirac cone at the $K$ point.
(Y) and (Z) depict the band structure when the conduction-band topological invariants are $C=\pm1$ at the phase points $\phi_0=\pm\pi/2$ and $M_0=2.54t_2$. (Y) shows how the mid-infrared laser-source oscillations (red-solid line) drive the topological material; this can be with both linear and circular polarizations.
We also depict a physical cartoon of the electron-hole pair dynamics driven {\color{alexis}by a linearly polarized} laser, i.e.\ creation, propagation and annihilation or recombination by the black-dashed lines with arrows, and finally the subsequent harmonic emission (violet oscillations).~The dashed lines in (a) indicate the cuts used for the parameter scans below.~Panels (b) and (c) show the Brillouin zone and the real-space lattice of the Haldane model with the couplings in use.}
\label{figure1}
\end{center}
\end{figure*}

Interestingly, Tran {\it et al.}~\cite{Tran2017} considered Chern insulators in {\color{alexis}an} ultra-cold atom system in circularly shaken lattices, corresponding to  impinging  a circularly polarized driver.~They demonstrated  that the depletion rates  corresponding to {\color{alexis}left- and right-polarized drivers show} dependence on the orientation of the circular shaking.~This~{\it circular dichroism} was then directly related to the topological invariant, the so-called Chern number. The latter was recently measured in a cold atom experiment~\cite{Asteria_2018}. 
In this work we show that the circular dichroism produced with HHG can be used to herald the topological phase transition in a 2D topological Chern insulator material, {\color{alexis}analogous} to its role in magnetic materials~\cite{Zhang2018}.
We demonstrate this in detail for the paradigmatic Haldane model (HM, illustrated in Fig.~\ref{figure1})~\cite{Haldane1988}.~We derive, verify and apply the theory of HHG driven by both linearly and circularly polarized light in a two-band model, fully including the effects of the Berry curvature and its associated connection.
We characterize and analyze the HHG spectrum, via both the intra- and inter-band contributions, and using both numerical and analytical approaches.

Moreover, we develop a theoretical formalism which is gauge invariant with respect to both the electromagnetic and the Bloch wavefunction gauges~\cite{YueGaarde2020}. In topological phases, no continuous single-gauge charts exist~\cite{Kohmoto1985} which would allow a correct and regular description of dipole transition matrix elements over the entire Brillouin zone (BZ): for every gauge, there is at least one point in $k$-space where the dipole moment (and with it the Berry connection) has a discontinuity.
To deal with this inevitable discontinuity, we develop a multi-gauge approach using separate Bloch-gauge charts for different regions of the BZ. 
We also perform a crystal-momentum saddle-point analysis of the obtained expressions for the inter-band and intra-band currents. This allows us to derive expressions for the semi-classical electron-hole trajectories --- including the effect of the band topology and Berry curvature and connection --- and also to preserve gauge invariance (in contrast to e.g.\ Ref.~\cite{Yang2013}) accounting explicitly for the phase of the transition dipole moments.

Our theoretical approach {\color{alexis}provides} a complete model, which predicts the following:~(i) reported features of experiments for both linearly-and circularly-polarized driving lasers~\cite{Liu2017, Luu2018, Saito2017}, and novel behaviors for topologically non-trivial systems;
(ii) that HHG is extremely sensitive to inversion symmetry (IS) and, in addition, to breaking time-reversal symmetry (TRS); and
(iii) the HHG spectrum intensity depends on the topological phases, and the circular dichroism of high harmonic orders is sensitive to the crossing of a topological phase transition boundary.
{\color{alexis} Our work is mainly focused on the Haldane model, but we expect that the broad features are generally applicable to {\it Chern insulators}}.

After this Introduction {\color{alexis}we review the Haldane model} in Section~\ref{sec:haldane-model}.
Next, in Section~\ref{sec:dipole_discont}, we describe in detail how to handle the singularities of the dipole transition moment which are required to appear in topologically-nontrivial phases. 
In Section~\ref{sec:SBEs} we discuss the semiconductor Bloch equations and methods of their treatment, as well as our observables, the inter- and intra-band charge currents. 
Section~\ref{sec:KeldyshSFA} is devoted to the discussion of the Keldysh approximation, as well as a quasi-classical analysis based on the saddle point approximation applied to the gauge-invariant action, from which we can analyze the effect of the topology on the electron-hole trajectories. 
In Section~\ref{sec:HHG-in-practice} we begin to examine applications of our theory and consider the first HHG spectra in practice, as a benchmark for our theory against known results for $\alpha$-quartz. 
We go on, in Section~\ref{sec:signatures}, to discuss the question whether HHG can provide signatures of topological phases in any sense. 
We argue that the answer to this question is positive, and this is particularly spectacular when we consider circular dichroism, which we discuss in detail in Section~\ref{sec:dichroism}. 
Additional support for our results can be gathered by analyzing the effects of dephasing on HHG spectra: the harmonics that exhibit strong dichroism are particularly robust with respect to dephasing, which we explore in Section~\ref{sec:dephasingt2}. 
Finally, we conclude and present an outlook in Section~\ref{sec:conclusions}.

\section{The Haldane model}
\label{sec:haldane-model}
The Haldane model (HM)~\cite{Haldane1988}, originally introduced as a toy model, represents the first example of an anomalous quantum Hall effect~\cite{Cooper2018}, and it captures the essential features of a number of materials~\cite{bernevig2006, konig2007} --- in particular, the quantized transverse (spin) conductivities of the quantum spin Hall effect~\cite{Kane2005}, where the transport {\color{alexis} takes place} along protected edge states.
Moreover, this model remains solvable and implementable in quantum simulators~\cite{Jotzu2014}, which makes it a flexible tool for understanding a wide range of phenomena.
Specifically, the HM describes a tight-binding Hamiltonian of spinless fermions on a 2D hexagonal lattice with a real {\color{alexis}nearest-neighbor hopping~$t_1$}, an on-site staggering potential $M_0$, and a complex {\color{alexis}next-to-nearest-neighbor hopping} $t_2e^{i\phi_0}$. The HM belongs to the class of Chern insulators~\cite{Ching-Kai2016} and is characterized by a topological invariant, i.e. the Chern number~{\color{alexis}(see Eq.~\eqref{eqn:ChernNumber}).}
In this section, {\color{alexis}we provide the details of} the Hamiltonian, its eigenvalues and eigenvectors, its Berry connection, the Berry curvature and Chern number that derive from it, and the transition dipole moments, as well as the relationships between these quantities.

\subsection{Hamiltonian for {\color{alexis} the} Haldane model}
The HM is described by a tight-binding Hamiltonian of spinless fermions with a real nearest-neighbor (NN) hopping $t_1$, a complex next-nearest-neighbor hopping (NNN) $t_2e^{i\phi_0}$, and an on-site staggering potential $M_0$. 
The Hamiltonian in momentum space reads
\begin{align}
\label{HaldaneHamiltonian}
H_0({\bf k})
&=
B_{0,{\bf k}} \, \sigma_0
+{\bf B}_{\bf k}\cdot {\bm \sigma},
\end{align}
where the set of ${ B}_{0,\bf k}$ and ${\bf B}_{\bf k}=\{B_{1,{\bf k}},B_{2,{\bf k}},B_{3,{\bf k}}\}$ {\color{alexis}is} known as a ``pseudomagnetic field'' describing the tight-binding components of the Hamiltonian in momentum space, which read
{\allowdisplaybreaks%
\begin{subequations}%
\begin{align}
B_{0,{\bf k}}
& =
2t_2\cos\phi_0\sum_{i=1}^3 \cos(\mathbf{k} \cdot \mathbf{b}_{i}),
\\
B_{1,{\bf k}}
& =
t_1 \sum_{i=1}^3 \cos(\mathbf{k} \cdot \mathbf{a}_{i}),
\\
B_{2,{\bf k}}
& =
t_1 \sum_{i=1}^3 \sin(\mathbf{k} \cdot \mathbf{a}_{i}),   
\\
B_{3,{\bf k}}
& =
M_0-2t_2 \sin\phi_0\sum_{i=1}^3 \sin(\mathbf{k} \cdot \mathbf{b}_{i})
.
\end{align}
\label{eqn:HamCohef}%
\end{subequations}%
}%
Here the ${\bf a}_i$ are the NN vectors of the lattice (from atom A to B, see Fig.~\ref{figure1}(c) for more details), 
${\bf a}_1=\left(0,a_0\right)$, 
${\bf a}_2=\tfrac12 \left(-{\sqrt{3}},-1\right)a_0$ and 
${\bf a}_3=\tfrac12\left(\sqrt{3},-1\right)a_0$, 
and the ${\bf b}_i$ are the NNN vectors (from atom A to A, or B to B) given by 
${\bf b}_1=\left(\sqrt{3},0\right)a_0$, 
${\bf b}_2=\tfrac12\left(-\sqrt{3},+3\right)a_0$ and 
${\bf b}_3=\tfrac12 \left(-\sqrt{3},-3 \right)a_0$, 
with $a_0$ denoting the distance between the atoms A and~B.
In \eqref{HaldaneHamiltonian}, moreover, $\sigma_0$ is the identity matrix, and ${\bm\sigma }=\{\sigma_1,\sigma_2,\sigma_3\}$ is the vector of Pauli matrices, which is taken on a reference frame such that the eigenstates of $\sigma_3$ are Bloch waves that are localized to sites A and B, respectively, within each unit cell.
The symmetries of the model are obtained as follows: (i) a non-zero magnetic flux $\phi_0$  breaks the time-reversal symmetry (TRS) and (ii)~a non-zero staggering potential $M_0$ breaks the inversion symmetry  (IS).
\vspace{-0.15cm}

\subsection{Eigenvalues and eigenvectors}
As the Hamiltonian is described by a $2\times 2$ matrix in the Bloch basis, the eigenvalues and the eigenvectors of the Hamiltonian can be found analytically, using standard tools: the Hamiltonian is essentially a Pauli matrix along the ${\widehat{\bf B}}({\bf k}):= {\bf B}_{\bf k}
\big/|{\bf B}_{\bf k}|$ direction in the same space as the~$\bm\sigma$ Pauli vector, and its eigenvectors are the corresponding spin states as usual.
The energy spectrum thus reads
\begin{equation}
\varepsilon_{c/v}({\bf k}) = B_{0,{\bf k}} \pm |{\bf B_k}|, \label{eqn:EigenE}
\end{equation}
with the upper (lower) sign corresponding to the conduction (valence) band.
We define also, for future use, the polar coordinates $\phi_{\bf k}$ and $\theta_{\bf k}$ of ${\widehat{\bf B}}({\bf k})$, as
$
\tan \phi_{\bf k} :=
\frac{\widehat{B}^{\ }_{2,\bf k}}{\widehat{B}^{\ }_{1,\bf k}}$ and
$
\cos \theta_{\bf k} := \widehat{B}^{\ }_{3,\bf k}
$.

The eigenvectors of $H_0$ at crystal momentum $\bf k$ are the Bloch waves
\begin{equation}
|\Phi_{m,{\bf k}}\rangle
=
e^{i{\bf k}\cdot{\bf r}}
|u_{m,{\bf k}}\rangle
,
\end{equation}
{\color{alexis}where $|u_{m,{\bf k}}\rangle$ are the lattice-periodic functions which are given, in the basis of A- and B-Bloch waves} that the Hamiltonian~\eqref{HaldaneHamiltonian} acts in, by
{\allowdisplaybreaks %
\begin{align}
|u_{+,{\bf k}}\rangle
& =
\frac{1}{N_{+,{\bf k}}}
\begin{pmatrix}
B_{3,{\bf k}} +  |{\bf B}_{\bf k}|
\\
B_{1,{\bf k}} +  {\mathrm i}B_{2,{\bf k}}
\end{pmatrix}
=
\begin{pmatrix}
e^{-\mathrm{i}\phi^{\ }_{\bf k}/2 }
\cos\frac{\theta^{\ }_{\bf k}}{2}
\\
e^{+\mathrm{i}\phi^{\ }_{\bf k}/2}
\sin\frac{\theta_{\bf k}}{2}\label{eq:SpectralDecompositionH0a}
\end{pmatrix}
\\
|u_{-,{\bf k}}\rangle
& =
\frac{1}{N_{-,{\bf k}}}\begin{pmatrix}
{\mathrm i}B_{2,{\bf k}} -  B_{1,{\bf k}}
\\
B_{3,{\bf k}} +  |{\bf B}_{\bf k}|
\end{pmatrix}
=
\begin{pmatrix}
e^{-\mathrm{i} \phi^{\ }_{\bf k}/2}
\sin\frac{\theta^{\ }_{\bf k}}{2}
\\
-e^{+\mathrm{i}\phi^{\ }_{\bf k}/2}
\cos\frac{\theta^{\ }_{\bf k}}{2}\label{eq:SpectralDecompositionH0b}
\end{pmatrix},
\end{align}%
}%
with ${N_{\pm,{\bf k}}}$ a suitable normalization constant.
As usual, these eigenstates are only defined up to a (possibly $\bf k$-dependent) phase. Changes to this phase are the Bloch gauge transformations that we will examine in depth in Section~\ref{sec:dipole_discont}.~A gauge transformation of the wavefunction~{\color{alexis}(\ref{eq:SpectralDecompositionH0a}) and (\ref{eq:SpectralDecompositionH0b}) can} be performed by a unitary transformation and would lead to a new vector ${\bf B_{\bf k}}'=\{B_{1,{\bf k}}',B_{2,{\bf k}}',B_{3,{\bf k}}'\}$ (for example ${\bf B}' = \{B_2,B_3,B_1\}$).  Such a transformation keeps invariant the energy dispersions and Berry curvatures, but modifies the Berry connection and the transition dipole moments.
This gauge control will be implemented numerically to manipulate the discontinuities and singularities of the dipole matrix element and Berry connections in the BZ of topological phases.

\subsection{Berry Connection, Berry Curvature, and Chern number}
To calculate the radiation-interaction properties of these eigenstates, we require the dipole transition matrix elements between them. {\color{alexis} These can be singular functions of~$\bf k$~\cite{Blount1962,VladPRBVG_LG0} and read},
\begin{align}
\langle \Phi_{m',{\bf k}'}| {\bf x}|\Phi_{m,{\bf k}}\rangle
& =
-\mathrm{i}\nabla_{\bf k} \left( \delta_{m'm}\delta({\bf k}-{\bf k}')\right) 
\nonumber \\ & \qquad \qquad 
+ \delta({\bf k}-{\bf k}'){\bf d}_{m'm}({\bf k})
,
\label{eqn:Blount}
\end{align}
where
\begin{align}
{\bf d}_{m'm}({\bf k}) &= \mathrm{i}\langle u_{m',{\bf k}}| \nabla_{\bf k}|u_{m,{\bf k}}\rangle
\label{eqn:DME}
\end{align}
is a regular function which encodes the momentum gradient of the periodic part of the Bloch functions.
In Section~\ref{sec:dipole_discont} below we will examine in detail the off-diagonal matrix elements ${\bf d}_{cv}({\bf k})$, which must always exhibit singularities when the material is in the topological phase, but for now we focus on the diagonal elements.
In 1D systems with trivial topology, the diagonal elements of the transition dipoles can always be made to vanish, but in higher dimensionality this is impossible: these diagonal elements give the Berry connection of the band,
\begin{align}
{\bm\xi}_m(\mathbf {k})& = {\bf d}_{mm}({\bf k})
\nonumber \\
& =
\mathrm{i}\langle u_{m,{\bf k}}| \nabla_{\bf k}|u_{m,{\bf k}}\rangle,
\label{eqn:BerryConnection}
\end{align}
which is responsible for the parallel transport of wavefunction phase around the band.
This parallel transport is measured by the Berry curvature, given by the gauge-invariant curl
\begin{equation}
{\bm \Omega}_{m}({\bf k})
= \grad_{\bf k} \times {\bm \xi}_{m}({\bf k})
\end{equation}
of the connection, whose integral over the entire band,
\begin{equation}
C_m
\coloneqq
\tfrac{1}{2\pi} \int_\mathrm{BZ} {\bm \Omega}_m(\mathbf k) \mathrm\cdot d^2\mathbf k
,
\label{eqn:ChernNumber}
\end{equation}
is the topological invariant of the system, known as the Chern number.

For the specific case of the Haldane-model eigenstates with the gauge fixed by Eq.~(\ref{eq:SpectralDecompositionH0a},\ref{eq:SpectralDecompositionH0b}), the Berry connection is given explicitly by
\begin{equation}
{\bm \xi}_{v/c}({\bf k})
=
\mp \frac{1}{2}(\cos\theta_{\bf k})(\nabla_{\bf k}\phi_{\bf k}),
\label{eqn:BConn1}
\end{equation}
so that the Berry curvature reads
\begin{equation}
{\bm \Omega}_{v/c}({\bf k})
= \grad_{\bf k} \times {\bm \xi}_{v/c}({\bf k})
= \mp \frac{1}{2}(\nabla_{\bf k}\cos\theta_{\bf k})\times(\nabla_{\bf k}\phi_{\bf k}),
\label{eqn:BCurva1}
\end{equation}
and the Chern number results
\begin{equation}
C_{c/v}
=
 \mp  \frac{1}{4\pi}\int_{\rm BZ} d^2{\bf k}\,\cdot\,\big[\sin\theta_{\bf k}(\nabla_{\bf k}\theta_{\bf k})\times(\nabla_{\bf k}\phi_{\bf k})\big].
\label{eqn:ChernNo1}
\end{equation}
In the HM, the Chern number of the conduction band $C_c = C$ can take the values of $-1$, $0$ and $+1$ depending on $M_0$ and $\phi_0$ (see Fig.~\ref{figure1}). The trivial phase has a zero Chern number and the two topological phases are characterized by Chern numbers $C=\pm1$.
Figure~\ref{figure1}(a) depicts the phase diagram in terms of $\phi_0$ and $M_0$, showing three distinct phases: a trivial phase with Chern number $C=0$ and two different topological phases, with Chern numbers $C=\pm 1$. At the boundary between the phases, the bandgap closes, as shown for the point~(Q).

\subsection{Dipole moment and the Berry curvature in the Haldane model}
We now turn to the dipole matrix element, as initially defined in Eq.~\eqref{eqn:DME}, and which is here given by
\begin{equation}
{\bf d}_{cv}({\bf k})
=
\frac{1}{2} \left[ (\sin\theta_{\bf k}) (\nabla_{\bf k}\phi_{\bf k}) + i \nabla_{\bf k}\theta_{\bf k} \right].
\label{eqn:DME1}
\end{equation}
On the one hand, we  find that the cross product of the HM dipole Eq.~\eqref{eqn:DME1} yields
\begin{align}
\Im(d_{cv}^{x}d_{cv}^{y*})
& =
\frac{1}{4}\sin \theta_{\bf k}
\left[
    \partial_{k_y}\!\theta_{\bf k} \ \partial_{k_x}\!\phi_{\bf k}
    -\partial_{k_x}\!\theta_{{\bf k}} \ \partial_{k_y}\!\phi_{{\bf k}}
    \right].
\label{eqn:HMCrossDipole}
\end{align}
Expanding the Berry curvature in the HM, given by Eq.~\eqref{eqn:BCurva1}, one obtains
\begin{align}
{\bf \Omega}_{v/c}({\bf k})
& =
\mp \frac{1}{2}\sin\theta_{\bf k}(\nabla_{\bf k}\theta_{\bf k}\times\nabla_{\bf k}\phi_{\bf k})
\\
\nonumber 
& =
\mp \hat{\bf z}\frac{1}{2}\sin\theta_{\bf k}
\left[
    \partial_{k_y}\!\theta_{\bf k} \ \partial_{k_x}\!\phi_{\bf k}
    -\partial_{k_x}\!\theta_{{\bf k}} \ \partial_{k_y}\!\phi_{{\bf k}}
    \right].
\end{align}
We conclude, thus, that
\begin{equation}
\Omega_{v/c}  
= 
\mp 2\Im[d_{cv}^{(x)}d_{cv}^{*(y)}]
,
\end{equation}
which demonstrates the close relationship between dipole matrix elements and the Berry curvature in the HM. This  confirmation of the relation of the dipole product with the Berry curvature is extremely important, since this leads to a direct connection of the inter-band transition current of Eq.~\eqref{eqn:inter2} and the topological invariant for this model.

\section{Gauge-induced discontinuities in the transition dipoles}
\label{sec:dipole_discont}
In general, it is desirable for the phase in front of the eigenstates in Eq.~(\ref{eq:SpectralDecompositionH0a},\ref{eq:SpectralDecompositionH0b}), together with the Berry connections and the dipole matrix elements that derive from it, to be continuous over the entire Brillouin zone (BZ).
{\color{alexis}However, this is not always possible:~indeed, this is the {\it core distinction} between the topologically trivial and nontrivial phases}. In the nontrivial phase, it is provably impossible to find a {\color{alexis}Bloch wavefunction gauge} that will work smoothly for all momenta in the BZ~\cite{Kohmoto1985}.

In the trivial phase, on the other hand, globally-smooth gauges are viable~\cite{YueGaarde2020} -- but they are not guaranteed~\cite{VladPRBVG_LG0}, either, so that one must always be prepared to handle discontinuities and singularities in these quantities.
Moreover, as the {\color{alexis}semiconductor Bloch equations} (SBEs) directly incorporate these two quantities (see bellow~Eqs.~\eqref{eqn:SBEs1}~and~\eqref{eqn:SBEs2}), the numerical calculations of the currents can inherit those artificial singularities, leading to numerical instabilities and noise in the harmonic plateau or cut-off.~Here we show an example of how this numerical instability arises, and the method we use to solve it. This method relies on the control of the localization of the dipole discontinuity in the BZ. Such control is realized by choosing different gauges for different regions (in the language of differential geometry, for different charts~\cite{Spivak1999}) of the BZ.

\begin{figure*}[t]
\begin{center}
{
\setlength{\tabcolsep}{0mm}
\begin{tabular}{cccc}
\includegraphics[width=0.24\textwidth]{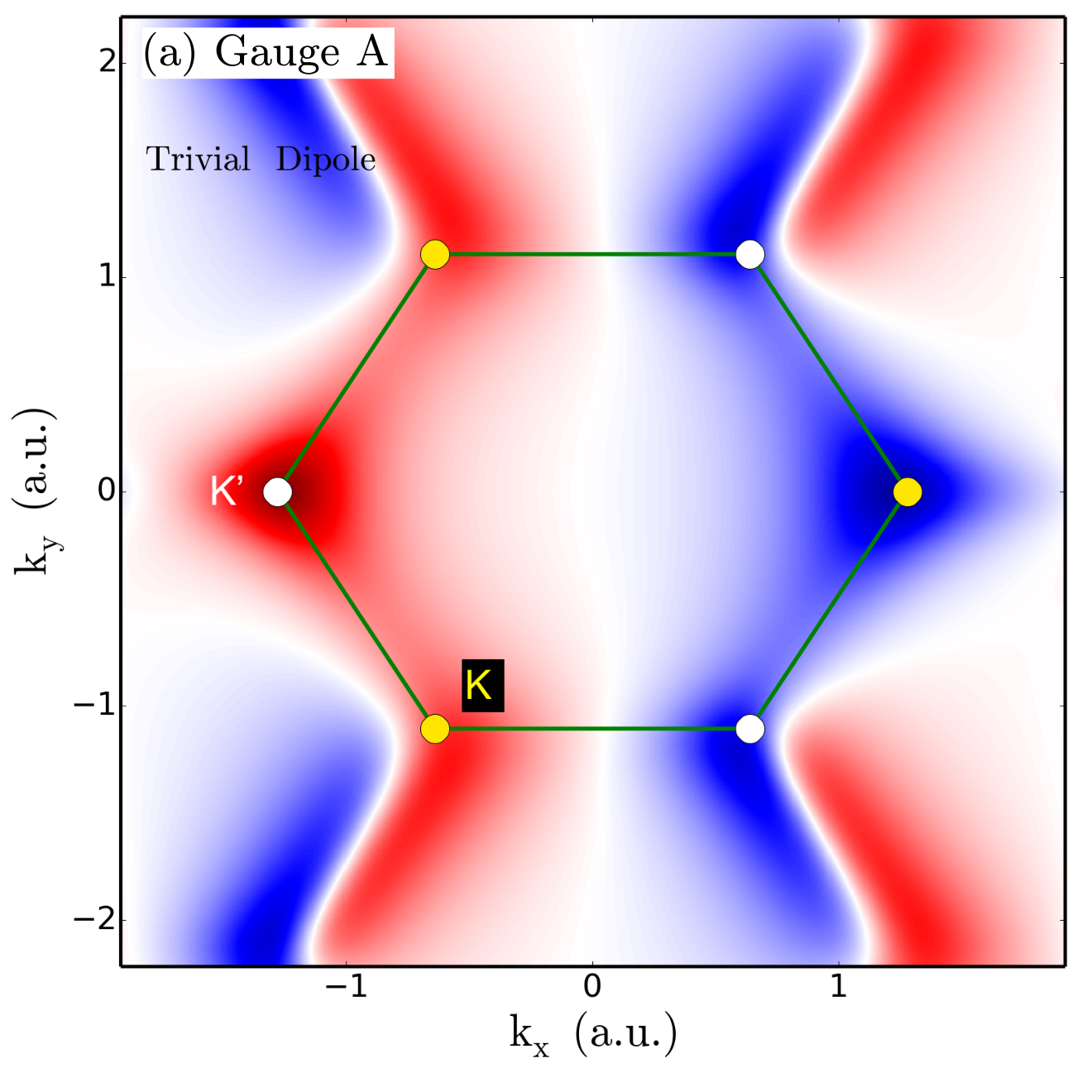} &
\includegraphics[width=0.24\textwidth]{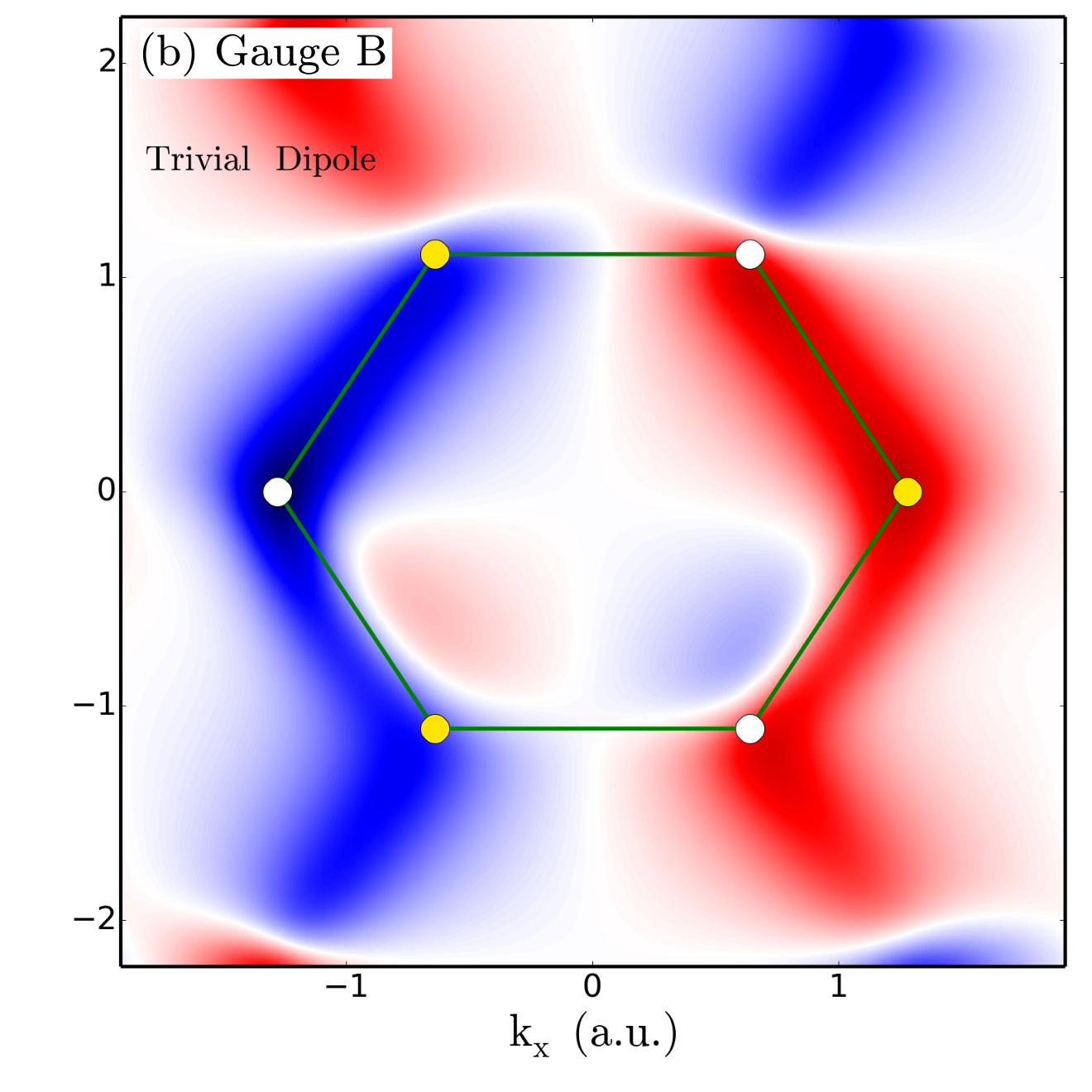} &
\includegraphics[width=0.24\textwidth]{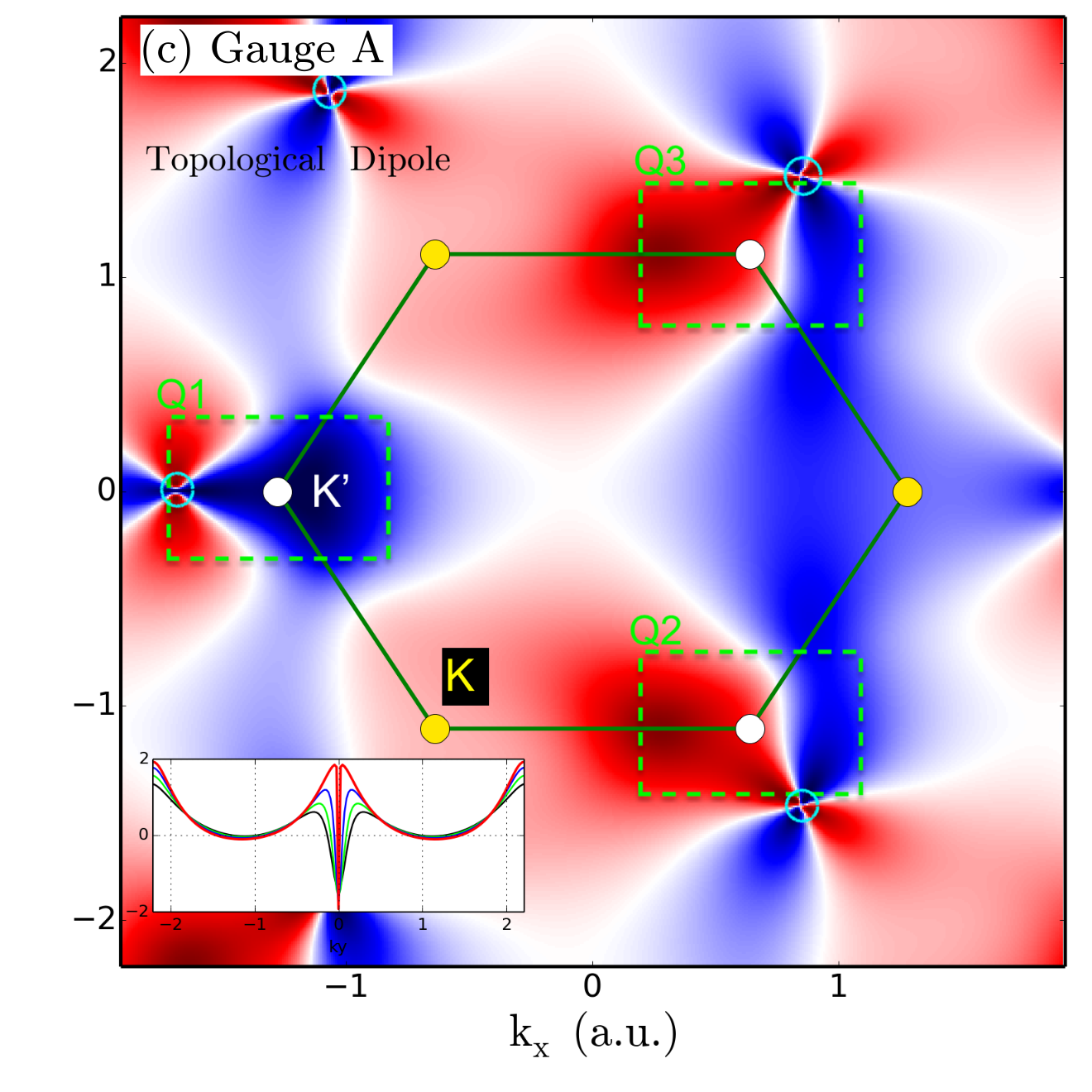} &
\includegraphics[width=0.24\textwidth]{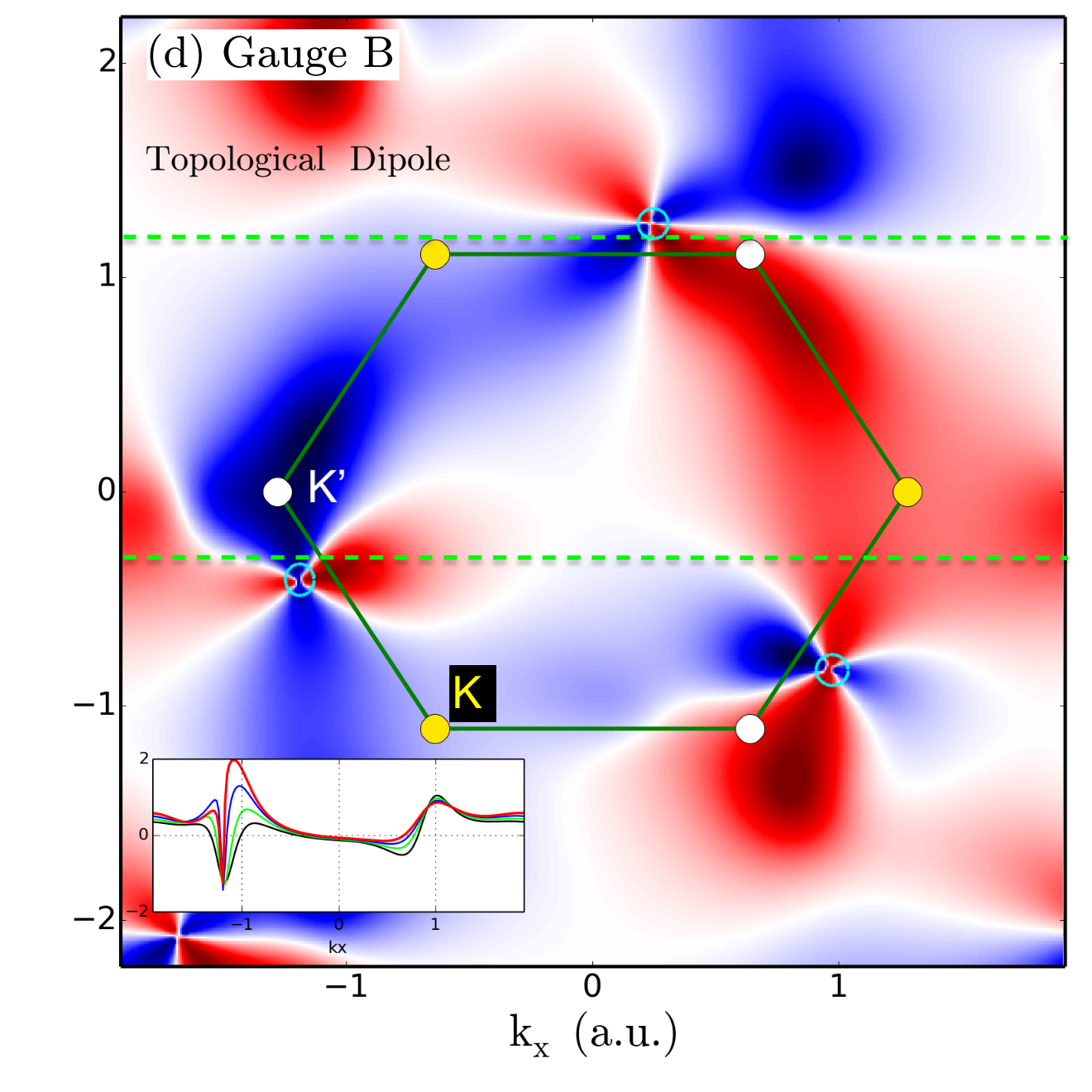} \\
\includegraphics[width=0.24\textwidth]{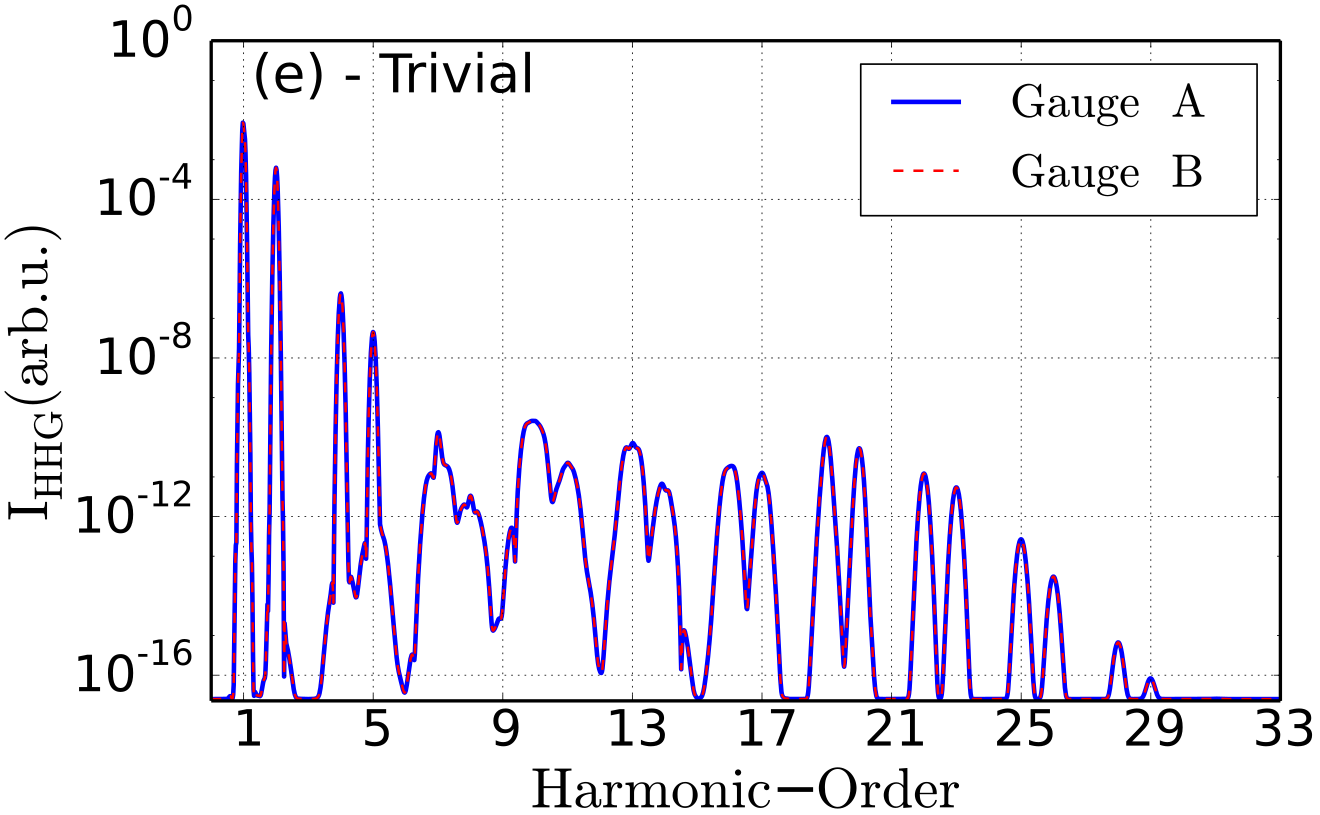} &
\includegraphics[width=0.24\textwidth]{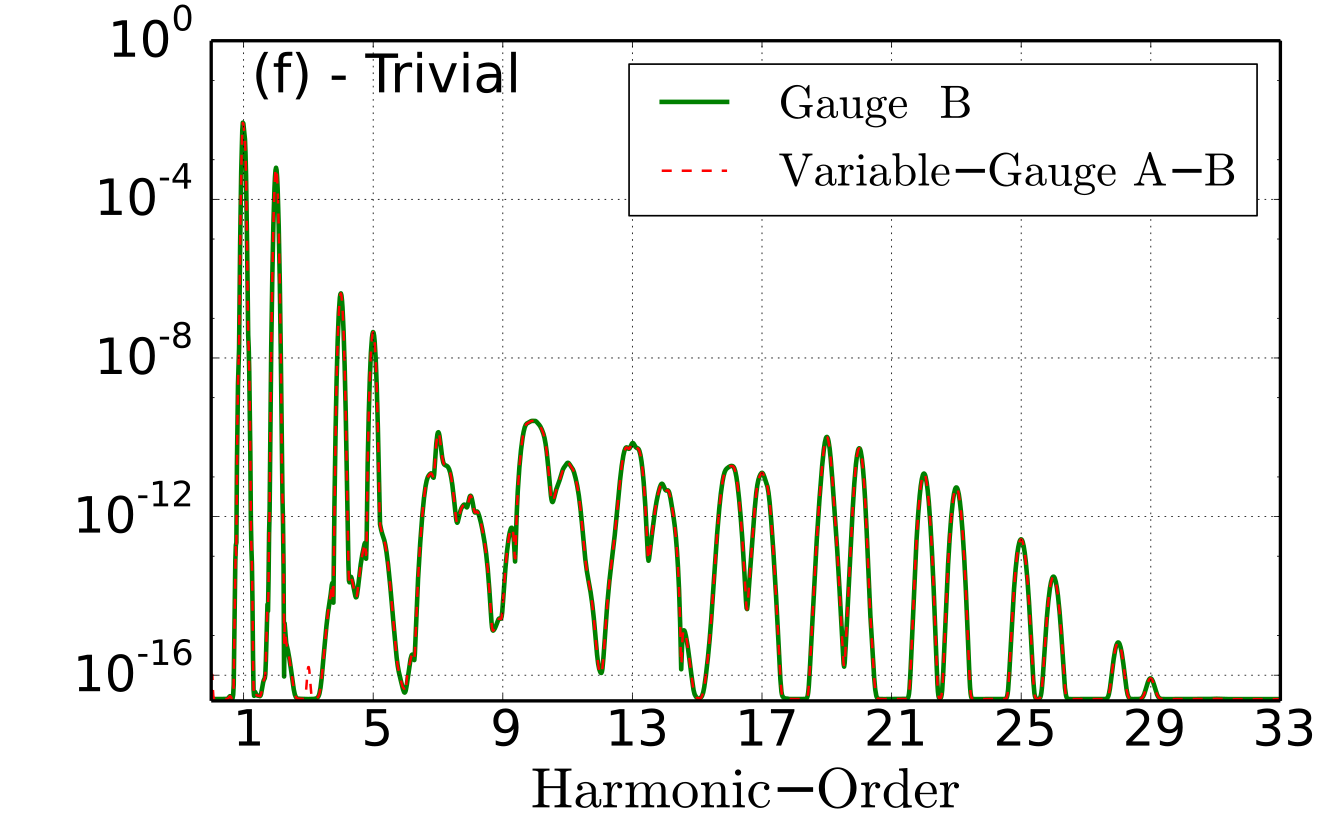} &
\includegraphics[width=0.24\textwidth]{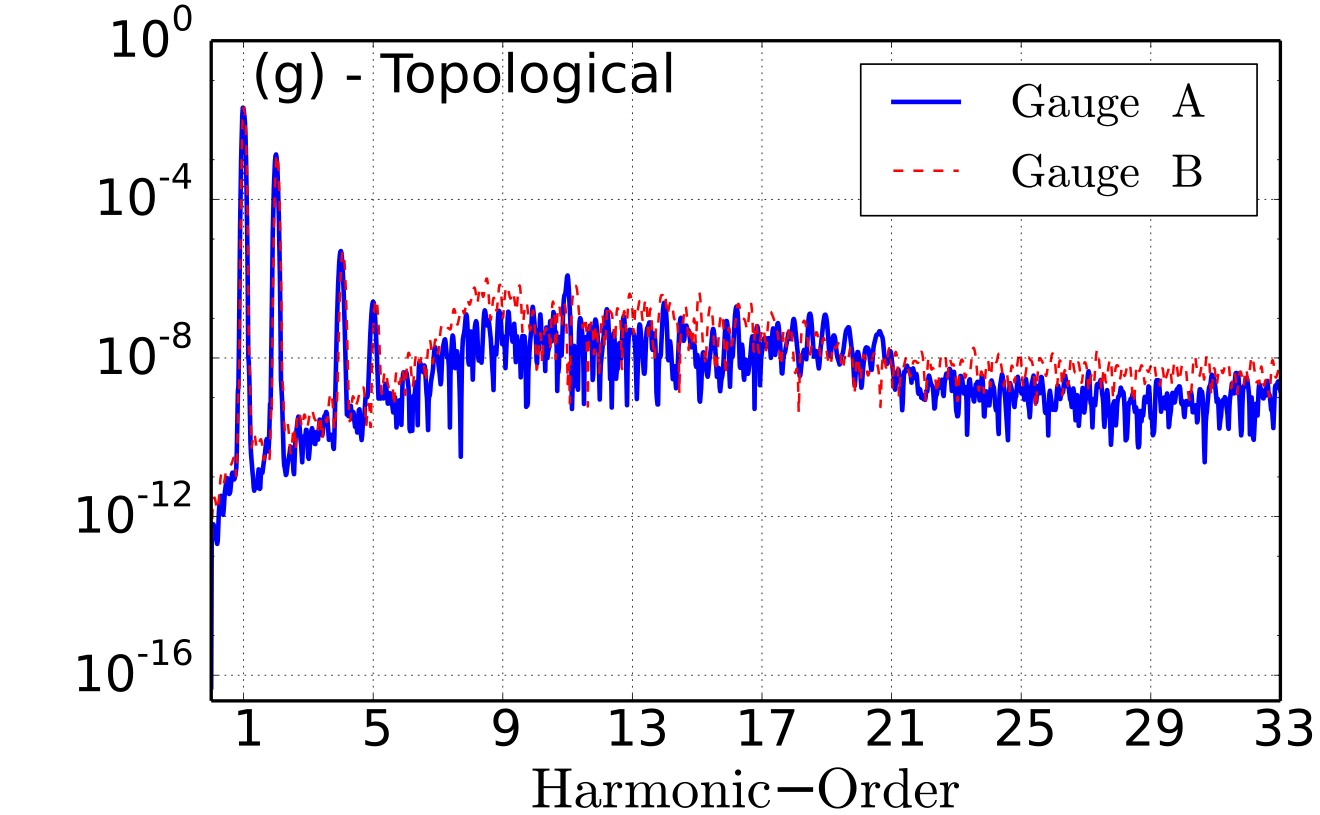} &
\includegraphics[width=0.24\textwidth]{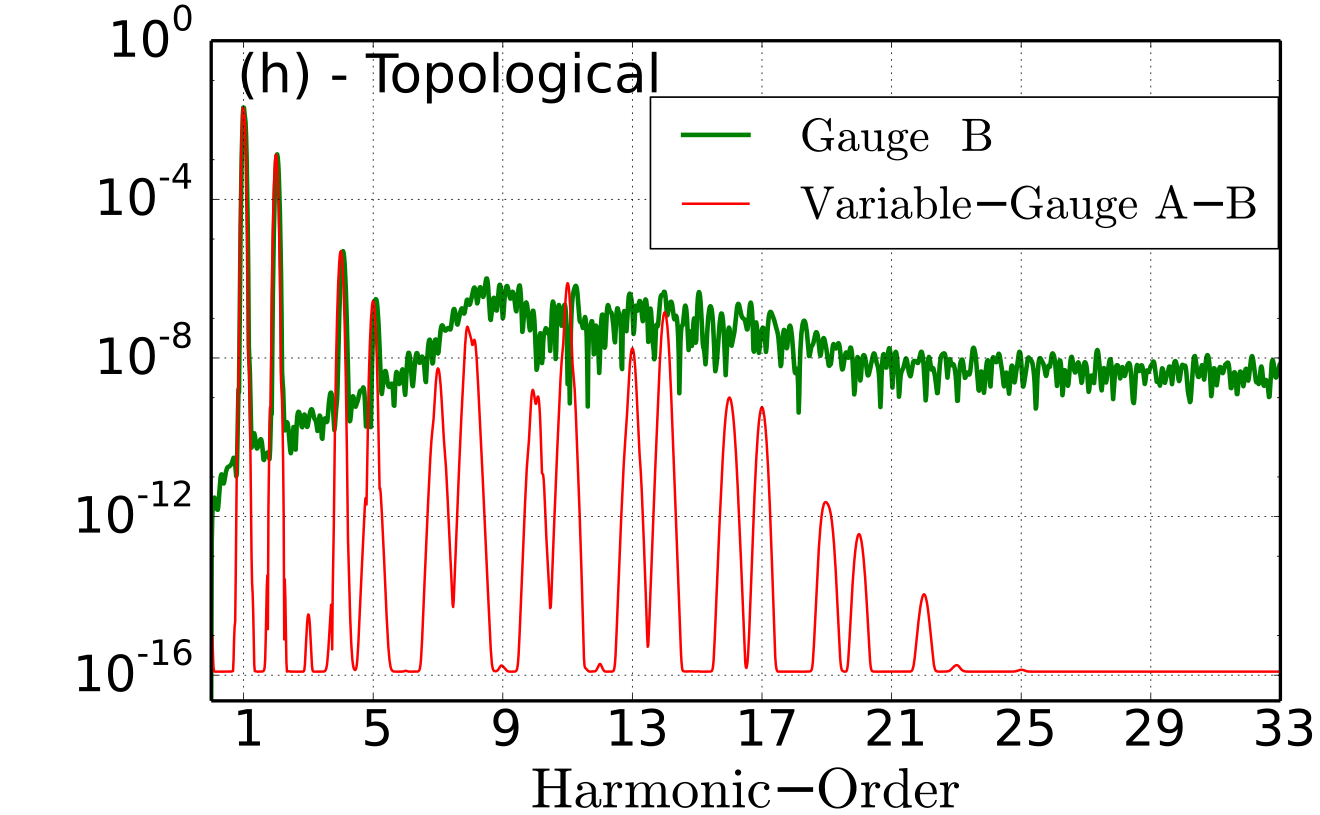}
\end{tabular}
}
\caption{%
(a-d) 
Dipole transition matrix elements over the Brillouin zone for trivial and topological phases for different gauges. We show the $x$ component in a color map ranging from dark blue to red, with white at zero.
The hexagon shows the first BZ, with the $\rm K$ ($\rm K'$) points denoted by the white (yellow) circles.
The insets in (c,d) show cuts through the singularities: 
(c) along $k_y$, at fixed $k_x=-1.85$~(black), $-1.80$~(green), $-1.75$~(blue) and $-\SI{1.70}{\au}$~(red), and 
(d) in vertical cuts at fixed $k_y=-0.55$~(black), $-0.50$~(green), $-0.45$~(blue) and $-\SI{0.40}{\au}$~(red).
The cyan circles highlight the discontinuities in the dipoles.
(e-h) 
The resulting HHG spectra under an LCP laser driver, for gauges A and B~(e,g) and our variable-gauge method~(f,h), for both phases.
We set $t_1=\SI{0.075}{\au}$, $t_2=t_1/3$, $M_0=\SI{0.0635}{\au}= 2.54t_2$ and $a_0=\SI{1}{\angstrom}$, giving a bandgap of $E_g=\SI{3.0}{eV}$, and we use a magnetic flux of
(a,b) $\phi_0=\SI{0.06}{rad}$, giving $C=0$, and
(c,d) $\phi_0=\SI{1.16}{rad}$, {\color{alexis}giving $C=+1$}.
For the HHG spectra we use the {\color{alexis}pulse~Eq.~\eqref{eq:alaser}} with $E_0=\SI{0.0045}{\au}$, $\omega_0=\SI{0.014}{\au}=\SI{0.38}{eV}$ (corresponding to $\lambda=\SI{3.2}{\micro m}$), and with a gaussian envelope lasting $N_c=14$ cycles at FWHM;
we use a dephasing time of $T_2=\SI{220}{\au}$ 
We integrate over a momentum grid of $N_x=873$ by $N_y=511$ points using a two-dimensional Simpson rule, with boundaries at $k_{x,{\rm max}} =\frac{2\pi}{\sqrt{3}a_0}$ and $k_{y,{\rm max}}=\frac{2\pi}{3a_0}$; for the SBE time integration our time step is $dt = \SI{0.1}{\au}$ 
}
  \label{DipoleMoments}
  \end{center}
\end{figure*}

\indent Figures~\ref{DipoleMoments}(a) and~\ref{DipoleMoments}(b) show the real part of the $x$ component of the dipole matrix element, $d_{cv}^{(x)}({\bf k})$, of a trivial topological phase, taken in two different gauges: A, where ${\bf B}'_{{\bf k},\rm A} =({B_{2,{\bf k}},B_{3,{\bf k}},B_{1,{\bf k}}})$, and B, where ${\bf B}'_{{\bf k},\rm B} = (B_{3,{\bf k}},B_{1,{\bf k}},B_{2,{\bf k}})$.
As can be seen in the figure, both gauges exhibit substantially different dipoles, but their corresponding {\color{alexis}high-harmonic emission} -- shown in Fig.~\ref{DipoleMoments}(e) -- are absolutely the same, i.e.\ {\it \color{alexis} the harmonic emission is gauge invariant},~as expected.~This is a simple and obvious test, but it demonstrates a powerful mathematical property that will help us to {\color{alexis}circumvent the numerical noise, which can} compromise our {\color{alexis}(or any)} interpretations of the harmonic emission from topological phases, {\color{alexis}if the singularities are not well handled or numerically integrated. Here, we find a simple method to solve this issue.}

\indent Figures~\ref{DipoleMoments}(c,d,g) {\color{alexis} show} similar results for a to\-po\-lo\-gi\-cally-nontrivial phase.~Most prominently, in the dipole maps of Figs.~\ref{DipoleMoments}(c) and \ref{DipoleMoments}(d),~we notice pronounced discontinuities in the dipole for both gauges, which are highlighted by the cyan circles {\color{alexis}(see inset plots)}.~Secondly, the HHG calculations for gauges A and B, shown in Fig.~\ref{DipoleMoments}(f), show numerical noise around the plateau {\color{alexis} at harmonic-orders (HOs) $>5^{\rm th}$} and, more troublingly, a lack of a well-defined cut-off.~These features signal a numerical instability of the calculation, which is independent of the numerical convergence of our 5\textsuperscript{th}-order Runge--Kutta method.

To solve this problem, we use the fact that the $k$-space discontinuities in the dipole can be shifted within the BZ (as shown in Figs.~\ref{DipoleMoments}(c) and \ref{DipoleMoments}(d)) by choosing different gauges: the existence of singularities is inevitable~\cite{Kohmoto1985}, but their location in $\bf k$-space is not fixed and it can be moved by switching phase conventions.~Therefore, we use different gauges to calculate the {\color{alexis}semiconductor Bloch equations (SBEs, see Eqs.~\eqref{eqn:SBEs1}-\eqref{eqn:SBEs2})} evolution in different patches of the BZ, and then coherently combine the currents at the end of the calculation.~Thus, in our variable-gauge technique, we apply the two gauges described above, A and B, for the different BZ regions where each gauge is regular, using the following three criteria:
\begin{itemize}
\item
Gauge A for the BZ region which obeys the constraints $-k_{y,{\rm max}}\leq k_y\leq\epsilon_{y,1}$, where $\epsilon_{y,1}=-0.35$~a.u. is marked with the horizontal green dashed line in Fig.~\ref{DipoleMoments}(d).

\item
Since all physical quantities are periodic within the whole BZ, the regions Q1, Q2 and Q3 (marked by green boxes in Fig.~\ref{DipoleMoments}(c)) are physically 
{\color{alexis}equivalent}, we use gauge~A shifted to the Q3 region when $k_y$ satisfies $ \epsilon_{y,1}< k_y\leq\epsilon_{y,2}$, where $\epsilon_{y,2}=0.02$~a.u.

\item
We use gauge B for momenta which satisfy $\epsilon_{y,2}<k_y\leq k_{y,{\rm max}}$.
\end{itemize}

\noindent We compute HHG with this variable-gauge method for both the trivial and topological phases.~For the trivial phase we show the resulting harmonic spectrum in Fig.~\ref{DipoleMoments}(h), where the red dashed line shows the variable-gauge  results; we find very good agreement between this approach and the fixed-gauge reference (using gauge B), which works well for the trivial phase, as discussed above, and thus serves as a benchmark for the variable-gauge approach.~{\color{alexis}Silva {\it et al.}~have developed an interesting alternative to circumvent the dipole matrix elements singularities in trivial and topological materials by integrating numerically the SBEs or the reduced density matrix, namely, the maximally-localized Wannier-approach (MLW) framework~\cite{SilvaPRB2019}.~Our {\it variable gauge approach} was compared to the MLW.~This variable-gauge Bloch wavefunction method perfectly agrees with MLW too.}

With this in place, we can confidently apply the same technique to the calculation of HHG for the topological phase, the results of which are shown in the red line of Fig.~\ref{DipoleMoments}(h).~The HHG spectrum for this topological phase shows a clean plateau with well-defined harmonics and a clear cut-off region. Moreover, there is a clear numerical threshold, in comparison to the fixed-gauge method using gauge B.
Our numerical threshold is fixed at {\color{alexis}$I_{\rm HHG}\approx10^{-16}$~arb.~units}, but we have found that this threshold can be reduced up to {\color{alexis}$10^{-25}$~arb.~units}\ for the HHG yield, depending on the parameters used for the BZ grid and the time integration.

That said, it is important to note that, while our variable-gauge method can map the plateau and cut-off very well for topological phases, our method finds an undefined harmonic cutoff, as shown in Fig.~\ref{figure4} below. This occurs at the critical point of the HM, where the bandgap closes, at the topological phase transition.~This behavior is expected, since the closing of the bandgap creates an undetermined dipole matrix element at the Dirac cone, where the population is most easily transferred to the conduction band.

\section{Semiconductor Bloch equations}
\label{sec:SBEs}
{\color{alexis}We introduce} and examine the equations governing the  microscopic time-dependent response of the solid due to the laser-matter interaction, the semiconductor Bloch equations~\cite{HaugKoch2004}.~We follow the approach of Refs.~\cite{GhimireNatPhy2011,VampaJPB2017}, generalizing it to include the effects of possible non-trivial topology.

\subsection{Semiconductor Bloch equations}
\label{SecSBEs}
If we write down the electronic-crystalline wavefunction in the form
\begin{equation}
|\Psi(t)\rangle 
= \sum_m \int_{\rm BZ} d{\bf k}\,\, a_m({\bf k},t)|\Phi_{m,{\bf k}}\rangle
,
\end{equation}
the time evolution of the Bloch-space probability amplitudes $a_m({\bf k},t)$, as determined by the Schr\"o\-din\-ger equation, can be transformed into the form~\cite{Blount1962}
\begin{align}
\mathrm{i} \dot{a}_{m}({\bf k}',t)
& = 
\varepsilon_{m}({\bf k}')a_{m}({\bf k}',t)
\label{eqn:A0} 
\\ & \qquad 
\nonumber
+ {\bf E}(t)\cdot\sum_{m'} \int_{\rm BZ}\langle \Phi_{m,{\bf k}'}|{\bf x}|\Phi_{m',{\bf k}}\rangle 
a_{m'}({\bf k}',t)
.
\end{align}
Here $\varepsilon_{m}({\bf k})$ denotes the energy dispersion for the valence/conduction band $m=v/c$, and $m'$ also ranges over both bands.
We then compute the second and third terms on the right-hand side of the above equation by using Eq.~(\ref{eqn:Blount}), and we find
\begin{align}
\dot{a}_{m}({\bf k},t)
&=
-\mathrm{i}
\big[
  \varepsilon_{m}({\bf k}) 
  + {\bf E}(t){\cdot}{\bm \xi}_{m}({\bf k})
  - \mathrm{i} {\bf E}(t)\cdot{\nabla_{\bf k}} 
  \big]
  {a}_{m}({\bf k},t)
\nonumber \\ \qquad
&-
\mathrm{i} 
{\bf E}(t)
\cdot
\sum_{m'\neq m} 
{\bf d}_{mm'}({\bf k}) \,
a_{m'}({\bf k},t)
. 
\label{eqn:A2}
\end{align}
This differential equation mixes time derivatives with the momentum gradient $\nabla_{\bf k}$, which also occurs for strong field approximation (SFA) treatment of atoms~\cite{Symphony2019}. Thus, we apply the same transformation, shifting the momentum over time to ${\bf K} = {\bf k} - {\bf A}(t)$ -- the canonical crystal quasi-momentum, which is a {\color{alexis}constant of motion} -- giving rise to a shifted Brillouin zone, denoted $\overline{\rm BZ}$.
More formally, by applying the substitution $a_m({\bf k},t) = e^{{\bf A}(t)\cdot{\nabla}_{\bf K}}\,b_m({\bf K},t)$, one finds that
\begin{align}
\label{eqn:B1}
\dot{b}_{m}({\bf K},t)
&=
-\mathrm{i} 
\big[
  \varepsilon_{m}({\bf K}+{\bf A}(t))
  \\ & \qquad \qquad 
  \nonumber
  + {\bf E}(t)\cdot{\bm \xi}_{m}({\bf K}+{\bf A}(t))
  \big]
  b_{m}({\bf K},t)
\nonumber \\ & \qquad 
- \mathrm{i} 
{\bf E}(t)
\cdot
\sum_{m'\neq m}
{\bf d}_{mm'}({\bf K}+{\bf A}(t)) \,
b_{m'}({\bf K},t)
.
\nonumber
\end{align}
We then transform the transition amplitude $b_m$ to the density matrix operator $\hat{\rho}$, i.e. the population $n_m=\rho_{mm}$ and coherence $\pi=\rho_{cv}$, given explicitly by $n_m({\bf K},t)=b_m^*({\bf K},t)b_m({\bf K},t)$, and $\pi({\bf K},t)=b_c({\bf K},t)b_v^*({\bf K},t)$. 
Thus, this leads to the so-called semiconductor Bloch equations (SBEs), which describe the laser-electron interaction in the lattice~\cite{GoldePRB2008}, in terms of the band population and coherence:
\begin{align}
\dot{n}_m({\bf K},t)
& =
\mathrm{i}
(-1)^m
{\bf E}(t)\cdot{\bf d}_{cv}^*({\bf K}+{\bf A}(t)) \,
{\pi}({\bf K},t) 
\nonumber \\ & \qquad 
+ \mathrm{c.c.},
\label{eqn:SBEs1}
\\
\dot{\pi}({\bf K},t)
& =
-\mathrm{i}
\bigg[
  \varepsilon_g({\bf K}+{\bf A}(t))
  + {\bf E}(t)\cdot{\bm \xi}_g({\bf K}+{\bf A}(t))
  \nonumber\\ & \qquad \quad 
  -\mathrm{i} \frac{1}{T_2} 
  \bigg]
  {\pi}({\bf K},t)
\nonumber\\ & \qquad 
-\mathrm{i} {\bf E}(t)\cdot{\bf d}_{cv}({\bf K}+{\bf A}(t))\,w({\bf K},t)
.
\label{eqn:SBEs2}
\end{align}
Here $\varepsilon_g({\bf k }) = \varepsilon_c({\bf k }) - \varepsilon_v({\bf k })$ is the energy gap between the conduction and valence bands, as a function of the crystal momentum $\bf k$, ${\bm \xi}_g({\bf k })= {\bm \xi}_{c}({\bf k }) - {\bm \xi}_{v}({\bf k })$ is the difference in the Berry connection between the conduction and valence bands, and the sign $(-1)^m$ takes the values $(-1)^c=1$ and $(-1)^v=-1$.
We also introduce a phenomenological dephasing term written in terms of the dephasing time~$T_2$~\cite{VampaJPB2017}.
Finally, $w({\bf k },t)= n_v({\bf k },t) - n_c({\bf k },t)$ denotes the momentum and time dependence of the population difference between the valence- and conduction-band or inversion population.

In our calculations, we describe the driving laser using the vector potential
\begin{align}
{\bf A}(t) 
=
\frac{E_0}{\omega_0}f(t)
\bigg(
\frac{1}{\sqrt{1+\epsilon^2}} 
\cos(\omega_0(t-t_0))\,{\bf e}_x 
\qquad & \nonumber \\
+
\frac{\epsilon}{\sqrt{1+\epsilon^2}} 
\sin(\omega_0(t-t_0))\,{\bf e}_y 
\bigg)
&
,
\label{eq:alaser}
\end{align}
where $E_0$ is the laser field strength, $\omega_0$ is the central photon frequency, and $f(t)$ is the laser-pulse envelope, given by $f(t) = \text{exp} ((t-t_0)^2/(2\sigma^2))$, where $\sigma$ is the pulse width.
Here $\epsilon$ denotes the ellipticity of the laser, given by $\epsilon=-1$ (+1) for right-handed (left-handed) circularly polarized laser fields, abbreviated (RCP) and (LCP), respectively, and by $\epsilon=0$ for linearly-polarized laser drivers.
The electric field is given by ${\bf E}(t) =-\partial_t {\bf A}(t)$.

We numerically solve the SBEs described by Eqs.~\eqref{eqn:SBEs1} and \eqref{eqn:SBEs2} with a 5\textsuperscript{th}-order \texttt{Runge--Kutta} scheme implemented in {\texttt{C++}}, using the Message Passing Interface (\texttt{MPI}) for parallelization.
Additionally, we validate our numerical findings in \texttt{C++} by comparing our numerical outcomes with the numerical solution obtained by the built-in solvers on \texttt{Wolfram}~\texttt{Mathematica}. We find very good agreement between the \texttt{C++} and~\texttt{Mathematica} results.

\subsection{Microscopic currents}
The HHG spectra are obtained by {\color{alexis}Fourier analyzing} microscopic charge currents induced by the driving laser pulse.  Here we derive the total microscopic current ${\bf J}(t)={\bf J}_\mathrm{ra}(t)+{\bf J}_\mathrm{er}(t)$, by focusing on its two components:  the intra-band ${\bf J}_\mathrm{ra}(t)$ and the inter-band ${\bf J}_\mathrm{er}(t)$ currents~\cite{GhimireNatPhy2011,VampaJPB2017}.

\subsubsection[B]{Inter-band current}
We start by writing the inter-band current ${\bf J}_\mathrm{er}(t)=\frac{d}{dt}{\bf P}_\mathrm{er}(t)$ in terms of the polarization
\begin{align}
{\bf P}_\mathrm{er}(t) 
& \equiv  
e\langle \Psi(t)| {\bf x} |\Psi(t)\rangle
\nonumber\\
&=
e
\int_{\rm BZ} \!\! d{\bf k}'\,
\int_{\rm BZ} \!\! d{\bf k}\,
a_{c}({\bf k}',t)\,a^*_{v}({\bf k},t)\langle \Phi_{v,{\bf k}'} |  {\bf x} |\Phi_{c,{\bf k}}\rangle
\nonumber \\ & \qquad 
+ \mathrm{c.c.}
,
\end{align}
where $e=\SI{-1}{\au}$ is the electron charge. 
Next, using the identities above, the current simplifies to
\begin{equation}
{\bf J}_\mathrm{er}(t)
=
e{\frac{d}{dt}} \int_{\rm  \overline{BZ}} {\bf d}_{cv}^{*}\left({\bf K}+ {\bf A}(t)\right)\pi({\bf K},t) \, d^2 {\bf K}
+\mathrm{c.c.},
\label{eqn:inter-SM}
\end{equation}
indexed by the constant canonical crystal quasi-mo\-men\-tum $\bf K$.
This result coincides with the ones found by Vampa {\it et al.} in Ref.~\cite{VampaPRL2014}.

\subsubsection{Intra-band current}
We write the intra-band current, defined as ${\bf J}_\mathrm{ra}(t)= e\sum_m \langle \Psi(t)| {\bm v}_m |\Psi(t)\rangle$, in terms of the intra-band velocity operator ${\bm v}_m = -\mathrm{i}\,[{\bm x}_m,H_0] -\mathrm{i}\,[{\bm x}_m,{\bm x}_m\cdot {\bf E}(t)]$. We use the electromagnetic length gauge with the single-active-electron Hamiltonian and the dipole approximation of the laser-crystal system, i.e.\ $H(t)=H_0+{\bf x}\cdot{{\bf E}(t)}$.
For simplicity, our derivation is focused on the conduction band $m=c$ of the intra-band current, which reads
\begin{align}
{\bf J}_\mathrm{ra}(t)
& =
e\int d{\bf k}' \int d{\bf k}\, a^*_{c,{\bf k}'}(t)a_{c,{\bf k}}(t)\,\langle \Phi_{c,{\bf k}'}|{\bm x}_{c}|\Phi_{c,{\bf k}}\rangle\varepsilon_c({\bf k})
\nonumber \\
& \qquad + (c\to v) 
\\ & =
\nonumber 
e\int d{\bf k}'\, \left[n_{c}({\bf k}',t)\,{\bf v}_{c}({\bf k}') + n_{v}({\bf k}',t)\,{\bf v}_{v}({\bf k}')\right].
\end{align}
Here, ${\bf v}_c({\bf k})={\bf v}_{\mathrm{gr},c}({\bf k}) + {\bf v}_{a,c}({\bf k})$ is written in terms of the group velocity ${\bf v}_{\mathrm{gr},c}({\bf k}) = \nabla_{\bf k}\varepsilon_c({\bf k})$, and the anomalous velocity ${\bf v}_{a,c}({\bf k})=-{\bf E}(t)\times{\bm \Omega}_c({\bf k})$, where $\bm\Omega_m({\bf k})=\nabla_{\bf k}\times{\bm\xi}_m({\bf k})$ is the Berry curvature.
The anomalous-velocity term comes from considering that the intra-band component ${x}_m^{(j)}$ of the position operator does not commute with ${x}_m^{(i)}$ component, with $i,j=x,y$ for 2D~\cite{Blount1962}.
This intra-band current also can be expressed on the shifted Brillouin zone $\overline{{\rm BZ}}$, as
\begin{equation}
{\bf J}_\mathrm{ra}(t)
=
e
\sum_m 
\int_{\rm \overline{BZ}} 
{\bf v}_{m}\mathopen{}\left({\bf K}+ {\bf A}(t)\right)\mathclose{}
n_m({\bf K},t) \,d^2 {\bf K}.
\label{eqn:intra-SM}
\end{equation}
{\color{alexis}To} calculate the HHG spectrum from a crystal, we have to evaluate Eqs.~(\ref{eqn:inter-SM}) and~(\ref{eqn:intra-SM}). Hence, our main task in the following will be to compute the coherence $\pi({\bf k},t)$ and the occupations $n_m({\bf k},t)$, which are related to the transition amplitude $a_m({\bf k},t)$.
In our numerical calculations, we determine the radiated power of each harmonic frequency by coherently superposing the time derivatives of the intra- and inter-band currents and then taking the squared modulus of its Fourier transform,
i.e.\ $I_{\rm HHG}(\omega)= \omega^2|{\rm FT}\left[\, {\bf J}_\mathrm{er}(t) +  {\bf J}_\mathrm{ra}(t)\,\right]|^2$.

\section{Keldysh approximation and quasi-classical analysis}
\label{sec:KeldyshSFA}
We may gain more insight into the physics of HHG  by applying the so-called Keldysh approximation, as  discussed for instance by Vampa {\it et al.}~\cite{VampaPRL2014}.
This approximation in a crystal solid reads: $w({\bf k},t)=n_v({\bf k},t)-n_c({\bf k},t) \approx 1$. Essentially, this means that the population transferred to the conduction band is very small compared to that remaining in the valence band. This approximation is very similar, {\color{alexis}but never equal}, to the one used in the Strong Field Approximation (SFA), developed originally for atoms and molecules~\cite{Corkum1993,Lewenstein1994,Symphony2019} -- we will thus term it SFA in the following.  We focus on the discussion of the inter-band current in this section.

\subsection{Inter-band current}
In  this approximation, Eq.~\eqref{eqn:SBEs1} decouples from \eqref{eqn:SBEs2},  and we obtain a closed   expression for the {\color{alexis} $i$\textsuperscript{th} vectorial-component ($i=x,\,y$) for 2D materials} of the inter-band current,
\begin{align}
J^{(i)}_\mathrm{er}(t)
&  =
-\mathrm{i}\sum_{j}{\frac{d}{dt}}
\int_{t_0}^{t} dt'
\int_{\rm  \overline{BZ}} d^2 {\bf K}\,
d^{(i)}_{cv}\left({\bf K}+ {\bf A}(t)\right)
\nonumber \\ & \qquad \quad  \times
 d^{(j)}_{cv}\left({\bf K}+ {\bf A}(t')\right) E^{(j)}(t')
\nonumber \\ & \qquad \quad \times 
e^{-\mathrm{i}S({\bf K},t,t')-(t-t')/T_2}
+\mathrm{c.c.},
\label{eqn:interML}
\end{align}
where $S({\bf K},t,t')$ is the so-called quasi-classical action for the electron-hole, which is defined as
\begin{align}
S({\bf K},t,t')
& =
\int_{t'}^{t}
\left[
  \varepsilon_g({\bf K} + {\bf A}(t'')) 
  \right. \nonumber \\ & \qquad \qquad \left.
  + \, {\bf E}(t'')\cdot{\bm \xi}_g({\bf K}+{\bf A}(t'')) 
  \right]dt'' .
\label{eqn:ActionML}
\end{align}
{\color{alexis}Here, $j=x,y$ indicates the component of the electric field and transition-dipole product which depends on the polarization of the driving laser.}

This integral form of the inter-band current has a nice physical interpretation {\color{alexis}in terms of the electron-hole (e-h) pair:}
\begin{enumerate}[label=(\roman*),itemsep=0em,parsep=0.3em,topsep=0.3em, partopsep=0.3em	]
\item at time $t'$ the {\color{alexis}e-h pair} is excited by the driving laser from the valence band to the conduction band through the dipolar interaction $ {\bf E}(t') \cdot {\bf d}_{cv}({\bf K}+ {\bf A}(t')) $ at the canonical crystal quasi-momentum ${\bf K}$;
\item the {\color{alexis}e-h pair propagates} in the conduction band and valence band, respectively, between $t'$ and $t$, and {\color{alexis}modifies their} trajectories and energies according to Eq.~(\ref{eqn:Action1}) below; and
\item at time $t$ the electron has a probability to recombine (or annihilate) with the hole, at which point it emits its excess energy as a high energy-photon. 
\end{enumerate}
The expressions \eqref{eqn:interML} and \eqref{eqn:ActionML} are the direct analogues of the Landau-Dykhne formula for HHG in atoms, derived in Ref.~\cite{Lewenstein1994}, following the idea of the simple man's model~\cite{Corkum1993} (see also \cite{Symphony2019} for a recent review). 
Below we will analyze {\color{alexis}these expressions} using the saddle point approximation over crystal momentum to derive the effects of the Berry curvature on the relevant trajectories.

\subsubsection{\color{alexis}Bloch wave gauge invariance: Quasi-classical action}
{\color{alexis} Before turning to the quasi-classical analysis, let us first demonstrate that the expression for the inter-band current~(\ref{eqn:interML}), together with the action~(\ref{eqn:ActionML}), is gauge invariant under the Bloch wavefunction transformation (BWT).~This refers to the phase freedom of the band eigenstates $|u_{m,\bf k}\rangle$, whose phases can be chosen independently of each other.~As such, a local phase change of the form
\begin{equation}
|u_{m,{\bf k}}\rangle 
\to 
|\tilde{u}_{m,{\bf k}}\rangle 
=
e^{-\mathrm{i}\varphi_m({\bf k})}|u_{m,{\bf k}}\rangle
\end{equation}
should leave the physics unchanged.~However, this is not that simple, as multiple relevant quantities change in response to such a BWT}.~Most notably, the off-diagonal matrix elements $d^{(i)}_{cv}({\bf k}) = |d^{(i)}_{cv}({\bf k})|e^{-\mathrm{i} \phi^{(i)}_{cv}({\bf k})}$ ($i=x,y$) of the dipole moment,
\begin{equation}
\tilde{\bf d}_{cv}({\bf k}) 
= 
e^{\mathrm{i}\varphi_c({\bf k})} {\bf d}_{cv}({\bf k})e^{-\mathrm{i}\varphi_{v}({\bf k})}
,
\label{eqn:DGT1}
\end{equation}
acquire a nontrivial phase shift
\begin{equation}
\tilde{\phi}^{(i)}_{cv}({\bf k})=  \phi^{(i)}_{cv}({\bf k}) - \varphi_g({\bf k})
\label{eqn:DGT4}
\end{equation}
for $\varphi_g({\bf k}) = \varphi_c({\bf k}) - \varphi_v({\bf k})$, while the diagonal elements --- the Berry connection, ${\bm\xi}_m(\mathbf {k}) = {\bf d}_{mm}({\bf k})$~--- shift by the gradient of the phase function
\begin{equation}
\tilde{{\bm \xi}}_m ({\bf k}) =  {\bm \xi}_m ({\bf k}) + \nabla_{\bf k}\varphi_m({\bf k})
.
\label{eqn:BCT1}
\end{equation}
The Berry curvature ${\bm \Omega}_m({\bf k}) = \nabla_{\bf k}\times{\bm \xi}_m({\bf k})$, of course, is not affected.


\indent Since our expressions~(\ref{eqn:interML}) and~(\ref{eqn:ActionML}) for the inter-band current depend explicitly on these gauge-dependent quantities, we are obliged to confirm that the current itself (a physical observable) is not affected by the transformation, as well as to examine the transformation properties of the internal components of those expressions.
To do this, we rewrite the expressions above by explicitly taking into account the phases of the transition dipole moment,
\begin{align}
J^{(i)}_\mathrm{er}(t)
& =
-\mathrm{i}\sum_{j}{\frac{d}{dt}}
\int_{t_0}^{t} dt'
\int_{\rm  \overline{BZ}} d^2 {\bf K}\,
\left\lvert d^{(i)}_{cv}\left({\bf K}+ {\bf A}(t)\right)\right\rvert
\nonumber \\ & \qquad \times
\left\lvert d^{(j)}_{cv}\left({\bf K}+ {\bf A}(t')\right)\right\rvert E^{(j)}(t')
\nonumber \\ & \qquad \times 
e^{-\mathrm{i}S({\bf K},t,t')-(t-t')/T_2 + \mathrm{i}\left(\phi_{cv}^{(j)}({\bf K},t)-\phi_{cv}^{(i)}({\bf K},t)\right) }
\nonumber \\ & \qquad 
+ \mathrm{c.c.},
\label{eqn:inter2}
\end{align}
where the action is ``corrected'' $\tilde{S}^{(j)}({\bf K},t,t')$ by the time-dependent dipole phases as
\begin{align}
\tilde{S}^{(j)}({\bf K},t,t')
& =
\int_{t'}^{t} \!
\bigg[
  \varepsilon_g({\bf K} {+} {\bf A}(t'')) 
  +{\bf E}(t'')\cdot{\bm \xi}_g({\bf K} {+} {\bf A}(t'')) 
  \nonumber \\ & \qquad \qquad
  - \frac{d}{dt''}\phi_{cv}^{(j)}({\bf K}+{\bf A}(t'')) 
  \bigg]dt''
.
\label{eqn:Action1}
\end{align}
Here $j=x,y$ determines the component of the dipole moment whose phase gets incorporated into the action, determined in part by the polarization of the driving laser.
Thus, it is important to note that for nontrivial 2D polarizations this action can split into several (though related) versions, each for a different component of the transition.
Note that the dipole phase in the action has the same index as the electric field $j$. The additional term $\phi_{cv}^{(j)}(t)-\phi_{cv}^{(i)}(t)$ only contributes in the perpendicular-emission configuration $i\neq j$, if one assumes the laser field is linearly polarized along the $x$ direction. 

The total action  $\tilde{S}$ can be written as
\begin{align}
\tilde{S}^{(j)}({\bf K},t,t')
& =
\int_{t'}^{t} \!
\bigg[
  \tilde{\varepsilon}_g({\bf K} + {\bf A}(t'')) 
  +  {\bf E}(t'')\cdot\tilde{{\bm \xi}}_g({\bf K}+{\bf A}(t'')) 
  \nonumber \\ & \qquad \qquad 
  - \frac{d}{dt''}\tilde{\phi}^{(j)}_{cv}({\bf K}+{\bf A}(t'')) 
  \bigg]dt''
,
\label{eqn:Action2}
\end{align}
where $\tilde{\varepsilon}_g = \varepsilon_g$ is gauge invariant, but $\tilde{{\bm \xi}}_g$ and $\tilde{\phi}^{(j)}_{cv}$ are not. However, we can use the transformation rules \eqref{eqn:DGT4} and~\eqref{eqn:BCT1}, for the dipole phase and the Berry connection:
\begin{align}
\tilde{S}^{(j)}({\bf K},t,t')
& =
\int_{t'}^{t}
\bigg[
  \varepsilon_g 
  +  {\bf E}(t'')\cdot\left({\bm \xi}_g + \nabla_{\bf K}\varphi_g\right) 
  \nonumber \\ & \qquad \qquad 
  - \frac{d}{dt''}(\phi^{(j)}_{cv}-\varphi_g) 
  \bigg]dt''
.
\label{eqn:Action2b}
\end{align}
(For simplicity, we have dropped the dependence on the variable ${\bf K} + {\bf A}(t)$ here.) Moreover, by using the attribute that for any function  $f({\bf K} + {\bf A}(t))$, 
\begin{equation}
\frac{d}{dt}f({\bf K} + {\bf A}(t))=-{\bf E}(t)\cdot\nabla_{\bf K}f({\bf K} + {\bf A}(t))
,
\end{equation}
as well as the property $\nabla_{\bf k} = \nabla_{\bf K}$~\cite{ShouCheng2011}, we have 
\begin{equation}
\left({\bf E}(t'')\cdot \nabla_{\bf K} + \frac{d}{dt''}\right)\varphi_g=0
,
\end{equation} 
which leads finally to the relation
\begin{equation}
\tilde{S}^{(j)}({\bf K},t,t')=\int_{t'}^{t}\left[\varepsilon_g +  {\bf E}(t'')\cdot{\bm \xi}_g  - \frac{d}{dt''}\phi^{(j)}_{cv} \right]dt''. \nonumber
\label{eqn:Action3}
\end{equation}
Since this expression is equal to Eq.~(\ref{eqn:Action1}), we find that ${\tilde S}^{(j)}({\bf K},t,t')$ is thus Bloch-gauge invariant. 
Both the HHG spectrum and the currents, as physical observables,  should be gauge-invariant quantities. Our analysis shows that this is indeed  the case in our theoretical model, even in the presence of a nontrivial Berry connection.

In addition, we can also rewrite the semiclassical action~as
\begin{align}
{\tilde S}^{(j)}({\bf K},t,t') 
& = 
\!
\int_{t'}^{t} \!
\bigg[
  \varepsilon_g({\bf K} {+} {\bf A}(t'')) 
  +  {\bf E}(t'')
  {\cdot}\bigg(
    {\bm \xi}_g({\bf K} {+} {\bf A}(t'')) 
  \nonumber \\ & \qquad \qquad 
    + \nabla_{\bf K}\phi^{(j)}_{cv}({\bf K} + {\bf A}(t''))
    \bigg)
  \bigg]dt''
,
\label{eqn:Action4}
\end{align}
by explicitly calculating the time derivative of the dipole phase, $\frac{d}{dt''}\phi^{(j)}_{cv}$, to bring to the front the explicit factor of the electric field, ${\bf E}(t'')$.
Here the action phase includes geometric and topological properties, such as the Berry connection and the dipole phase $\phi^{(j)}_{vc}({\bf k})$.
Given this form, it is probable that the electron-hole pair trajectories will be modified by these geometric properties, thereby affecting the non-linear HHG process.
The total phase of Eq.~(\ref{eqn:inter2}) has an extra dipole-phase contribution, $\phi_{cv}^{(i)}-\phi_{cv}^{(j)}$ regarding the ``perpendicular'' inter-band current to the electric field. From the phase transformation given by Eq.~(\ref{eqn:DGT4}), this phase difference term is also gauge invariant.

\begin{figure*}[htbp]
\begin{center}
\includegraphics[width=0.45\textwidth]{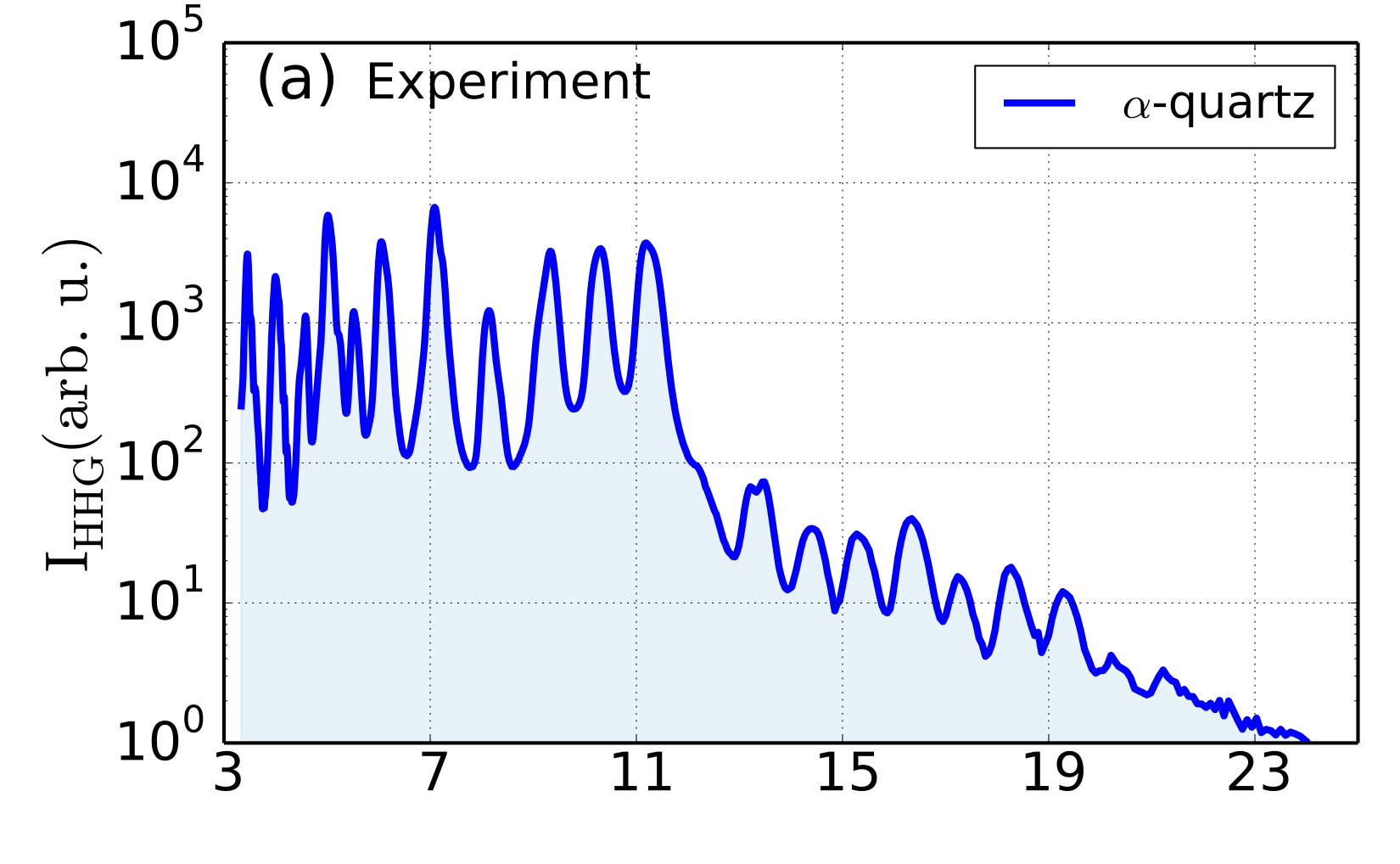}
\includegraphics[width=0.45\textwidth]{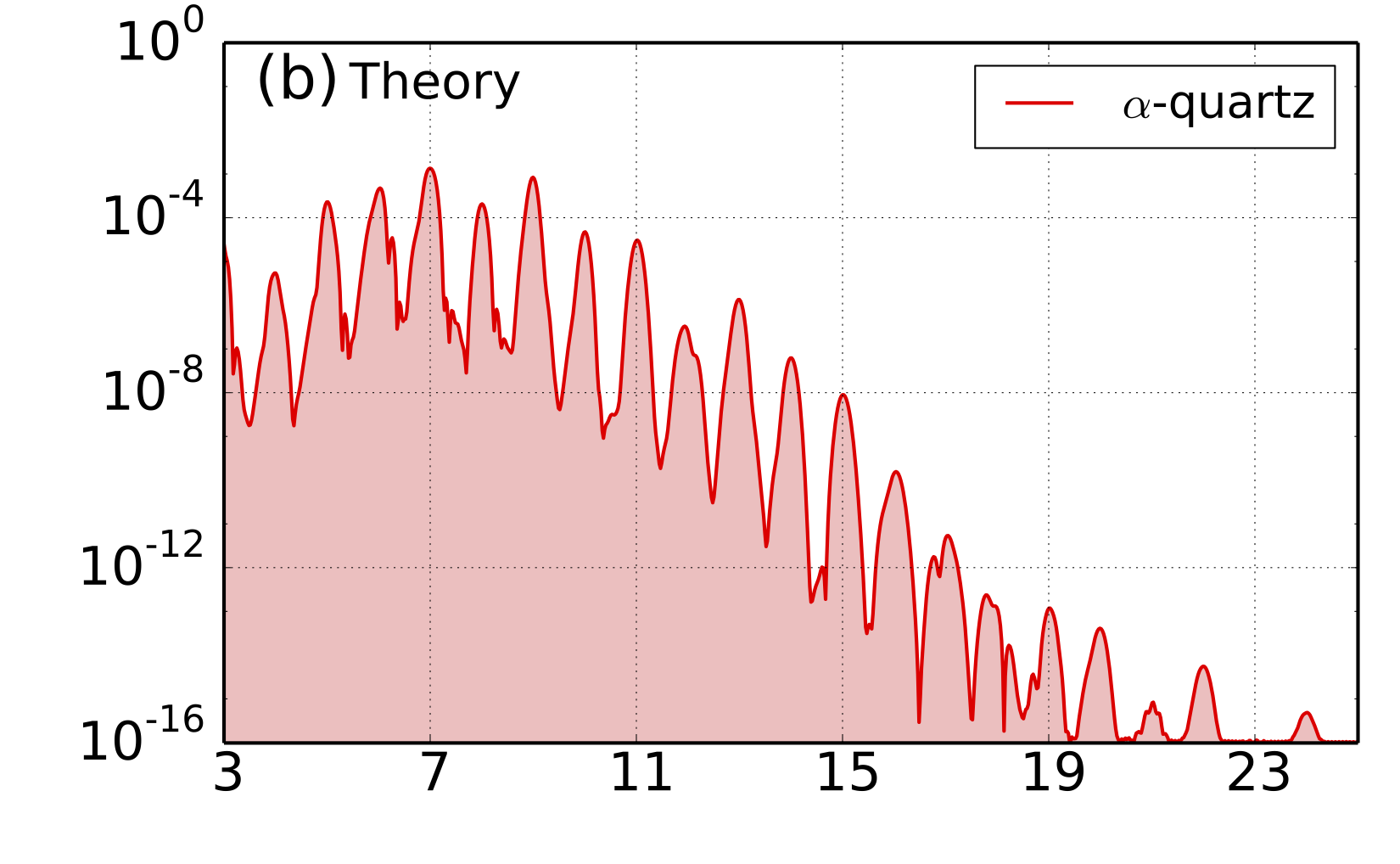}\\
\includegraphics[width=0.46\textwidth]{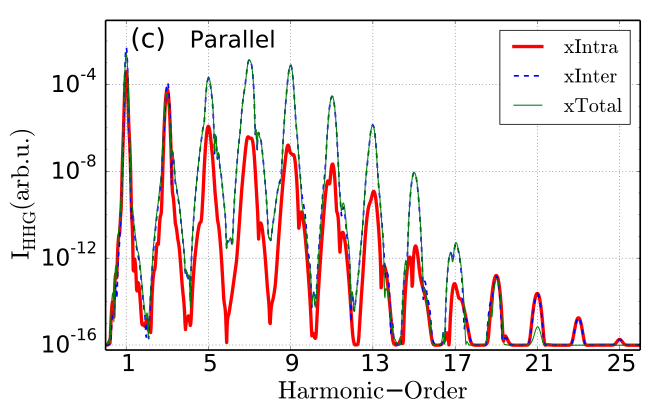}
\includegraphics[width=0.46\textwidth]{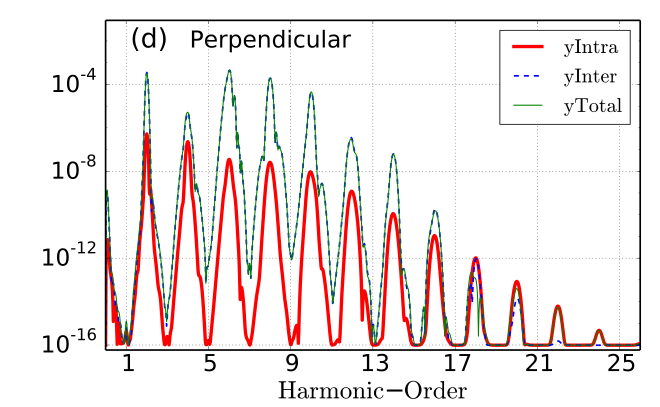}
\caption{%
(a) Experimental~\cite{Luu2018} and (b) theoretical HHG spectra from an $\alpha$-quartz crystal. Both the theoretical and experimental data are averaged along the parallel and perpendicular harmonic emission with respect to a MIR $\SI{800}{nm}$ laser linearly polarized along the $\Gamma$-$K$ crystal orientation (cf.\ Fig.~\ref{figure1}(b)).
(c) Parallel and (d) perpendicular harmonic emission from our $\alpha$-quartz toy model, showing the inter-band (blue), intra-band (red) and total (green) currents.
We use laser parameters following Ref.~\cite{Luu2018}, with 
a wavelength $\lambda_0=\SI{800}{nm}$ giving a central frequency {\color{alexis}$\omega_0=\SI{0.056}{\au}~(\SI{1.53}{eV})$},
FWHM duration of $10$ cycles under a Gaussian envelope, and
peak intensity of $I_0=\SI{3.2e12}{W/cm^2}$.
We use a honeycomb lattice with
lattice constant $a_0\sim\SI{4.8}{\angstrom}$,
NN hopping $t_1 = \SI{2.1}{eV}$, NNN hopping amplitude $t_2=\SI{0.36}{eV}$ with magnetic flux $\phi_0=0$, and staggering potential $M=\SI{4.5}{eV}$, with a band gap $E_g\sim\SI{9}{eV}$,
and a dephasing time of $T_2=\SI{220}{\au} = \SI{5.3}{fs}$.
}
\label{figure2}
\end{center}
\end{figure*}

\subsection{Quasi-classical approach and electron-hole pair trajectories}\label{SPASec}
Assuming that the exponentiated quasi-classical action $e^{-iS}=e^{-iS({\bf K},t,t')}$ oscillates rapidly as a function of the crystal momentum $\bf K$, one can apply the saddle-point approximation to find the points ${\bf K}_s$  where the integrand's contributions to the inter-band current (\ref{eqn:inter2}) concentrate. These are solutions of the saddle-point equation $\nabla_{\bf K} S({\bf K},t,t')|_{{\bf K}_s}= {\bf 0}$, which can be rephrased as
\begin{equation}
\Delta{\bf x}_{c}({\bf K}_s,t,t')-\Delta{\bf x}_{v}({\bf K}_s,t,t')= {\bf 0}.
\label{eqn:SaddleP1}
\end{equation}
From the last equation two different trajectories are identified, the first one related to the excited electron $\Delta{\bf x}_{c}({\bf K}_s,t,t')$ in the conduction band, and the second one regarding the trajectory $\Delta{\bf x}_{v}({\bf K}_s,t,t')$, followed by the hole in the valence band.
We then obtain a general $m$\textsuperscript{th} trajectory for the electron ($m=c$) and hole ($m=v$), which reads
\begin{align}
\Delta{\bf x}_{m}({\bf K}_s,t,t') 
& =
\int_{t'}^{t}
\bigg[
  {\bf v}_{\mathrm{gr},m}
  + {\bf E}(t'')\times{\bm \Omega}_{m}
\label{eqn:SaddleP3}
\\ & \!\!
\nonumber 
  +\left({\bf E}(t''){\cdot}{\nabla_{\bf K}}\right)\!
  \bigg({\bm \xi}_{m}  
  {+}
  (-1)^m \frac{1}{2}\nabla_{\bf K}\phi^{(j)}_{cv}\bigg) 
  \bigg]dt''
,
\end{align}
where $(-1)^m$ is the alternating sign $(-1)^{c}=+1$ and $(-1)^v=-1$,
and the group velocity of the $m$\textsuperscript{th} band is ${\bf v}_{\mathrm{gr},m} = \nabla_{\bf K}\varepsilon_m$. Here we recognize the Berry curvature ${\bm \Omega}_m$ as well as the anomalous velocity ${\bf v}_{a,m}$, which are given by ${\bm \Omega}_m={\nabla}_{\bf K}\times{\bm\xi}_m$ and ${\bf v}_{a,m}={\bf E}(t)\times{\bm\Omega}_m$, respectively, for the electron-hole trajectories of Eq.~(\ref{eqn:SaddleP3}). We can rewrite the previous expression as
\begin{align}
\Delta{\bf x}_{m}({\bf K}_s,t,t')
& =
\int_{t'}^{t} \!
\bigg[
  {\bf v}_{\mathrm{gr},m}  
  + {\bf v}_{a,m} 
\label{eqn:SaddleETs}
\\ & \qquad 
\nonumber 
  -\frac{d}{d t''}\left({\bm \xi}_m{+}\frac{(-1)^m}{2}\nabla_{\bf K}\phi^{(j)}_{cv}\right)  
  \bigg]dt''
.
\end{align}
These electron-hole pair trajectories, together with the saddle-point condition of Eq.~(\ref{eqn:SaddleP1}), should produce com\-plex-valued solutions for ${\bf K}_s$, as is the case for HHG of gases.

However, finding the solutions ${\bf K}_s$ is not a trivial task, since it depends explicitly on the geometrical features, i.e.\ the Berry curvature and connection, and the phase of the dipole matrix elements.
Moreover, {\color{alexis}this gets further} complicated as the eigenstates and eigenvalues that make up the energy bands are expected to exhibit branch cuts and branch points connecting the two bands~\cite{Pechukas1976, Hwang1977}. Once the momentum is allowed to take on complex values (as it does in complex band structure theory~\cite{Reuter2016}), this leads to a nontrivial geometrical problem with a high dimensionality whose analysis requires detailed attention.

Nevertheless, these saddle points ${\bf K}_s$ should have a component perpendicular to the driving laser field ${\bf E}(t)$ (for the case of linear drivers), which appears as a consequence of anomalous-velocity features and in particular of the Berry curvature ${\bm \Omega}_m({\bf k})$.

\section{HHG in practice: the role of inter- and intra-band currents}
\label{sec:HHG-in-practice}
Having developed the theory of HHG in topological solids over the previous sections, we now turn to applying it to some examples of topologically trivial and non-trivial systems. In a semiconductor or insulator driven by mid-infrared lasers or THz sources, the harmonic emission is governed by the coherent sum of the intra-band ${\bf J}_{ra}(t)$ and inter-band ${\bf J}_{er}(t)$ current oscillations ${\bf J}(t)={\bf J}_{ra}(t)+{\bf J}_{er}(t)$~\cite{GhimireNatPhy2011, VampaJPB2017} given above by Eqs.~(\ref{eqn:inter-SM}) and~(\ref{eqn:intra-SM}).
We stress, as mentioned previously, that the inter-band current contains explicit information about the Berry curvature through the cross product of the dipole moments. This is important, since in 2D materials the integral of the Berry curvature over the BZ is the topological invariant, i.e.\ the Chern number. In Fig.~\ref{figure1}(Y) we depict a cartoon of the HHG physical process which takes place in topological materials.

Via our quasi-classical saddle-point analysis we find the following:
(1) The electron-hole pair is more likely to be excited around the $K'$ point than the $\Gamma$ or $K$ points, since the energy gap is larger at those points compared to $K'$ for this HM.
(2) The electron-hole pair is propagated by the driving laser field along the path indicated with black dashed line of the BZ emitting intra-band harmonics in the process.
This propagation follows the laser vector potential in the momentum plane, with the real-space motion (and thus the intra-band current) governed by the group velocity produced by the energy-band dispersion as well as an anomalous velocity caused by the Berry curvature.
(3) Finally, the electron can recombine with its hole and release its energy as a photon.
This inter-band emission carries traces of the topological features of the material via Berry-curvature contributions to the action, which are accrued in the continuum traversal as well as via the dipole matrix elements that mediate the transition.

\begin{figure*}[htbp]
\begin{center}
\includegraphics[width=0.45\textwidth]{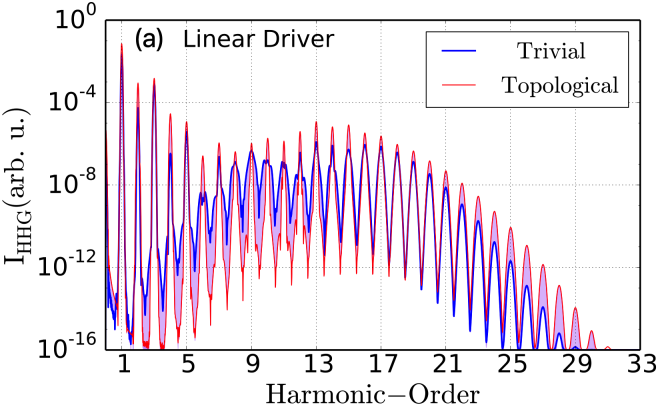}
\includegraphics[width=0.45\textwidth]{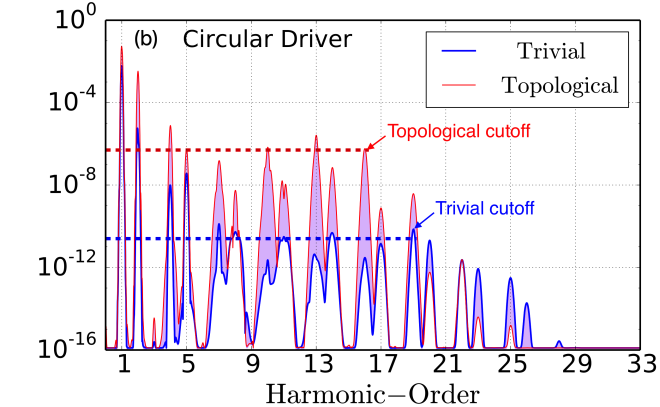}
\caption{%
High-order harmonics for trivial and nontrivial topological materials.
We show the harmonic emission from equal-bandgap instances from the trivial and topological phases driven by lasers with (a) linear polarization along $\Gamma$-$K$ and (b) right-circular polarization; the violet areas indicate the difference in the emission between the two phases.
We use $t_1=\SI{2.04}{eV}$, $t_2=t_1/3$ and $M_0= 2.54\,t_2$, with lattice constant $a_0=\SI{1}{\angstrom}$,
driven by a laser centered at $\SI{3.25}{\micro m}$ ($\omega_0=\SI{0.014}{\au}$) with a Gaussian envelope with 14-cycles FWHM and a peak electric-field strength of $E_0=\SI{0.0045}{\au}$ ($I_0=\SI{7.0e11}{W/cm^2}$),
and we use a dephasing time of $T_2=\SI{5.2}{fs}$.
The magnetic flux of the trivial and non-trivial phases, $\phi_0=0.06$ and $\phi_0=\SI{1.16}{rad}$, {\color{alexis}respectively}, are chosen to give equal bandgaps of $E_g\approx\SI{3.0}{eV}$.
}
\label{figure2-copy}
\end{center}
\end{figure*}

We start by comparing our theory to the results of the recent HHG experiment of Luu and W\"orner~\cite{Luu2018} on $\alpha$-quartz~(SiO$_2$).
Since a cut along the $z$-direction of this crystal has an approximate honeycomb lattice structure, we use the trivial limit of the HM to validate our theory.
Due to the breaking of IS in $\alpha$-quartz, the HHG spectrum contains even harmonic orders (HOs), shown in Fig.~\ref{figure2}(a), mainly along the perpendicular emission configuration.
As we detail below, our simulations show that the inter-band contribution has a larger magnitude than the intra-band current. 
Generally, the experimental features of Ref.~\cite{Luu2018} are qualitatively reproduced by our theory (see Fig.~\ref{figure2}(b)); moreover, this approach qualitatively agrees with Liu {\it et al.}'s experiment reported for MoS$_2$~\cite{Liu2017}.
That said, it is important to note that the calculations overestimate the intensity yield for the high-order harmonics (HOs) of the plateau.

The discrepancy between theory and experiment is related to the limitations of this toy model under the tight-binding approach, which will not recreate the whole experimental complexity for the following reasons:
(i) the lattice crystal is not a monolayer (the crystal thickness is $\SI{20}{\micro m}$~\cite{Luu2018}),
(ii) the crystalline structure of the hexagonal 3D lattice of SiO$_2$ is much more complex than our simplified 2D lattice,
(iii) the possibility of higher conduction bands and lower valence bands playing a role is not considered in our approximations~\cite{Hawkins2015}, and
(iv) the propagation effects in the $\SI{20}{\micro m}$-thick crystal can play a fundamental role~\cite{Floss2018}.
However, despite this discrepancy, we believe that physical insight can be extracted in terms of identifying dominant mechanisms for the harmonic emission -- (intra-band) nonlinear Bloch oscillations of charge  carriers~\cite{GoldePRB2008}, and (inter-band) electron-hole recombination processes~\cite{VampaJPB2017} -- and their relative importance.

We now turn to the parallel and perpendicular components of the harmonic emission when driven by a linearly-polarized laser, with respect to the driver's polarization, which we show in Fig.~\ref{figure2}(c,d) broken out into the intra- and inter-band components.
Firstly, our calculations show the same qualitative features of the experiments~\cite{Luu2018,Liu2017}: odd and even harmonics are emitted along the parallel and perpendicular emission directions, respectively.
Secondly, the inter-band current dominates over the intra-band one for both parallel and perpendicular configurations, at least for these excitation conditions, and with the usual caveats regarding the nontrivial coupling between the two components~\cite{GoldePRB2008}. This implies that a recollision mechanism is likely to be at play in a system where the IS is broken. We also note that along the perpendicular emission HHG shows only even harmonics for the intra-band analysis exhibiting dominant behavior for largest harmonic orders (HOs).
We believe that the appearance of even harmonics along the perpendicular direction is not only a matter of intra-band or Berry-curvature effects: they are produced by the inter-band process, due to a combination of the dipole phases, Berry curvature and Berry connection features. This explanation is apparent both in our saddle-point arguments for the inter-band mechanism above and the numerical calculations given here.
The HHG emission is not only the product of a single effective intra-band contribution; both inter-band and intra-band mechanisms are present in the rich physics of this non-linear harmonic emission process.

Finally, as additional validation of our simulations, we have studied the scaling of the harmonic cutoff with respect to the driving-field amplitude, which we report below in Appendix~\ref{sec:lin_cut}. As expected, the cutoff exhibits a linear scaling in good agreement with previously obtained theoretical and experimental results~\cite{GhimireNatPhy2011, VampaPRB2015}.

\section{Signatures of topological phases in HHG}
\label{sec:signatures}
We now address the question of whether HHG is sensitive to (i) different topological phases and (ii) topological transitions.
To this end, in Fig.~\ref{figure2-copy} we show the calculated full HHG radiation spectrum, $I_{\rm HHG}(\omega)=\omega^2\left(|{ J}_{x}(\omega)|^2+|{ J}_{y}(\omega)|^2\right)$ produced by linearly-polarized (Fig.~\ref{figure2-copy}(a)) and circularly-polarized (Fig.~\ref{figure2-copy}(b)) driving lasers, comparing instances from the trivial and the topologically-nontrivial phase at equal bandgaps. 
For a linearly-polarized driving laser, the harmonic intensity spectrum exhibits relatively small differences between the two phases (though signatures of the topological phase do appear in the polarization state of the harmonics~\cite{Misha2018}).
In our results we observe an enhancement of about one order of magnitude for even-order harmonics in the low spectral region, i.e.\ the $2^{\rm nd}$, $4^{\rm th}$ and $6^{\rm th}$ harmonics.

For a circularly-polarized driver, on the other hand, shown in Fig.~\ref{figure2-copy}(b), the spectra produced by the different topological phases exhibit far greater differences, in particular:
(1) a considerable enhancement about four orders of magnitude in {\color{alexis} plateau harmonics in }the topological phase as compared with the trivial phase,
(2) a shorter cut-off for the topological emissions in comparison to the trivial, and
(3) an asymmetric yield between the co-rotating ($3n+1$) and contra-rotating ($3n+2$) harmonic orders in plateau harmonics.
As such, it is clear that circularly-polarized drivers have significant potential in bringing out the signatures of the topological phase in the harmonic spectrum, and we will focus on them for the rest of this work.

The circularly-driven spectra of Fig.~\ref{figure2-copy}(b) are also noticeably different from the HHG spectra driven by linearly-polarized pulses, in that they exhibit a clear selection rule, where harmonics of order $3n$ are forbidden and $3n\pm1$ are allowed, for an integer $n$.
This selection rule is identical to that observed in the well-known `bicircular' field configuration -- two counter-rotating circularly-polarized drivers at frequencies $\omega_0$ and $2\omega_0$~\cite{Fleischer2014} -- and it stems from the same origin, a dynamical symmetry of the system, which combines a rotation by $\SI{120}{\degree}$ with a time delay~\cite{Alon1998}. This symmetry argument has been applied in solids~\cite{Saito2017, Sorngard2013, Chen2019, Zurron2019, JimenezGalan2019} but has also been applied to molecules~\cite{Mauger2016, Baykusheva2016, Yuan2018} and even plasmonic metamaterials~\cite{Konishi2014}.

In our case, however, the generating medium~-- the HM on a hexagonal lattice -- is symmetric under $\SI{120}{\degree}$ rotations, but it lacks reflection symmetry, which allows it to respond nontrivially to the different helicities of circularly-polarized driving lasers. This is much like $p$ states in noble-gas atoms~\cite{Medisauskas2015, Pisanty2017} and chiral molecules~\cite{Harada2018, Ayuso2018}, which points to the {\color{alexis}{\it circular dichroism} (CD)} of the optical response (i.e.\ the difference in response to left- and right-handed circularly-polarized drivers) as a natural place to look for signatures of the topological phase of the material.

\begin{figure*}[htbp]
\begin{center}
\begin{tabular}{cc}
\includegraphics[width=0.49\textwidth]{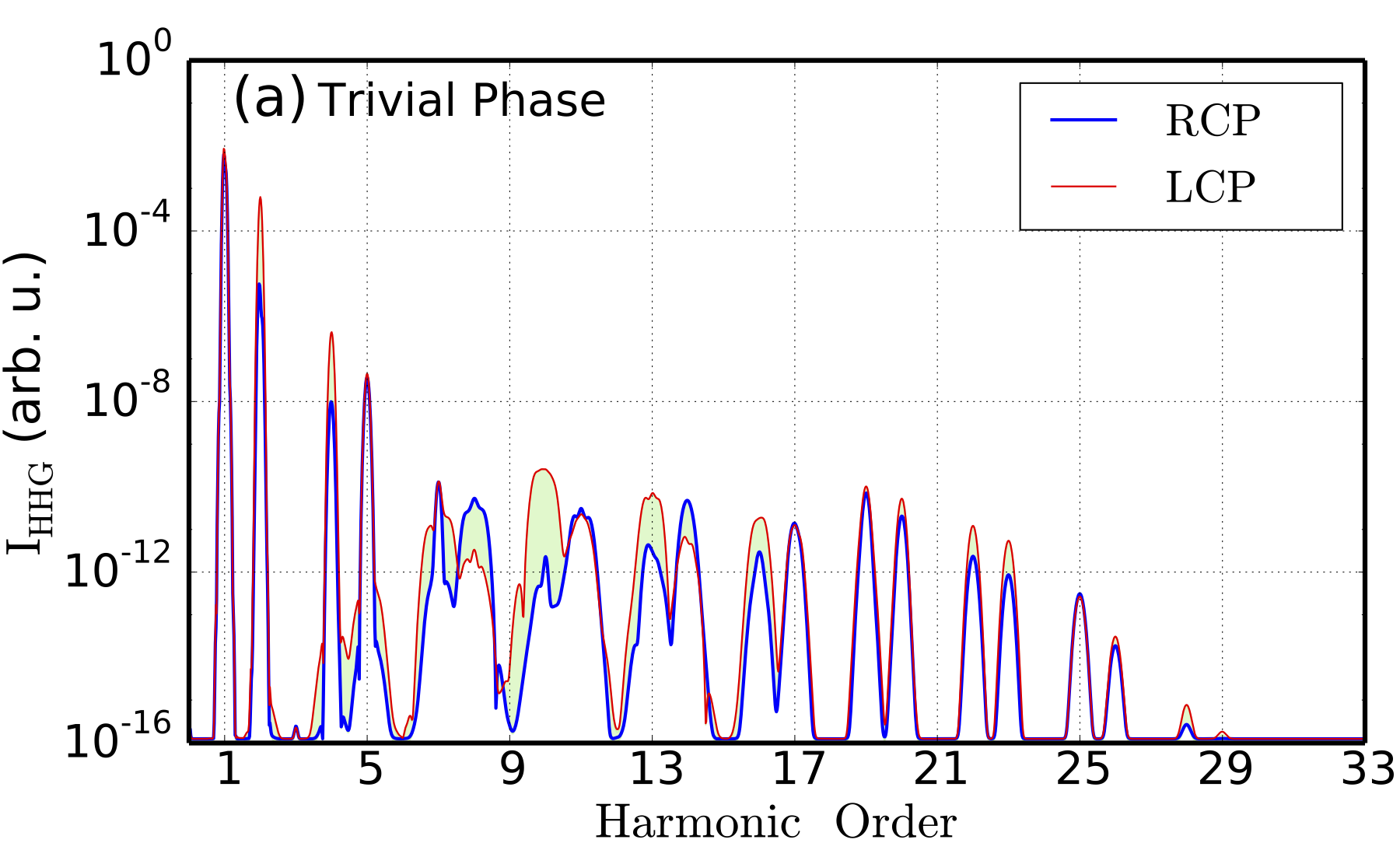} &
\includegraphics[width=0.49\textwidth]{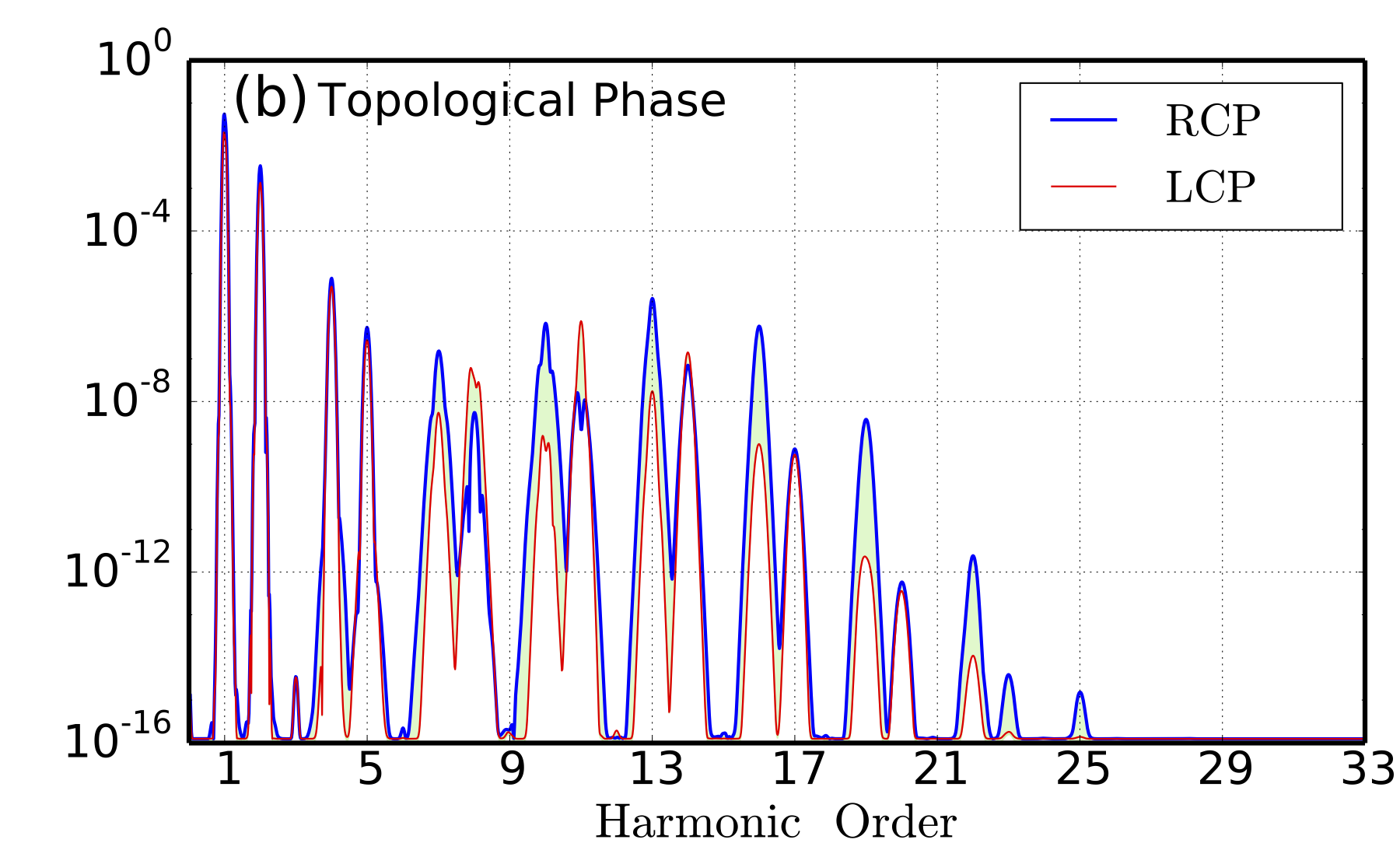} \\
\includegraphics[width=0.49\textwidth]{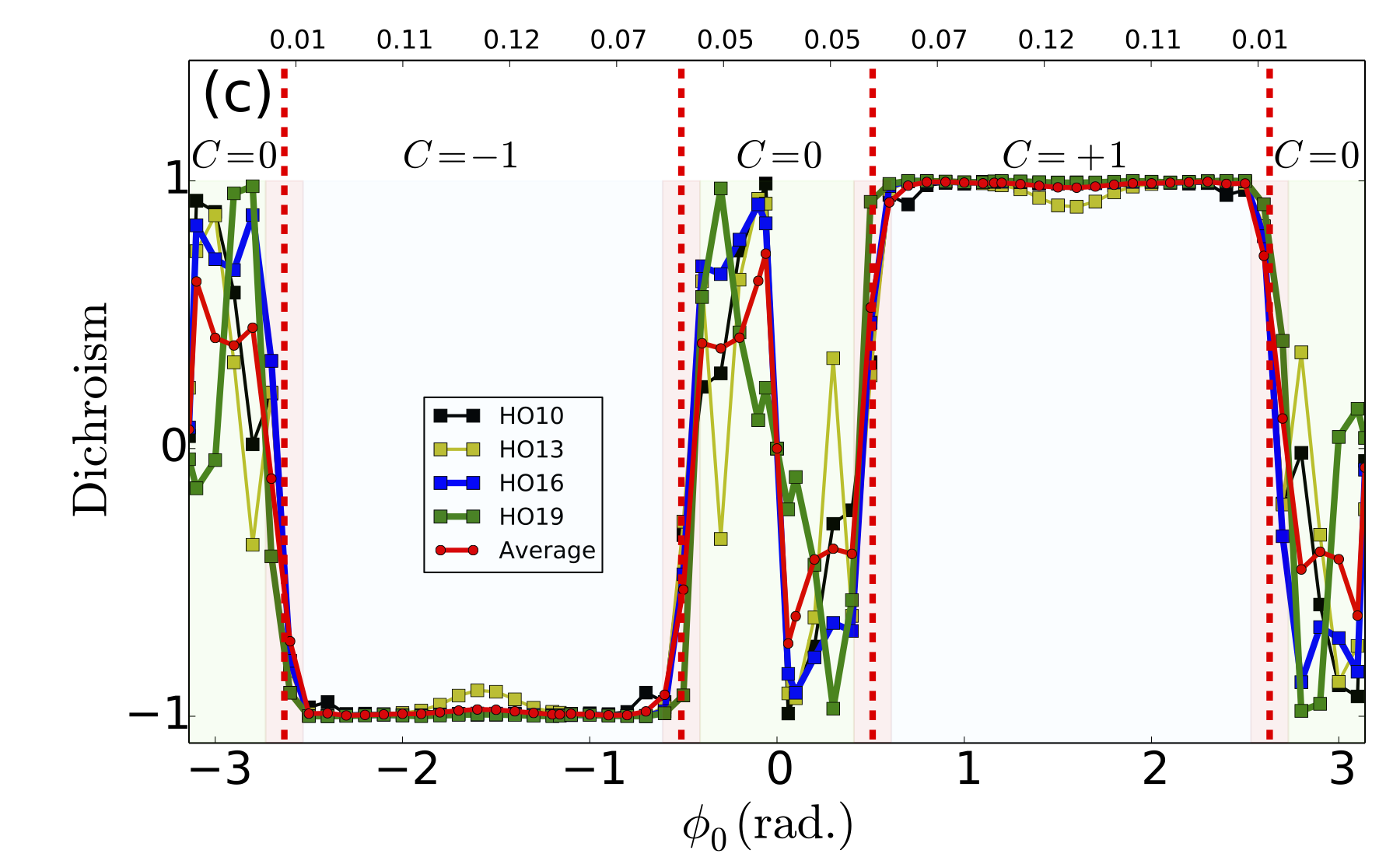} &
\includegraphics[width=0.49\textwidth]{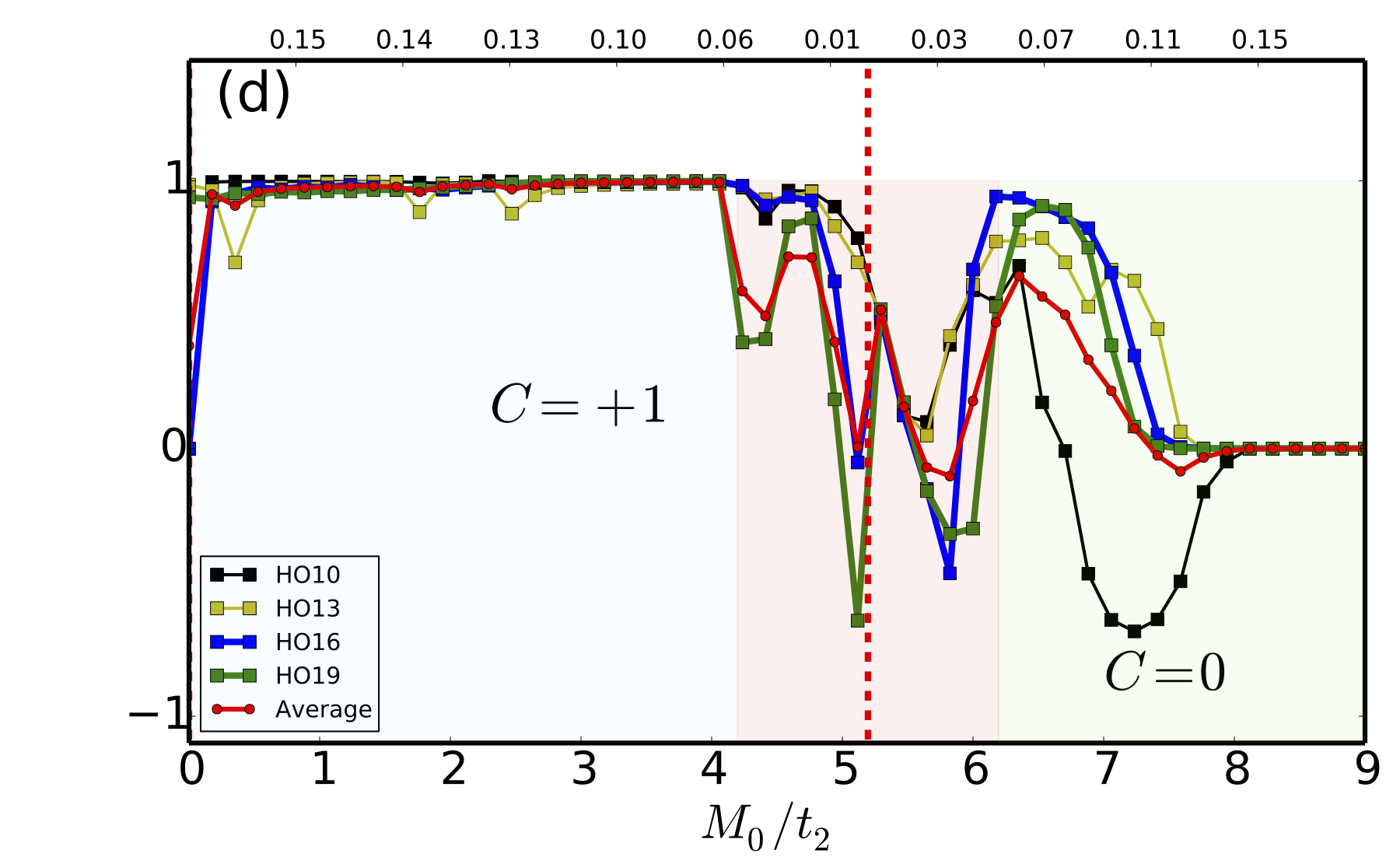}
\end{tabular}
\caption{
Circular dichroism as a signature of topological phase transitions in HHG.
(a,b) %
Harmonic spectra of equal-bandgap instances from the trivial and the topological phase, respectively, driven by right- and left-circularly polarized lasers, with the difference between the two emissions shaded in green.
(We use the same parameters as in Fig.~\ref{figure2}).
(c,d) %
Circular dichroism (CD), as defined in Eq.~\eqref{CD-definition}, for the magnetic-flux and staggering potential parameter scans marked with dashed lines in Fig.~\ref{figure1}, for the plateau harmonics. The top horizontal axes show the energy bandgap $E_g$ in atomic units as a function of $\phi_0$~(c) and $M_0/t_2$~(d), {\color{alexis} respectively.}
The trivial and topological regions are shaded green and blue, with the critical region shaded red.
}
  \label{figure3}
  \end{center}
\end{figure*}

\section{Circular dichroism}
\label{sec:dichroism}
As an initial test of this {\color{alexis}circular dichroism ($\CD$)}~idea, we show in Figs.~\ref{figure3}(a) and \ref{figure3}(b) the total harmonic spectra produced in topologically trivial and nontrivial phases, respectively, when driven by a right- and left-circularly polarized (RCP and LCP, shown in blue and red, respectively) laser pulses.
The topologically-nontrivial phase shows a distinct increase in harmonic order response of the form $k=3n+1$ (which co-rotates with the driving laser due to the usual selection rules) when driven by an RCP laser as compared with an LCP driver, while the topologically-trivial phase shows a small \textit{decrease} in the emission of those harmonics.
(In absolute terms, this combination of laser helicities matches that of {\color{alexis}the next-nearest-neighbor coupling} $t_2 e^{i\phi_0}$, as shown in Fig.~\ref{figure1}(c); if a negative $\phi_0$ is chosen instead, the effect reverses direction.)

This observation indicates that, indeed, the CD of the harmonic emission carries clear information about the topological phase performing the emission (the CD is invariant under the electromagnetic minimum-coupling gauge), and we focus on it for the rest of this work.
We define, in particular,
\begin{equation}
\CD_{k} = \frac{I_{k,\RCP}^{(+)} -I_{k,\LCP}^{(-)}}{I_{k,\RCP}^{(+)} +I_{k,\LCP}^{(-)}}
,
\label{CD-definition}
\end{equation}
that is, the normalized difference in harmonic response driven by left- and right-handed circularly-polarized dri\-vers, for harmonic orders $k=3n+1$.~(For the counter-rotating harmonics $k=3n+2$, the signs in the superscripts should be reversed.)
Here the CD is defined in terms of the circular components of the harmonic emission,
$I^{(\pm)}_{k,\nu}=(k\omega_0)^2|J_{x,{\nu }}(k \omega_0)\pm i J_{y,{\nu }}(k \omega_0)|^2$,
 $J_{x,\nu}$ and $J_{y,\nu}$ {\color{alexis}denoting the $x$} and $y$ components of the spectral current driven by~a $\nu=\RCP,\,\LCP$ laser.

{\it \color{alexis} Our main result is shown}~{\color{alexis} in Figs.~\ref{figure3}(c,d).~These plot the CD for the co-rotating plateau harmonics over the~{\color{alexis} three distinct} topological phases of the Haldane phase diagram (see Fig.~\ref{figure1}(a)).~(i) in Fig.~\ref{figure3}(c), we scan over $\phi_0$ with $M_0$ constant (see horizontal cut depicted in orange dashed line of Fig.~\ref{figure1}(a)).~And~(ii)~in Fig.~\ref{figure3}(d) at $\phi_0=\pi/2$, we also scan the CD over $M_0/t_2$ (vertical cut depicted in blue dashed line of Fig.~\ref{figure1}(a)).}~{\color{alexis}For the case (i), we observe two clear topological-plateau structures in the circular dichroism with values of $\CD\approx \mp1$ for the topological invariants $C=\mp1$, a clear {\it topological-plateau signature for the $\CD$ of the co-rotating harmonics}.~Particularly, high oscillatory behavior is {\color{alexis}observed in the CD} for the trivial phase $C=0$, which deviates from the robust CD topological-plateau structure, once the topological phase boundary is crossed.~In short, the CD observable shows clear distinct characteristics as a function of the Chern number $C=-1,\,0,\,+1$.}

{\color{alexis}In the case (ii), while the ${\rm CD}\approx+1$ for $C=+1$, once the topological phase boundary is crossed, the CD disappears relatively quickly as $M_0/t_2$ increases.}~This effect can be understood rather simply, since in the $M_0\gg t_2$ limit, the material's Hamiltonian is dominated by the staggering potential $M_0$, which is reflection-symmetric and does not support a circularly-dichroic response.
However, the most notable aspect of the CD spectra is that it still exhibits clear variations depending on the topological phase the system is in. This is true even for instances of the HM which are `equally chiral' at the level of the Hamiltonian (in the sense of having identical $M_0/t_2$ ratios).~In other words, the high-harmonic CD response {\color{alexis} shows} a clear signature of the topological phase of the system.

\begin{figure*}[htbp]
\begin{center}
  \includegraphics[width=8.4cm]{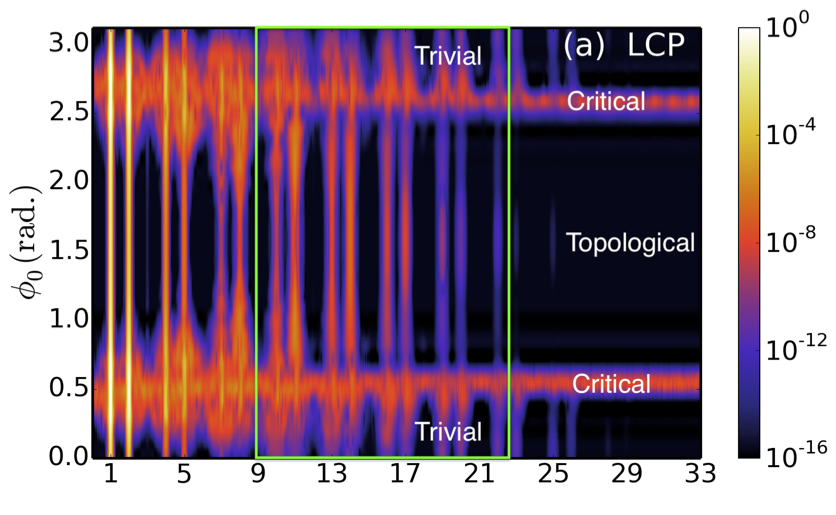}
  \includegraphics[width=8.4cm]{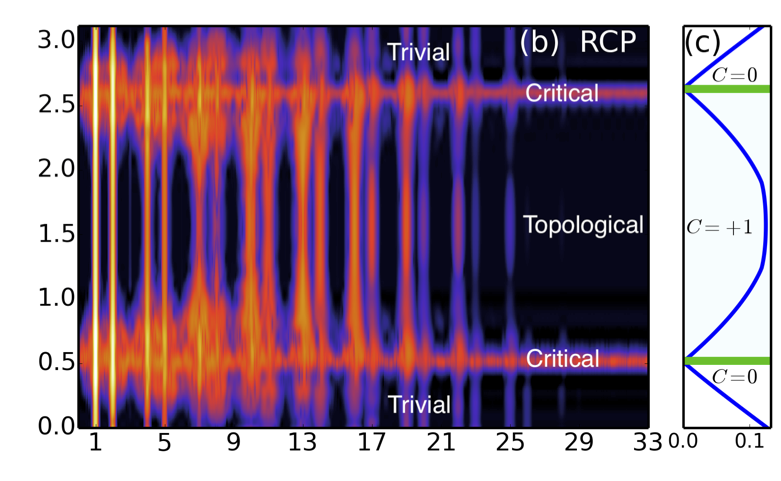}\\
  \hspace{0.14cm} \includegraphics[width=8.4cm]{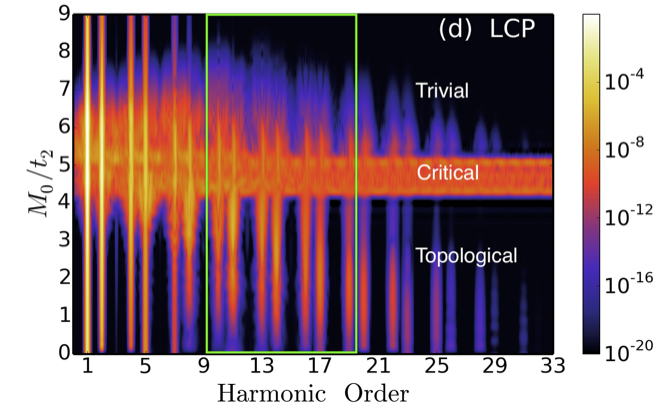}
  \hspace{0.25cm}  \includegraphics[width=8.4cm]{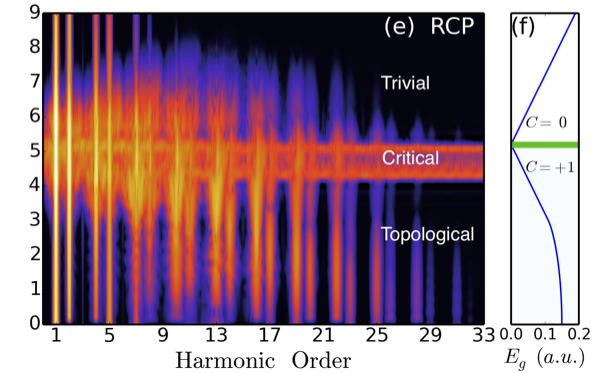}
  \caption{%
Variation of the harmonic spectrum for circularly-polarized drivers.
(a,b) Harmonic spectrum for the magnetic-flux parameter scan indicated with the red dashed line in Fig.~\ref{figure1}(a), driven by LCP and RCP light, respectively, with $\phi_0$ ranging over a 60-point scan of the interval $[0,\pi]$ and the staggering potential held fixed at $M_0 = 2.54 t_2$. At the critical region, where the bandgap closes between the two phases (as indicated in (c), which charts the bandgap), the harmonic emission increases, as carrier creation is much easier. The signature of the topological phase is in the circular dichroism (the difference between the RCP-driven signal in~(b) and the LCP-driven signal in~(a)), which we plot in Fig.~\ref{figure3} for the harmonics in the green box. (d,e,f) Identical plots, for the staggering-potential scan marked by the blue dashed line in Fig.~\ref{figure1}(a), varying $M_0/t_2$ at fixed magnetic flux of $\phi_0=\frac{\pi}{2}$.
}
  \label{figure4}
  \end{center}
\end{figure*}

There is also a clear transition regime between the two topological phases, shown shaded pink in Figs.~\ref{figure3}(c,d), where the CD (particularly in the $M_0/t_2$ scan) takes on irregular values.
At the topological phase boundary the bandgap closes, as it must do whenever the Chern number changes value, and at these points the material becomes a semimetal with one exact graphene-like Dirac cone per BZ.
In the critical region around those points (see panel (Q) in Fig.~\ref{figure1}), the material is a semiconductor with a negligible gap, and it is similarly permissive to nonadiabatic transitions.
Within this critical region, as in graphene~\cite{Zurron2019}, the lowered bandgap produces a much {\color{alexis} higher e-h population}, which naturally raises the harmonic emission. This is shown in Fig.~\ref{figure4}, which displays the total emission for both of the phase-space scans of Fig.~\ref{figure3} for each of the driving helicities.

This enhanced emission acts naturally as a litmus test that the topological phase boundary itself has been reached, to complement the CD as a signature of which topological phase the material is in. Moreover, the total-emission scans of Fig.~\ref{figure4} confirm that the features observed previously also hold more broadly over the phase diagram: for one, the topologically-nontrivial phase shows a higher emissivity than the trivial material, at similar bandgaps (shown in Figs.~\ref{figure4}(c,f)), and, similarly, there are noticeable changes in the harmonic cutoff as well.

\section{Dephasing time in the harmonic emission for topological phases}%
\label{sec:dephasingt2}%

It is also important to point out that the features previously shown both in the harmonic intensity and the CD depend (sometimes sensitively) on the dephasing time $T_2$ in our calculation; we now turn to examine this dependence.
Since the electron-hole trajectory mechanism is modified by the dephasing time, particularly for long trajectories, the harmonic intensity, and from it the CD, inherit this effect, as expected~\cite{VampaPRL2014, VampaPRB2015, VampaJPB2017}.~On the other hand, and somewhat surprisingly, in the topological phase the CD is more robust against dephasing than in the trivial phase, particularly for the $k=3n+1$ co-rotating harmonics, which again points to a clear role of the material's topology in its harmonic response, and cements the interaction between the two as a clear target for continued investigation.

\begin{figure*}[htbp]
\begin{center}
    \includegraphics[width=7.cm]{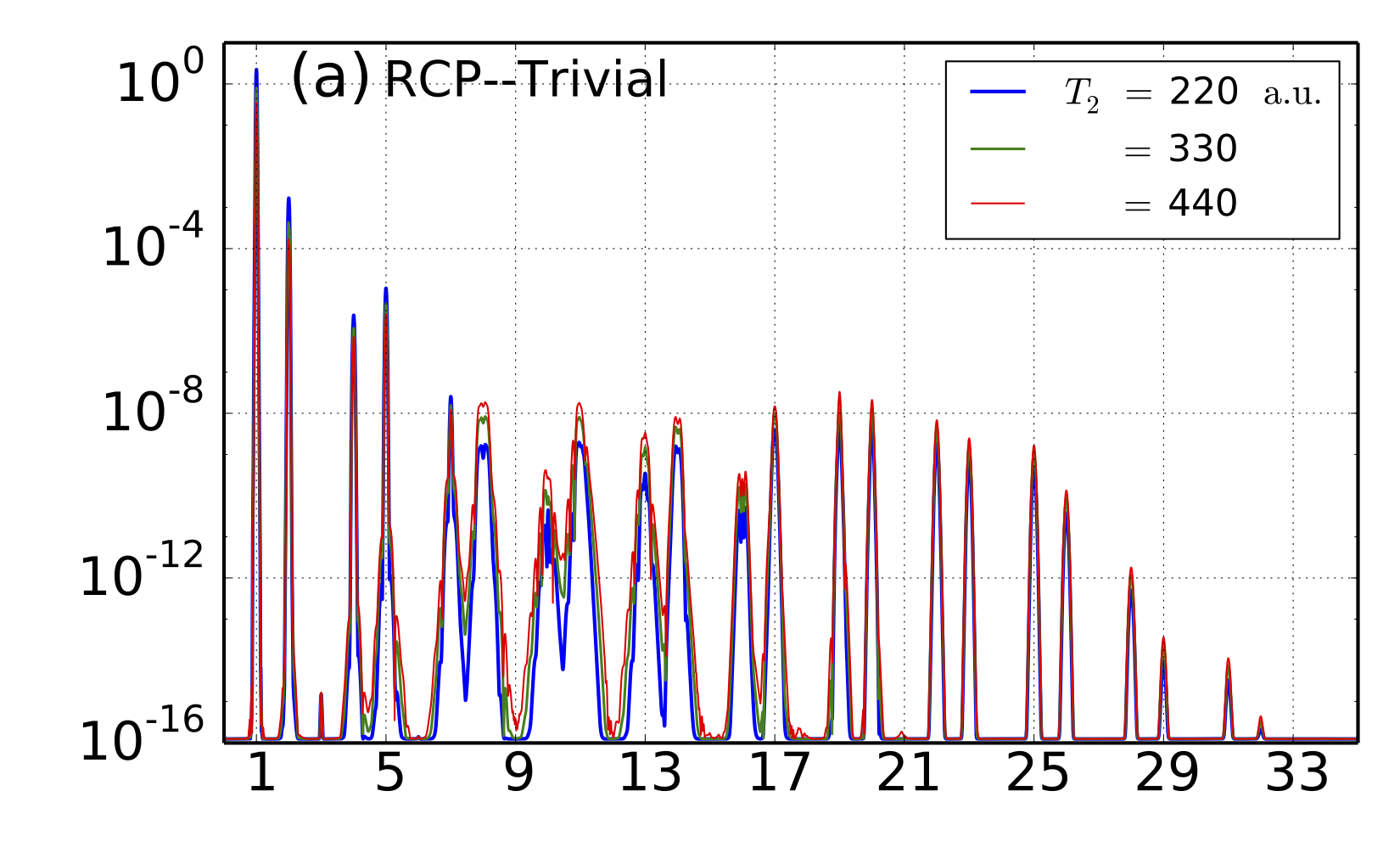}
    \includegraphics[width=7.cm]{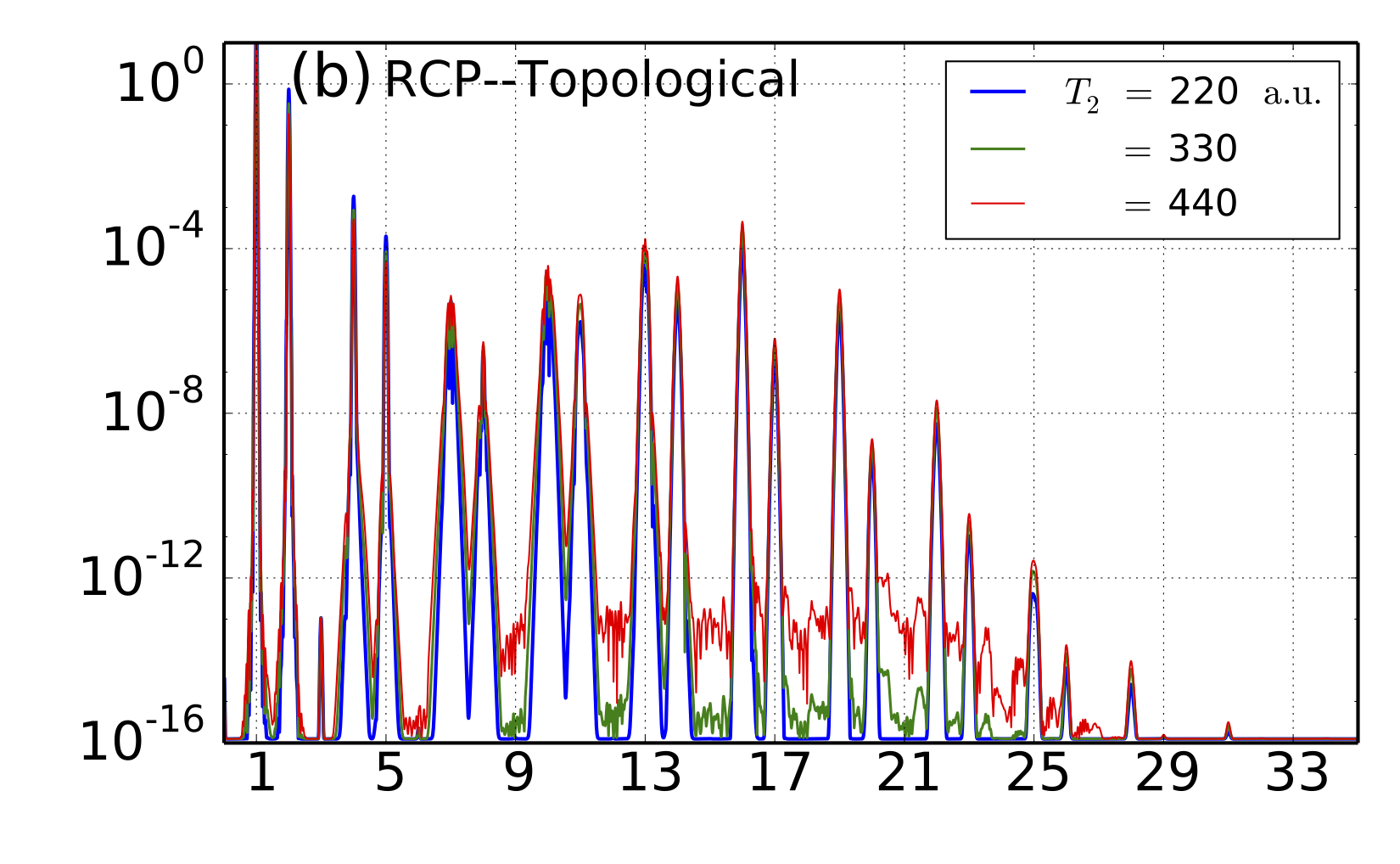}
    \includegraphics[width=7.cm]{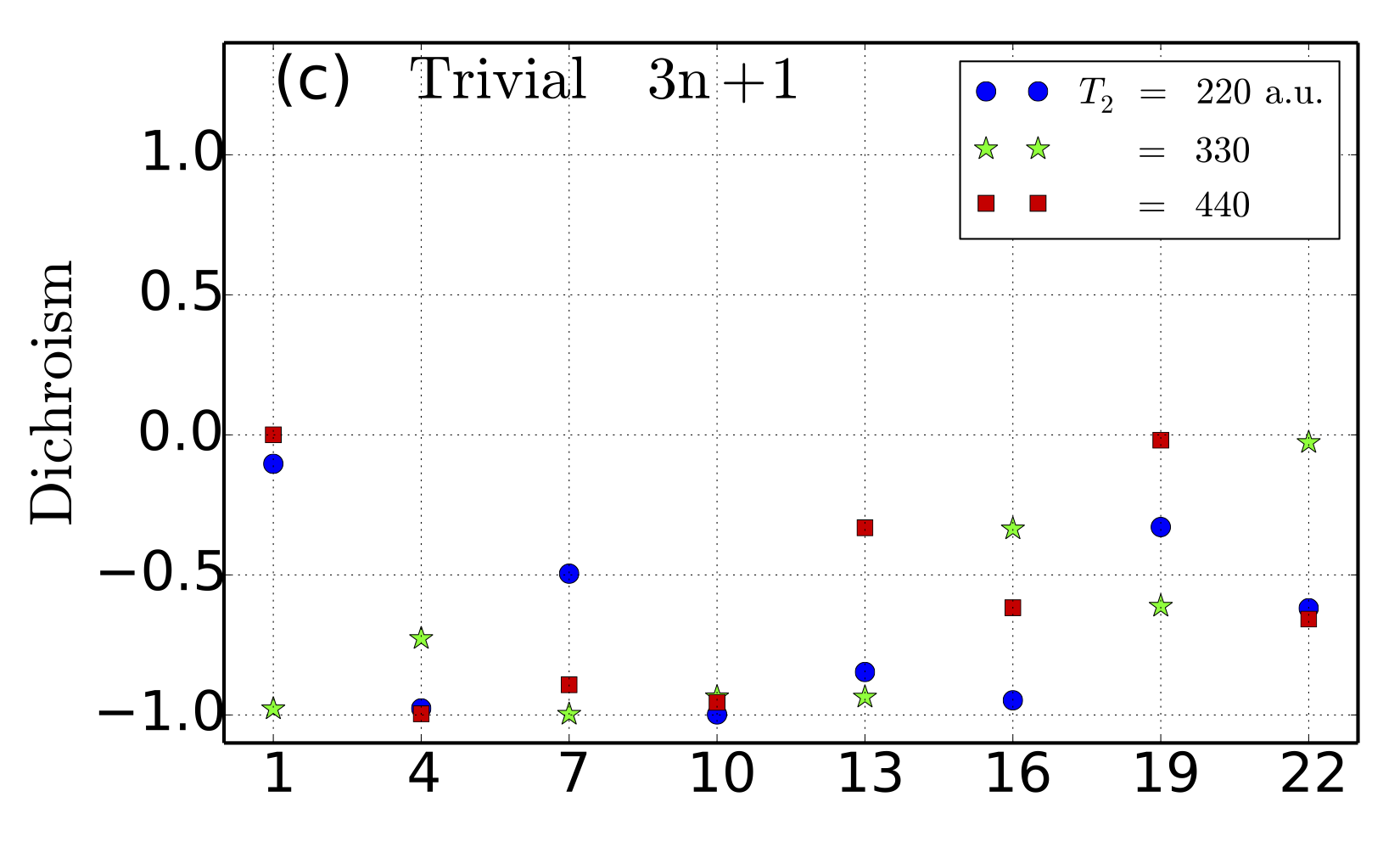}
    \includegraphics[width=7.cm]{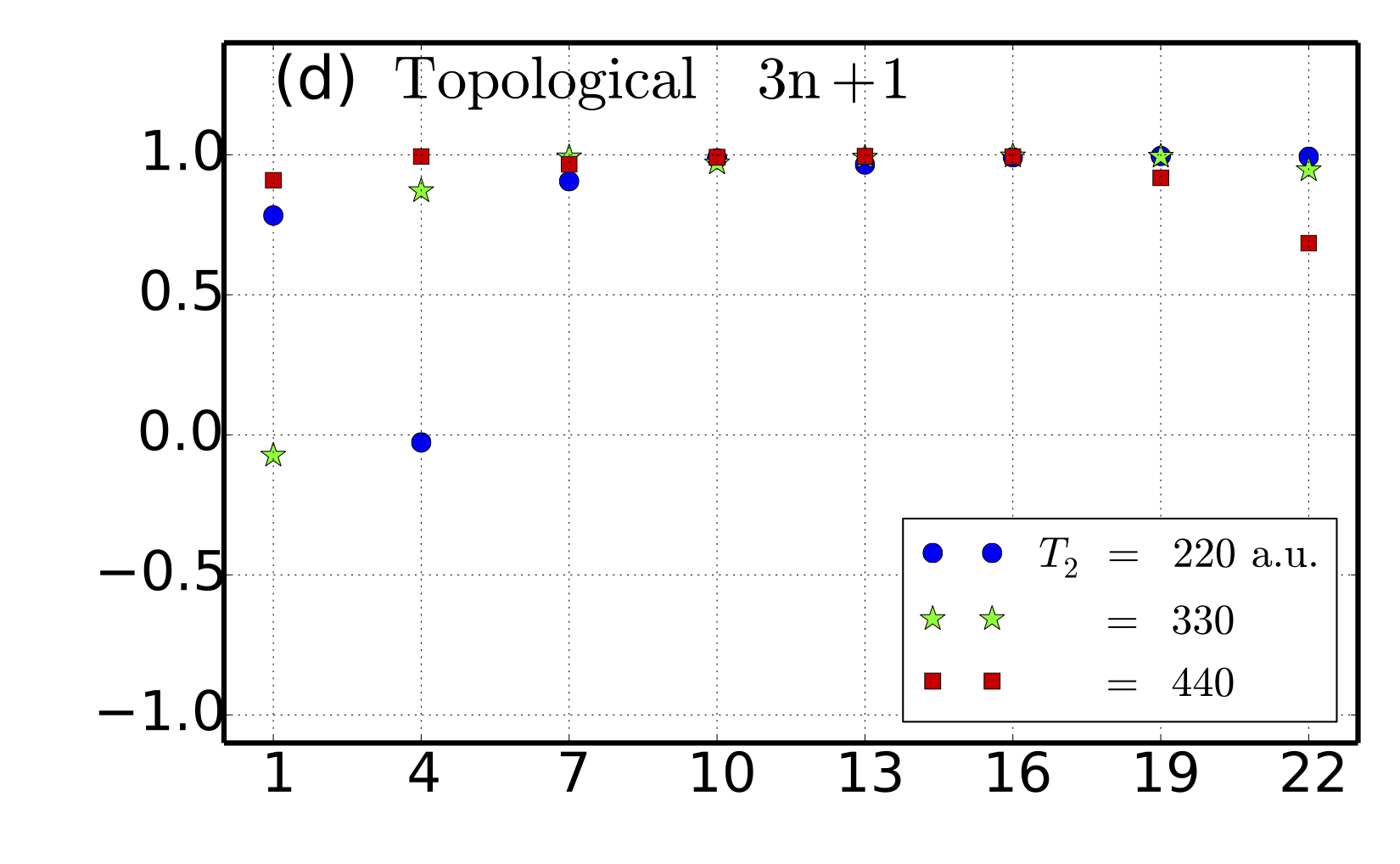}
    \includegraphics[width=7.cm]{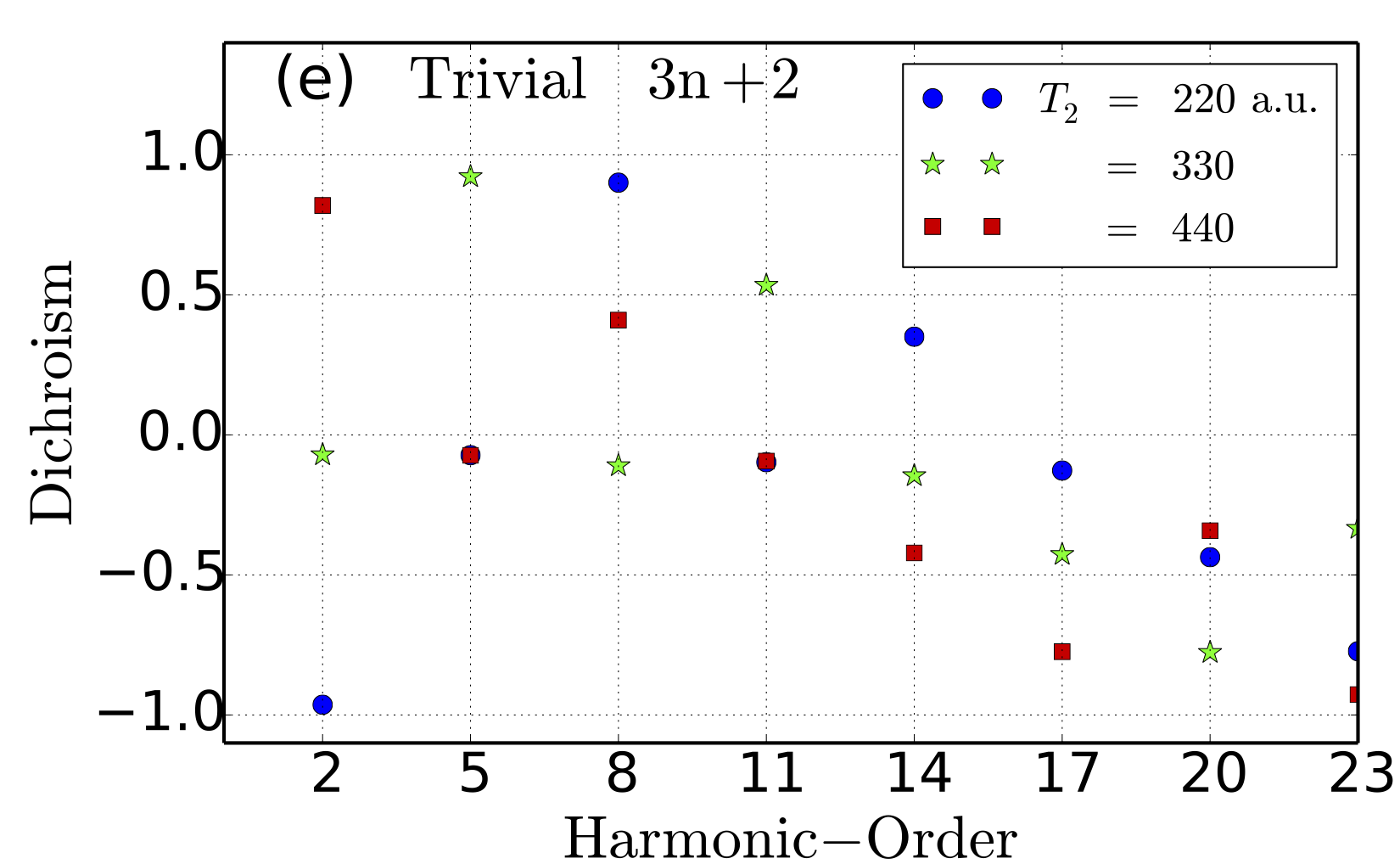}
    \includegraphics[width=7.cm]{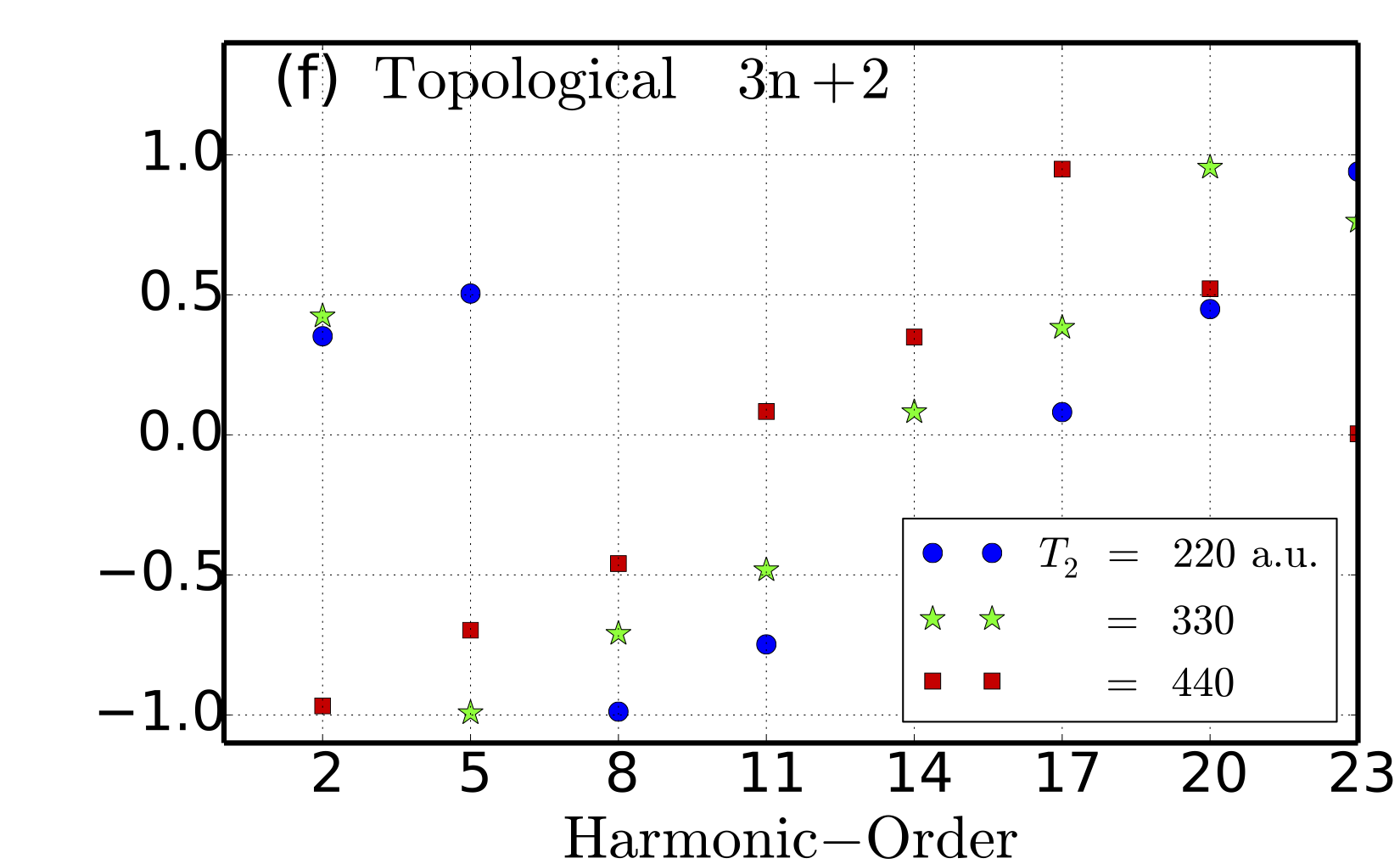}
  \caption{%
Role of the dephasing time in the high-harmonic emission and in its circular dichroism.
(a,b) {\color{alexis}High-harmonic generation} spectra for various dephasing times $T_2$, for trivial and topological phases, respectively, driven by a right-circularly polarized laser.
(c-f) Circular dichroisms for $3n+1$ (co-rotating) and $3n+2$ (contra-rotating) harmonic orders ((c,d) and (e,f), {\color{alexis}respectively}), for trivial (c,e) and topological (d,f) phases.
The parameters for the material and the driving laser are the same as in Fig.~\ref{DipoleMoments} with a slightly altered peak field strength, $E_0=\SI{0.005}{\au}$, and pulse duration, $N_c=16$.
}
  \label{Dephasing}
  \end{center}
\end{figure*}

We now study the role of the dephasing time $T_2$ on the CD for the trivial and topological phases. Figures~\ref{Dephasing}(a,b) show the HHG spectra as a function of $T_2$ for trivial and nontrivial topological phases, respectively. The emitted harmonic intensity is sensitive to the dephasing time, as shown also in literature for linearly polarized lasers~\cite{VampaPRL2014}. Here we also observe sensitivity of the HHG emission yields with respect to $T_2$ for both trivial and topological phases. $T_2$ usually reduces the coherence of the occupations and therefore also the electron-hole recombination mechanisms, more strongly for long trajectories~\cite{VampaPRL2014}.

These effects also appear to take place in topological phases under this simple two-band approximation, which would lead to a potential dependence of the dichroism for the different harmonic orders.
Additionally, we computed the HHG spectra for the case $T_2\rightarrow\infty$ (not shown, for simplicity), in which case the harmonic plateau becomes extremely noisy, making it impossible to extract any reliable insights from those simulations. 
Similar features can be also noticed in the numerical background noise of Fig.~\ref{Dephasing}(b) as the dephasing time increases.

For the trivial phase, Figs.~\ref{Dephasing}(c,e) show the dichroism as a function of $T_2$ for $3n+1$ (co-rotating) and $3n+2$ (counter-rotating) harmonic orders, respectively; we find a strong dependence of the CD on the dephasing time for each individual order.~Nevertheless, when averaging the CD for co-rotating orders over $T_2$, it remains negative, providing a systematic tendency around the plateau and, likely, the cut-off.
In contrast, for the counter-rotating orders, CD oscillates between positive and negative values depending on lowest or highest region of the spectrum.~{\color{alexis}This behavior} seems to be a natural consequence of the complex relationship between the electron-hole recombination mechanism, intra-band and inter-band mechanisms and the~$T_2$.

In contrast, the co-rotating CD for the topological phase shown in Fig.~\ref{Dephasing}(d) is much more robust against $T_2$, exhibiting values around +1, most noticeably when $\mathrm{HO}>4^{\rm th}$. For counter-rotating orders (see Fig.~\ref{Dephasing}(f)), the CD describes similar features across the whole spectrum as in the trivial phase. In that sense, the CD depends on the region of the harmonic spectrum at play. 
These observations show that the CD is a more robust quantity for the co-rotating than for the counter-rotating harmonic orders, which is our main reason for focusing on the CD of the former in the previous section.

\section{Conclusions and outlook}
\label{sec:conclusions}

\noindent
In summary, our theory shows that the {\color{alexis}high-harmonic generation (HHG) spectrum}~is
\begin{itemize}[itemsep=0em,parsep=0.3em,topsep=0.3em,partopsep=0.3em]
\item able to probe topological phase  transitions; 
\item able to test topological invariants by using {\color{alexis}the Circular Dichroism~(CD)} of the co-rotating harmonics; and
\item extremely sensitive to the breaking of both time-reversal and inversion symmetries (TRS, IS, respectively).
\end{itemize}
%
{\color{alexis}Our results are applicable to {\color{alexis}the Haldane model (HM)}, so they constitute clear evidences that the HHG process can trace topological phases and transitions in a topological Chern Insulator}.~But it is still necessary to explore other classes of Chern insulators as well as Topological Insulators to confirm that this HHG-circular dichroism observable iss {\color{alexis}{\it a broadly} defined quantity} to map topological transitions.~{\color{alexis}In particular, we show $\CD$ calculations for the so-called Wilson mass model~\cite{WilsonMassModel} (WMM) which suggests that the results obtained by the HM are generally extended to Chern Insulators. Since the main focus of this paper is on HM, we will show partially the regarding evidences to WMM in Appendix A}.~The formulation of relativistic fermions in lattice gauge theories~\cite{W1} is hampered by the fundamental problem of species doubling~\cite{W2a,W2b}, namely, the rise of spurious fermions that modify the physics at long wavelengths.~To prevent the abundance of such fermion doublings, a suitable tailoring of their masses is required, leading to the so-called Wilson fermions \cite{WilsonMassModel}.~The WMM with an inverted mass gives rise to a certain axion background \cite{W5a,W5b,W5c,W6a,W6b,W7a,W7b}, and thus may help explain the lack of experiments confirming the charge-parity violation in strong interactions (for quantum simulations of this model see Ref.~\cite{PRLNathan}).~When exposed to ultra-intense short laser pulses, the WMM shows similar behavior to the one we report here for the HM; these results will be reported in a future publication.
In that general sense, our theory can be applied to a larger range of topological materials with similar band-gap properties described in this manuscript, as well as extended to THz sources~\cite{RodriguezL2018}. 
Another possible application could be Bi$_2$Se$_3$, which is a good candidate for topological insulator; possible layer-thick modifications could be created in this material in order to control the topological transition as in Ref.~\cite{Neupane2014}. Also,  THz sources should open a path to access HHG in topological materials by pushing the driving photon energy well below the band-gap in two different topological orders; this would also raise new questions about exploring the dynamics in strongly-correlated systems with spin-orbit couplings and degeneracies, in particular in bilayer systems and twistronics \cite{Cao18-1,Cao18-2,Lu19, Balents2020, Avetissian2020bilayer}. 

Finally, it is important to note that all of these ideas can be applied and tested in quantum simulators~\cite{AcinRoadmap}, in particular using ultracold atoms and lattice shaking~\cite{LSA12}, as well as Rydberg atoms, polaritons, or circuit QED \cite{Bloch08, specialissueNaturePhys, specialissueNaturePhys1, specialissueNaturePhys2, specialissueNaturePhys3, specialissueNaturePhys4}. 
The use of quantum simulators to study strong-field phenomena is an emerging application of the former~\cite{Sala2017, Senaratne2018, Ramos2019}, and it is naturally suited to the study of systems in ordered lattices, so it should provide additional clarity to the role of material topology in high-harmonic emission.

\section*{Acknowledgements}
We thank T.T.\ Luu and H.J.\ W\"orner for providing experimental data in editable format.~A.C. thanks D.\ Baykusheva for running extra calculations of {\color{alexis}Figs.~\ref{figure3}~and~\ref{figure4}}, in the supercomputer Sherlock at SLAC, and for useful discussions of the CD results, also to N.\ Aguirre and A.\ Adedoyin for suggestions in technical details of numerical implementations.
We acknowledge the Spanish Ministry MINECO (National Plan 15 Grant: FISICATEAMO No. FIS2016-79508-P, SEVERO OCHOA No. SEV-2015-0522, FPI), European Social Fund, Fundaci\'o Cellex, Fundació Mir-Puig, Generalitat de Catalunya (AGAUR Grant No. 2017 SGR 1341 and CERCA/Program), ERC AdG OSYRIS and NOQIA, EU FETPRO QUIC, EU FEDER, European Union Regional Development Fund~- ERDF Operational Program of Catalonia 2014-2020 (Operation Code: IU16-011424), MINECO-EU QUANTERA MAQS (funded by The State Research Agency (AEI) PCI2019-111828-2 / 10.13039/501100011033), and the National Science Centre, Poland-Symfonia Grant No. 2016/20/W/ST4/00314.
E.P.\ acknowledges CELLEX-ICFO-MPQ Fellowship funding.
A.P.\ acknowledges funding from Co\-mu\-ni\-dad de Madrid through TALENTO grant ref.\ 2017-T1/{\allowbreak}IND-{\allowbreak}5432.
M.F.C.\ thanks support of the project Advanced research using high intensity laser produced photons and particles (CZ.02.1.01/0.0/0.0/16 019/0000789) from the European Regional Development Fund (ADONIS).
S.P.K.\ acknowledges financial support from the UC Office of the President through the UC Laboratory Fees Research Program, Award Number LGF-17- 476883.
A.S.M.\ acknowledges funding from the UK Engineering
and Physical Sciences Research Council (EPSRC) InQuBATE grant EP/P510270/1.

A.C., W.Z., S.K., C.T. and A.S.\ acknowledge support from LDRD No.\ 2160587 and the Science Campaigns, Advanced Simulation, and Computing and LANL, which is operated by LANS, LLC for the NNSA of the U.S.\ DOE under Contract No.\ DE-AC52-06NA25396 CNLS-LANL.
A.C., D.K. and D.E.K. acknowledge support from Max Planck POSTECH/KOREA Research Initiative Program [Grant No 2016K1A4A4A01922028] through the National Research Foundation of Korea (NRF) funded by Ministry of Science, ICT \& Future Planning, partly by Korea Institute for Advancement of Technology (KIAT) grant funded by the Korea  Government (MOTIE) (P00\-08763, The Competency Development Program for Industry Specialist).

\appendix
\section{Hamiltonian for the Wilson mass model}\label{sec:lin_cut}

{\color{alexis}
 In the main text, we focused on the Haldane model with inversion symmetry (IS) breaking. One question is, whether or not the observed Circular Dichroism ($\CD$) produced by high-harmonic generation is model dependent.
Here, we fully answer that question by introducing the simplest square lattice with two localized orbitals per atomic site (${\mathcal A}$ and ${\mathcal B}$), dubbed as {\it Wilson mass model} (WMM). The Hamiltonian has IS ${\mathcal P} H_0({\bf k}){\mathcal P}^{\dagger} = H_0(-{\bf k})$, in which {${\mathcal P}$} defines the IS operator for ${\bf k} \rightarrow -{\bf k}$.~Note, nevertheless, this WMM breaks the Time-Reversal Symmetry (TRS) as well as Haldane model does. The WMM Hamiltonian reads,
\begin{align}
\label{WilsonHamiltonian}
H_0({\bf k})
&=\Delta\sin\left(k_x\,a_0\right) \sigma_1 + \Delta\sin\left(k_y\,a_0\right) \sigma_2 \nonumber \\ 
 & - \left\{2t_1\left[ \cos(a_0\,k_x) + \cos(a_0\,k_y) \right] +\mu \right\} \sigma_3,
\end{align}
where $\Delta$ is the hopping parameter between ${\mathcal A}$ and ${\mathcal B}$. This denotes the overlap integral of the two nearest neighbors with localized states ${\mathcal A}$ and ${\mathcal B}$, which in principle can be purely imaginary. Here $\mu$ is the relative on-site potential between states ${\mathcal A}$ and ${\mathcal B}$ of the square Wilson lattice.~The lattice constant is $a_0$ and $t_1$ is the hopping parameter of the same two-consecutive-localized states ${\mathcal A}-{\mathcal A}$ or~${\mathcal B}-{\mathcal B}$~at the square lattice.~We define the so called Wilson mass ${\mathcal M}$ term~\cite{WilsonMassModel,W6a} by $\mu = 2 t_1 {\mathcal M}$ and re-scale $\Delta = 2t_1 \delta_0$ by $\delta_0$. Here,  ${\mathcal M}$ and $\delta_0$ are adimensional parameters which control the topological phases with $C=\pm1$ and $C=0$.	
The calculated high-harmonic spectrum for left and right circularly polarized laser-lights are shown in Fig.~(\ref{WilsonTopological}a) for two different topological phases, trivial $C=0$ and topological $C=+1$, respectively.
\begin{figure*}[htbp]
\begin{center}
    \includegraphics[width=7.5cm]{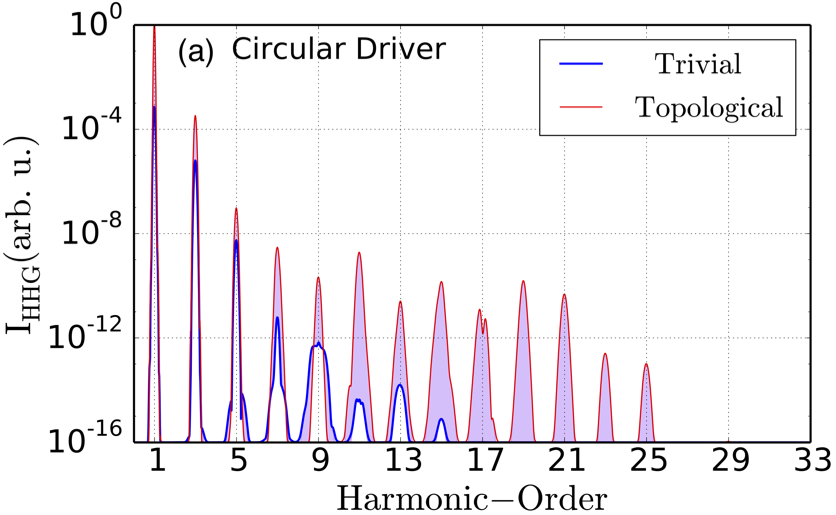}\hspace{1cm}
    \includegraphics[width=7.5cm]{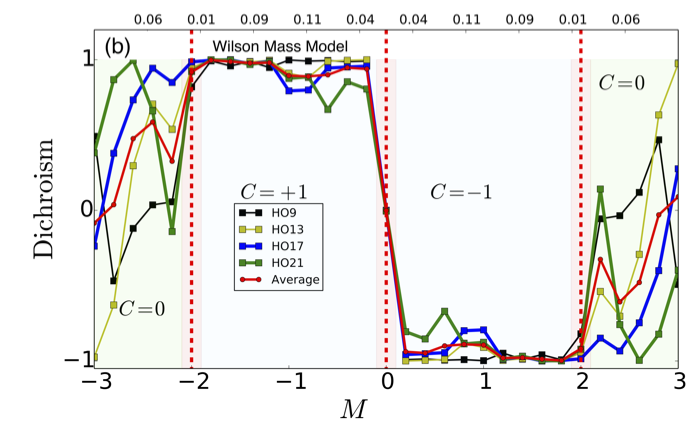}
  \caption{ \color{alexis}Trivial, topological phases and transitions with high-harmonic generation in Wilson mass model. (a)~Total high-harmonic generation emissions for trivial (blue) and topological (red with violet shadow) phases.~The Wilson mass model (WMM) parameters of lattice constant $a_0=3.33$~a.u. and others for trivial, $C=0$ (topological, $C=+1$) phase are, hopping parameter $t_1=0.021$~a.u. ($0.034$~a.u.) with Wilson mass term $\mathcal{M}=\frac{10}{3}$, ($\frac{6}{5}$) and hopping $\delta_0 =  0.0434$ ($0.0723$). Those WMM parameters produce equal band gaps of $E_g\approx3.0$~eV for trivial and the topological phases. (b) Depicts the topological phase transition for co-rotating harmonic orders ($3n+1$) by the normalized circular dichroism (CD) as a function of $\mathcal{M}$ at a fixed hopping parameter of $t_1=0.034$~a.u.~The top horizontal axis points to the energy gap $E_g$ (atomic units are used) as a function of $\mathcal{M}$. The $\CD$ shows two different plateaus of $\CD\approx+1$ and $\CD\approx-1$ for respectively the topological invariants $C=+1$ and $C=-1$. On the contrary, for regions where the Chern Number $C=0$, the CD is oscillating. The latter observations are absolutely consistent with the results provided by Haldane model in Fig.~(\ref{figure3}c). The laser parameters are the same as in Fig.~(\ref{figure3}).}
  \label{WilsonTopological}
  \end{center}
\end{figure*}
We observe a predominant behavior of the harmonic nonlinear responses from the topological phase, a clear enhancement (two or three orders of magnitude across the harmonic-plateau) and an harmonic cut-off shift too. 
Finally, we report in Fig.~(\ref{WilsonTopological}b) calculations of the $\CD$ as a function of the mass ${\mathcal M}$ in which three distinct topological phases exist, as well as phase transitions. The $\CD$ has values of $\CD\approx\pm1$ for the topological invariants $C=\pm1$, and $\CD$ oscillates for the trivial phase. The latter observations denote three different topological structures for the circular dichroism in the three corresponding topological phases.~These $\CD$ results produced by HHG  in the Wilson Mass Model are clearly consistent with the evidences obtained from Haldane model. Hence, we can conclude that the circular dichroism is a general tool which can map topological phases and transitions in Chern Insulators.}

\section{Linear field scaling of the high harmonic cut-off}\label{sec:lin_cut}

In order to validate our theoretical model and numerical variable-gauge methods, we investigate how the HHG emission cut-offs for the total, intra-band and inter-band contributions behave as a function of the laser field strengths, for both trivial and topological phases.

\begin{figure*}[htbp]
\begin{center}
    \includegraphics[width=6.5cm]{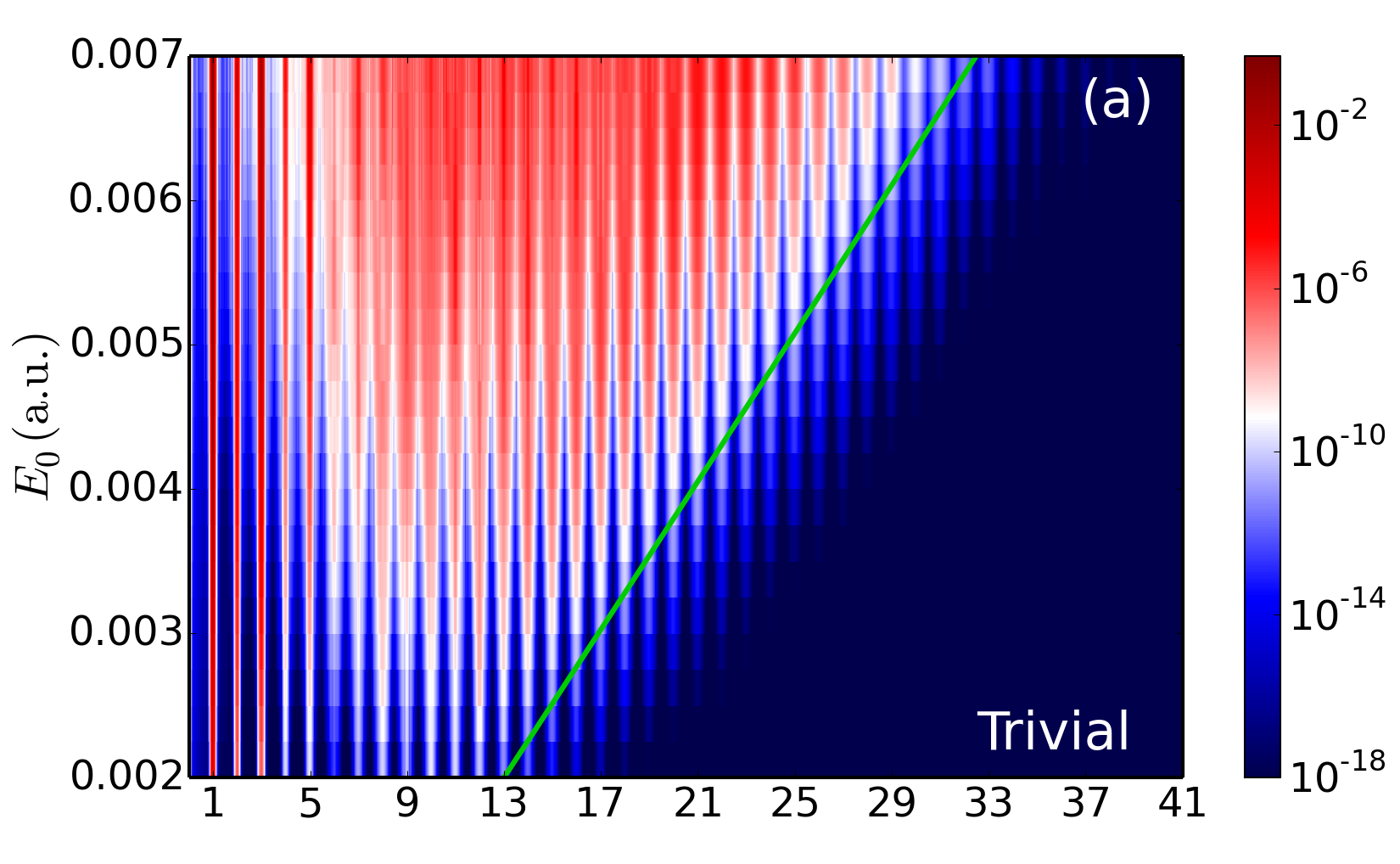}\hspace{1cm}
    \includegraphics[width=6.5cm]{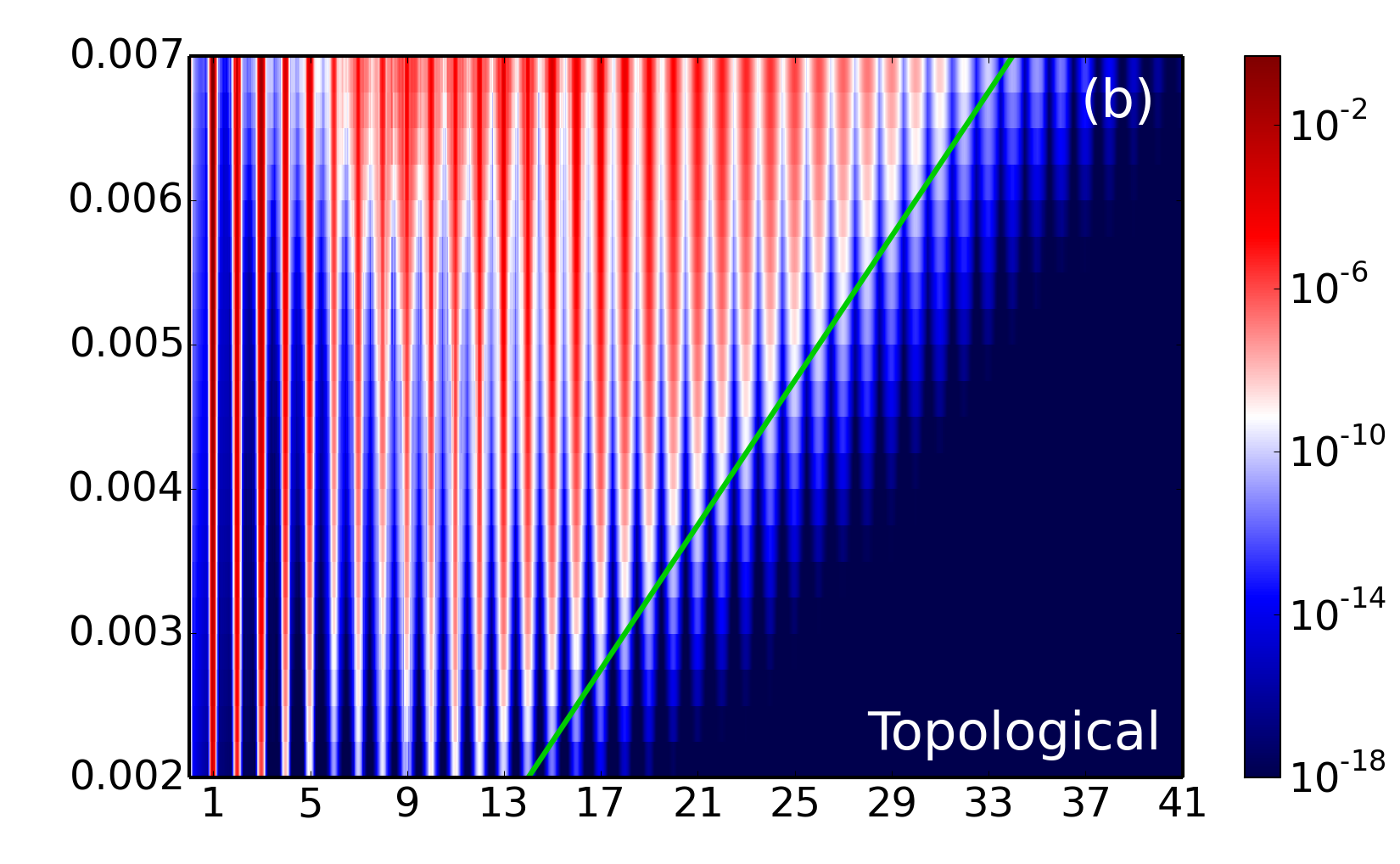}
    \includegraphics[width=6.5cm]{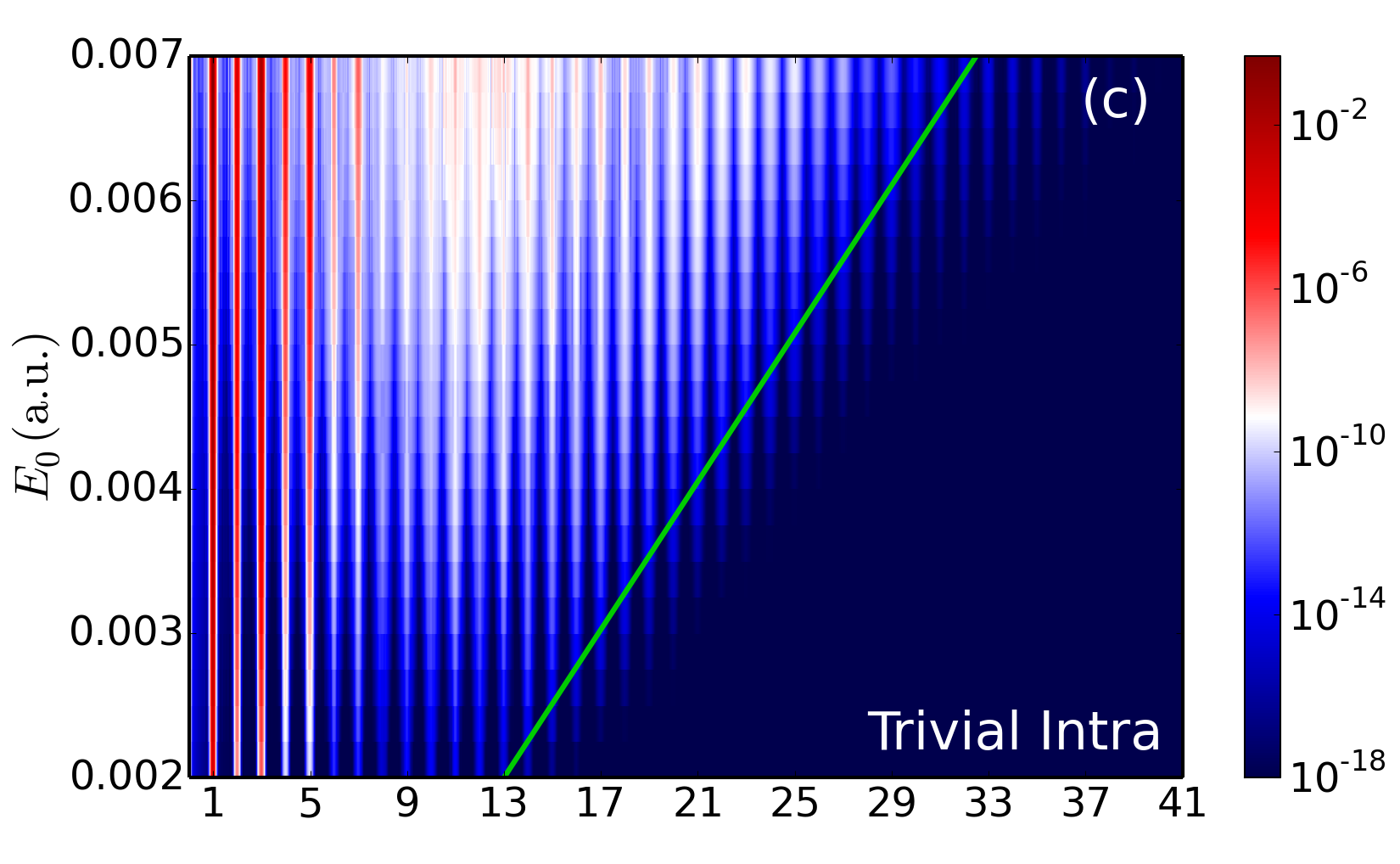}\hspace{1cm}
    \includegraphics[width=6.5cm]{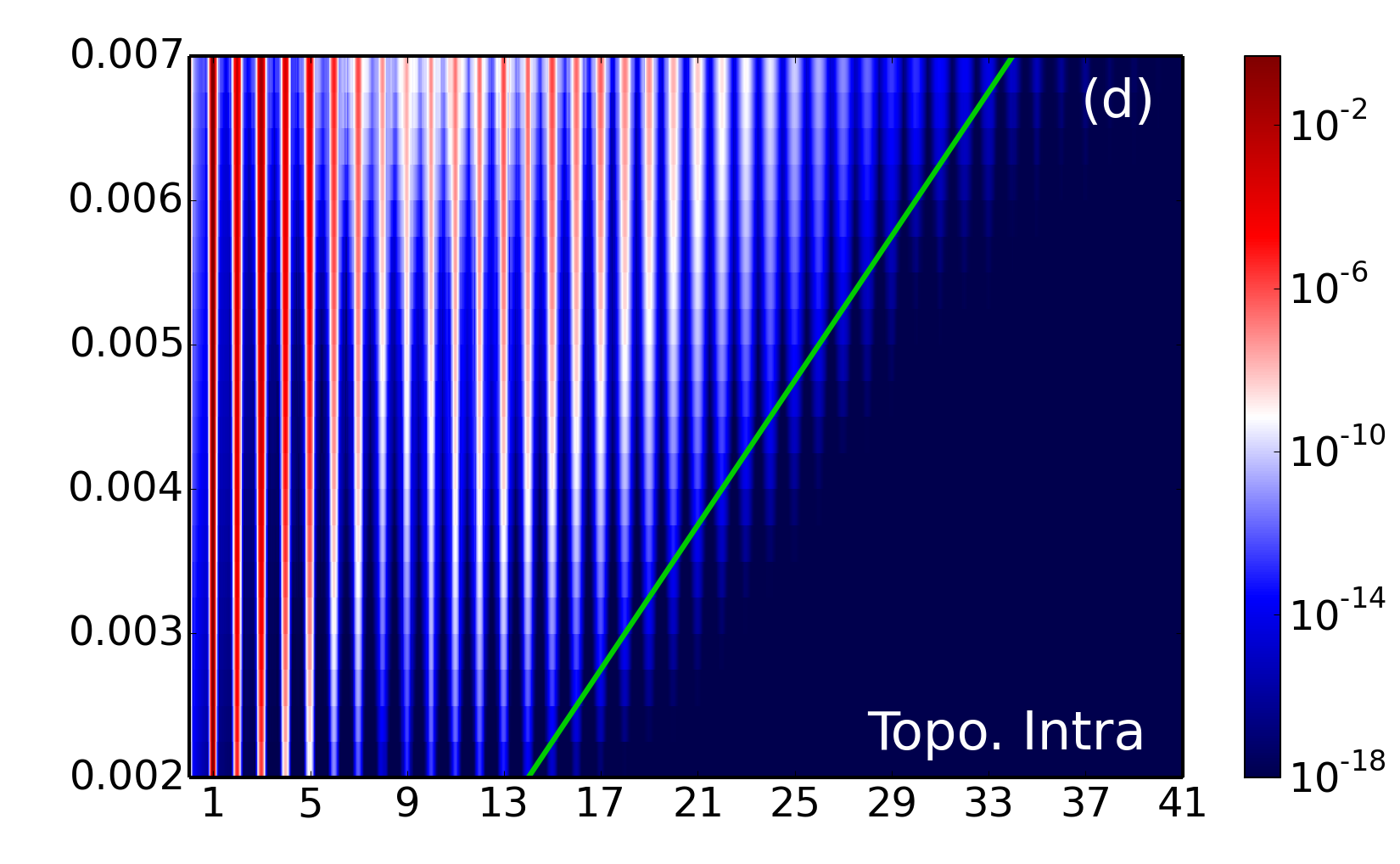}
    \includegraphics[width=6.5cm]{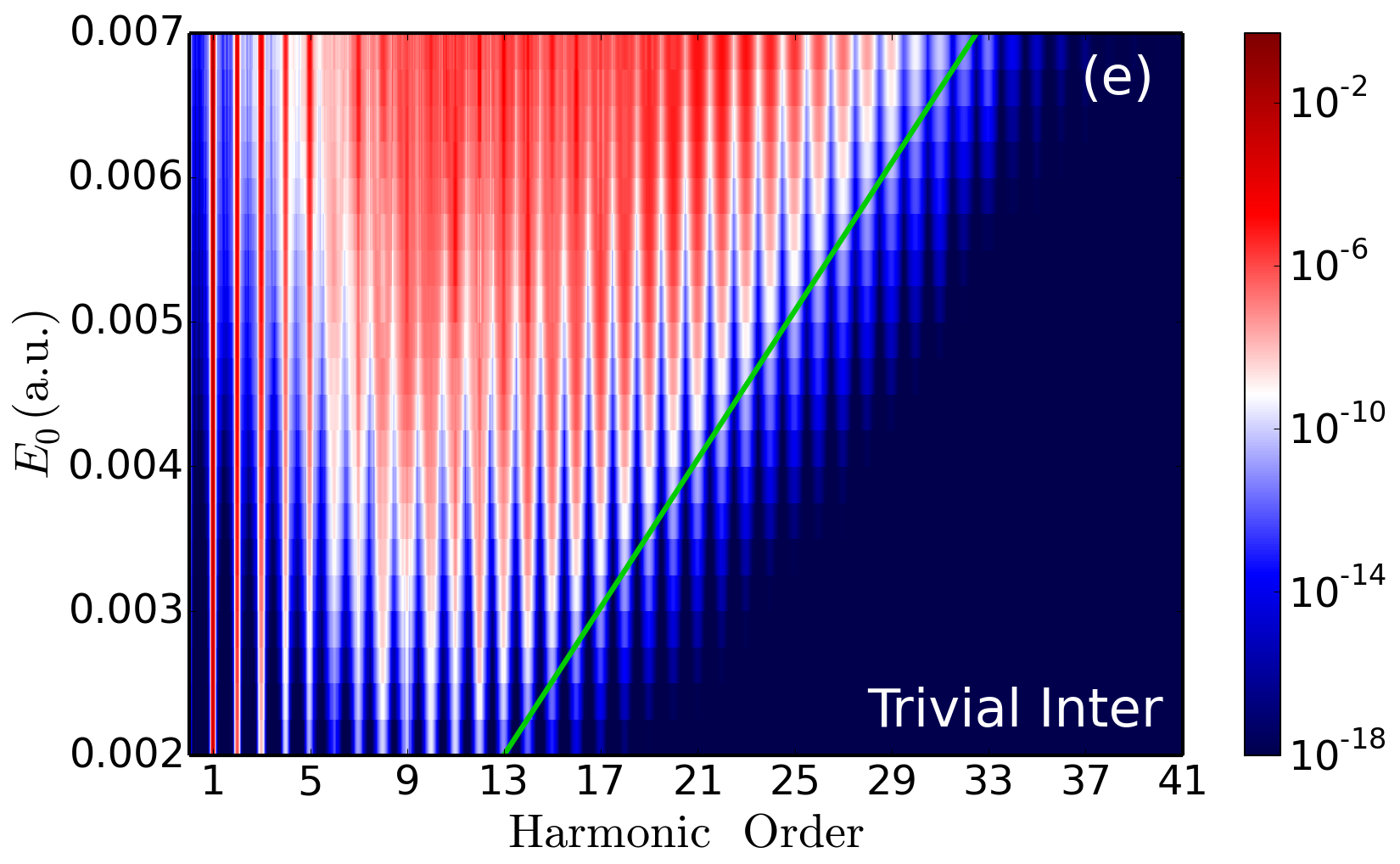}\hspace{1cm}
    \includegraphics[width=6.5cm]{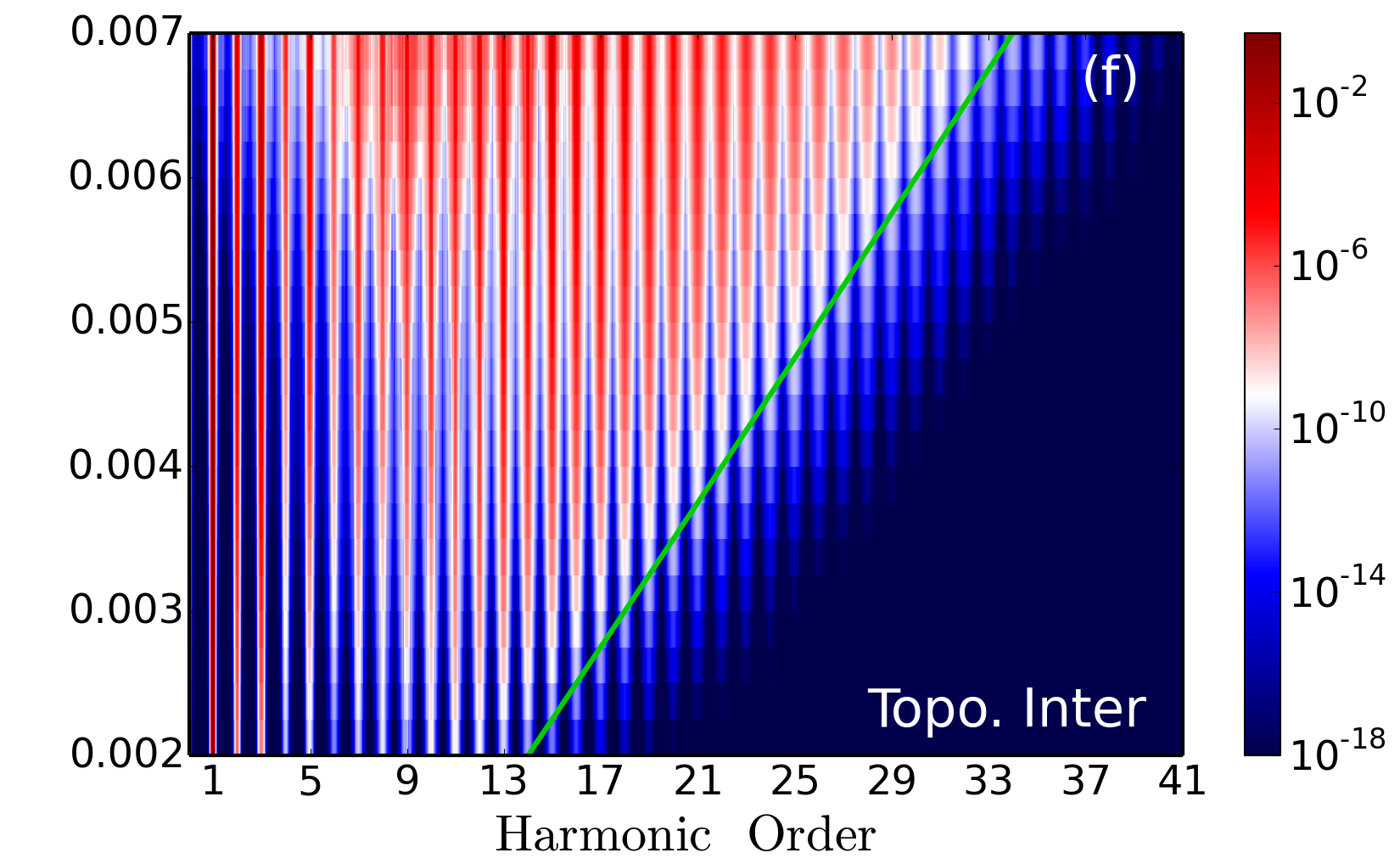}
  \caption{
  Trivial and topological high harmonic cut-off law for linear polarized driving fields. Total HHG emissions for trivial and topological phases, with the same HM parameters than those used in Fig.~\ref{DipoleMoments}, as a function of the laser field strength $E_0$ are depicted in (a,b), respectively. (c,e) The decomposition of the total harmonic emission for the intra-band and inter-band current emissions of the trivial phase, respectively. (d,f) The intra-band and inter-band {\color{alexis}current emissions} for the topological phase. Green lines mimic the linear-cut off for both trivial and topological phases. The laser field for these HHG calculations is linearly polarized along {\color{alexis} the} $\rm K'-\Gamma-K$ direction of the honeycomb hexagonal lattice.
  }
  \label{LinearScaling}
  \end{center}
\end{figure*}

Figures~\ref{LinearScaling}(a,b) show the results for the total harmonic emission $I_{\rm HHG}(\omega)=\omega^2 \left[|J^{(x)}(\omega)|^2+|J^{(y)}(\omega)|^2\right]$ of the trivial and nontrivial topological phases. First, we see a clear linear cut-off in terms of the laser strength for the topological trivial case. This tendency is in very good agreement with the pioneering experimental observation of Ref.~\citealp{GhimireNatPhy2011} and the theoretical confirmation by Vampa et al.\ in Refs.~\citealp{VampaPRL2014,VampaPRB2015}. In the case of the topological phase, we observe a linear cut-off scaling as a function of the laser field strength, similar to the trivial one.

We also depict, in Figs.~\ref{LinearScaling}(c,d), the corresponding intra-band harmonic emissions for trivial and {\color{alexis} nontrivial phases}, via the group and anomalous velocity emissions. We find {\color{alexis}that the intra-band case} for the topological emission seems to exhibit a much larger intensity yield than the trivial phase (though the coupling between the two also needs to be considered in more detail~\cite{GoldePRB2008}).

In contrast, Figs.~\ref{LinearScaling}(e,f) show that, for our excitation conditions, the inter-band contributions for the trivial and topological phases dominate around the plateau and cut-off in comparison to intra-band emissions.  One therefore needs to carefully consider and analyze the inter-band mechanism for high nonlinear emission, i.e.\ $\mathrm{HO}>5^{\rm th}$, and not uniquely the intra-band mechanism.

\bibliographystyle{arthur}
\bibliography{references}

\begin{thebibliography}{100}
\providecommand{\url}[1]{\texttt{#1}}
\providecommand{\urlprefix}{URL }
\expandafter\ifx\csname urlstyle\endcsname\relax
  \providecommand{\doi}[1]{doi:\discretionary{}{}{}#1}\else
  \providecommand{\doi}{doi:\discretionary{}{}{}\begingroup
  \urlstyle{rm}\Url}\fi
\providecommand{\selectlanguage}[1]{\relax}
\providecommand{\bibAnnoteFile}[1]{%
  \IfFileExists{#1}{\begin{quotation}\noindent\textsc{Key:} #1\\
  \textsc{Annotation:}\ \input{#1}\end{quotation}}{}}
\providecommand{\bibAnnote}[2]{%
  \begin{quotation}\noindent\textsc{Key:} #1\\
  \textsc{Annotation:}\ #2\end{quotation}}
\providecommand{\eprint}[2][]{\url{#2}}

\bibitem{KosterlitzNobel2016}
{J.~M. Kosterlitz}.
\newblock Nobel lecture: Topological defects and phase
  transitions\href{http://dx.doi.org/10.1103/RevModPhys.89.040501}{.
\newblock \emph{Rev. Mod. Phys.} \textbf{89} no.~4, p. 040\,501 (2017)}.
\newblock
  \href{https://www.nobelprize.org/nobel_prizes/physics/laureates/2016/kosterlitz-lecture.html}{Nobel
  Prize lecture}.
\bibAnnoteFile{KosterlitzNobel2016}

\bibitem{KlitzingNobel1985}
{K.~von Klitzing}.
\newblock The quantized {Hall}
  effect\href{http://dx.doi.org/10.1103/RevModPhys.58.519}{.
\newblock \emph{Rev. Mod. Phys.} \textbf{58} no.~3, pp. 519--531 (1986)}.
\newblock
  \href{https://www.nobelprize.org/nobel_prizes/physics/laureates/1985/klitzing-lecture.html}{Nobel
  Prize lecture}.
\bibAnnoteFile{KlitzingNobel1985}

\bibitem{StormerNobel1998}
{H.~L. Stormer}.
\newblock Nobel lecture: The fractional quantum {Hall}
  effect\href{http://dx.doi.org/10.1103/RevModPhys.71.875}{.
\newblock \emph{Rev. Mod. Phys.} \textbf{71} no.~4, pp. 875--889 (1999)}.
\newblock
  \href{https://www.nobelprize.org/nobel_prizes/physics/laureates/1998/stormer-lecture.html}{Nobel
  Prize lecture}.
\bibAnnoteFile{StormerNobel1998}

\bibitem{Haddad2016}
{D.~Haddad} et~al.
\newblock Invited article: A precise instrument to determine the {Planck}
  constant, and the future kilogram\href{http://dx.doi.org/10.1063/1.4953825}{.
\newblock \emph{Rev. Sci. Instrum.} \textbf{87} no.~6, p. 061\,301 (2016)}.
\bibAnnoteFile{Haddad2016}

\bibitem{ShouCheng2011}
{X.-L. Qi and S.-C. Zhang}.
\newblock Topological insulators and
  superconductors\href{http://dx.doi.org/10.1103/RevModPhys.83.1057}{.
\newblock \emph{Rev. Mod. Phys.} \textbf{83} no.~4, pp. 1057 --1110 (2011)}.
\newblock \href{https://arxiv.org/abs/1008.2026}{arXiv:1008.2026}.
\bibAnnoteFile{ShouCheng2011}

\bibitem{Hasan2010}
{M.~Z. Hasan and C.~L. Kane}.
\newblock Colloquium: Topological
  insulators\href{http://dx.doi.org/10.1103/RevModPhys.82.3045}{.
\newblock \emph{Rev. Mod. Phys.} \textbf{82} no.~4, pp. 3045--3067 (2010)}.
\newblock \href{https://arxiv.org/abs/1002.3895}{arXiv:1002.3895}.
\bibAnnoteFile{Hasan2010}

\bibitem{Goldman2014}
{N.~Goldman} et~al.
\newblock {Light-induced gauge fields for ultracold
  atoms}\href{http://dx.doi.org/10.1088/0034-4885/77/12/126401}{.
\newblock \emph{Rep. Prog. Phys.} \textbf{77} no.~12, p. 126\,401 (2014)}.
\newblock
  \href{https://arxiv.org/abs/1308.6533}{arXiv:\allowbreak{}1308.\allowbreak{}6533}.
\bibAnnoteFile{Goldman2014}

\bibitem{Ozawa2018}
{T.~Ozawa} et~al.
\newblock Topological
  photonics\href{http://dx.doi.org/10.1103/RevModPhys.91.015006}{.
\newblock \emph{Rev. Mod. Phys.} \textbf{91} no.~1, p. 015\,006 (2019)}.
\newblock
  \href{https://arxiv.org/abs/1802.04173}{arXiv:{\allowbreak}1802.04173}.
\bibAnnoteFile{Ozawa2018}

\bibitem{Huber2016}
{S.~D. Huber}.
\newblock Topological mechanics\href{http://dx.doi.org/10.1038/nphys3801}{.
\newblock \emph{Nat. Phys.} \textbf{12} no.~7, p. 621 (2016)}.
\bibAnnoteFile{Huber2016}

\bibitem{KortKamp2017}
{W.~J.~M. Kort-Kamp}.
\newblock Topological phase transitions in the photonic spin {Hall}
  effect\href{http://dx.doi.org/10.1103/PhysRevLett.119.147401}{.
\newblock \emph{Phys. Rev. Lett.} \textbf{119}, p. 147\,401 (2017)}.
\newblock \href{https://arxiv.org/abs/1709.02057}{arXiv:1707.02057}.
\bibAnnoteFile{KortKamp2017}

\bibitem{Tran2017}
{D.~T. Tran} et~al.
\newblock Probing topology by ``heating'': Quantized circular dichroism in
  ultracold atoms\href{http://dx.doi.org/10.1126/sciadv.1701207}{.
\newblock \emph{Sci. Adv.} \textbf{3} no.~8, p. e1701\,207 (2017)}.
\newblock \href{http://arxiv.org/abs/1704.01990}{arXiv:1704.01990}.
\bibAnnoteFile{Tran2017}

\bibitem{Itatani2004}
{J.~Itatani} et~al.
\newblock Tomographic imaging of molecular
  orbitals\href{http://dx.doi.org/10.1038/nature03183}{.
\newblock \emph{Nature} \textbf{432} no. 432, pp. 867--871 (2004)}.
\bibAnnoteFile{Itatani2004}

\bibitem{Symphony2019}
{K.~Amini} et~al.
\newblock Symphony on strong field
  approximation\href{http://dx.doi.org/10.1088/1361-6633/ab2bb1}{.
\newblock \emph{Rep. Prog. Phys.} \textbf{82} no.~11, p. 116\,001 (2019)}.
\newblock \href{https://arxiv.org/abs/1812.11447}{arXiv:1812.11447}.
\bibAnnoteFile{Symphony2019}

\bibitem{Corkum1993}
{P.~B. Corkum}.
\newblock Plasma perspective on strong field multiphoton
  ionization\href{http://dx.doi.org/10.1103/PhysRevLett.71.1994}{.
\newblock \emph{Phys. Rev. Lett.} \textbf{71}, p. 1994 (1993)}.
\bibAnnoteFile{Corkum1993}

\bibitem{Lewenstein1994}
{M.~Lewenstein} et~al.
\newblock Theory of high-harmonic generation by low-frequency laser
  fields\href{http://dx.doi.org/10.1103/PhysRevA.49.2117}{.
\newblock \emph{Phys. Rev. A} \textbf{49}, p. 2117 (1994)}.
\bibAnnoteFile{Lewenstein1994}

\bibitem{Zuo1996}
{T.~Zuo, A.~D. Bandrauk and P.~B. Corkum}.
\newblock Laser-induced electron diffraction: a new tool for probing ultrafast
  molecular dynamics\href{http://dx.doi.org/10.1016/0009-2614(96)00786-5}{.
\newblock \emph{Chem. Phys. Lett.} \textbf{259}, p. 313 (1996)}.
\newblock \href{https://hal.archives-ouvertes.fr/hal-00679964}{HAL eprint}.
\bibAnnoteFile{Zuo1996}

\bibitem{JensScience}
{B.~Wolter} et~al.
\newblock Ultrafast electron diffraction imaging of bond breaking in di-ionized
  acetylene\href{http://dx.doi.org/10.1126/science.aah3429}{.
\newblock \emph{Science} \textbf{354}, pp. 308--312 (2016)}.
\newblock \href{http://hdl.handle.net/2117/105958}{UPCommons eprint}.
\bibAnnoteFile{JensScience}

\bibitem{GhimireNatPhy2011}
{S.~Ghimire} et~al.
\newblock Observation of high-order harmonic generation in a bulk
  crystal\href{http://dx.doi.org/10.1038/nphys1847}{.
\newblock \emph{Nat. Phys.} \textbf{7} no.~2, pp. 138--141 (2011)}.
\newblock \href{http://www.dimauro.osu.edu/node/299}{OSU eprint}.
\bibAnnoteFile{GhimireNatPhy2011}

\bibitem{VampaJPB2017}
{G.~Vampa and T.~Brabec}.
\newblock Merge of high harmonic generation from gases and solids and its
  implications for attosecond
  science\href{http://dx.doi.org/10.1088/1361-6455/aa528d}{.
\newblock \emph{J. Phys. B: At. Mol. Opt. Phys.} \textbf{50} no.~8, p. 083\,001
  (2017)}.
\bibAnnoteFile{VampaJPB2017}

\bibitem{EOsikaPRX2017}
{E.~N. Osika} et~al.
\newblock Wannier-{Bloch} approach to localization in high-harmonics generation
  in solids\href{http://dx.doi.org/10.1103/PhysRevX.7.021017}{.
\newblock \emph{Phys. Rev. X} \textbf{7} no.~2, p. 021\,017 (2017)}.
\bibAnnoteFile{EOsikaPRX2017}

\bibitem{VampaNat2015}
{G.~Vampa} et~al.
\newblock Linking high harmonics from gases and
  solids\href{http://dx.doi.org/10.1038/nature14517}{.
\newblock \emph{Nature} \textbf{522} no. 7557, p. 462 (2015)}.
\newblock \href{http://www.attoscience.ca/pdf/Vampa_Nature_2015.pdf}{JASL
  eprint}.
\bibAnnoteFile{VampaNat2015}

\bibitem{VampaPRB2015}
{G.~Vampa} et~al.
\newblock Semiclassical analysis of high harmonic generation in bulk
  crystals\href{http://dx.doi.org/10.1103/PhysRevB.91.064302}{.
\newblock \emph{Phys. Rev. B} \textbf{91} no.~6, p. 064\,302 (2015)}.
\newblock \href{http://www.attoscience.ca/pdf/Vampa_PRB_2015.pdf}{JASL eprint}.
\bibAnnoteFile{VampaPRB2015}

\bibitem{VampaPRL2015}
{G.~Vampa} et~al.
\newblock All-optical reconstruction of crystal band
  structure\href{http://dx.doi.org/10.1103/PhysRevLett.115.193603}{.
\newblock \emph{Phys. Rev. Lett.} \textbf{115} no.~19, p. 193\,603 (2015)}.
\newblock \href{http://www.attoscience.ca/pdf/Vampa_PRL_2015.pdf}{JASL eprint}.
\bibAnnoteFile{VampaPRL2015}

\bibitem{GoldePRB2008}
{D.~Golde, T.~Meier and S.~W. Koch}.
\newblock High harmonics generated in semiconductor nanostructures by the
  coupled dynamics of optical inter- and intraband
  excitations\href{http://dx.doi.org/10.1103/PhysRevB.77.075330}{.
\newblock \emph{Phys. Rev. B} \textbf{77} no.~7, p. 075\,330 (2008)}.
\bibAnnoteFile{GoldePRB2008}

\bibitem{Lakhotia2020}
{H.~Lakhotia} et~al.
\newblock Laser picoscopy of valence electrons in
  solids\href{http://dx.doi.org/10.1038/s41586-020-2429-z}{.
\newblock \emph{Nature} \textbf{583} no. 7814, pp. 55--59 (2020)}.
\bibAnnoteFile{Lakhotia2020}

\bibitem{Liu2017}
{H.~Liu} et~al.
\newblock High-harmonic generation from an atomically thin
  semiconductor\href{http://dx.doi.org/10.1038/nphys3946}{.
\newblock \emph{Nat. Phys.} \textbf{13} no.~3, p. 262 (2017)}.
\newblock
  \href{https://web.stanford.edu/group/heinz/publications/Pub259.pdf}{Stanford
  eprint}.
\bibAnnoteFile{Liu2017}

\bibitem{Luu2018}
{T.~T. Luu and H.~J. W{\"o}rner}.
\newblock Measurement of the {Berry} curvature of solids using high-harmonic
  spectroscopy\href{http://dx.doi.org/10.1038/s41467-018-03397-4}{.
\newblock \emph{Nat. Commun.} \textbf{9} no.~1, p. 916 (2018)}.
\bibAnnoteFile{Luu2018}

\bibitem{Avetissian2020}
{H.~K. Avetissian and G.~F. Mkrtchian}.
\newblock High laser harmonics induced by the berry curvature in time-reversal
  invariant materials{
  }\href{https://arxiv.org/abs/2005.09449}{arXiv:2005.09449}.
\bibAnnoteFile{Avetissian2020}

\bibitem{AlNaib2014}
{I.~Al-Naib, J.~E. Sipe and M.~M. Dignam}.
\newblock High harmonic generation in undoped graphene: Interplay of inter- and
  intraband dynamics\href{http://dx.doi.org/10.1103/PhysRevB.90.245423}{.
\newblock \emph{Phys. Rev. B} \textbf{90} no.~24, p. 245\,423 (2014)}.
\bibAnnoteFile{AlNaib2014}

\bibitem{Dimitrovski2017}
{D.~Dimitrovski, L.~B. Madsen and T.~G. Pedersen}.
\newblock High-order harmonic generation from gapped graphene: Perturbative
  response and transition to nonperturbative
  regime\href{http://dx.doi.org/10.1103/PhysRevB.95.035405}{.
\newblock \emph{Phys. Rev. B} \textbf{95} no.~3, p. 035\,405 (2017)}.
\newblock \href{https://arxiv.org/abs/1612.05037}{arXiv:1612.05037}.
\bibAnnoteFile{Dimitrovski2017}

\bibitem{Yoshikawa2017}
{N.~Yoshikawa, T.~Tamaya and K.~Tanaka}.
\newblock High-harmonic generation in graphene enhanced by elliptically
  polarized light excitation\href{http://dx.doi.org/10.1126/science.aam8861}{.
\newblock \emph{Science} \textbf{356} no. 6339, pp. 736--738 (2017)}.
\bibAnnoteFile{Yoshikawa2017}

\bibitem{Plaja2018}
{O.~Zurr\'on, A.~Pic\'on and L.~Plaja}.
\newblock Theory of high-order harmonic generation for gapless
  graphene\href{http://dx.doi.org/10.1088/1367-2630/aabec7}{.
\newblock \emph{New J. Phys.} \textbf{20} no.~20, p. 053\,033 (2018)}.
\bibAnnoteFile{Plaja2018}

\bibitem{Tamaya2017}
{T.~Tamaya, S.~Konabe and S.~Kawabata}.
\newblock Orientation dependence of high-harmonic generation in monolayer
  transition metal dichalcogenides{
  }\href{https://arxiv.org/abs/1706.00548}{arXiv:\allowbreak{}1706.\allowbreak{}00548}.
\bibAnnoteFile{Tamaya2017}

\bibitem{Yoshikawa2019}
{N.~Yoshikawa} et~al.
\newblock Interband resonant high-harmonic generation by valley polarized
  electron--hole pairs\href{http://dx.doi.org/10.1038/s41467-019-11697-6}{.
\newblock \emph{Nat. Commun.} \textbf{10} no. 3709, pp. 1--7 (2019)}.
\bibAnnoteFile{Yoshikawa2019}

\bibitem{Guan2019}
{M.-X. Guan} et~al.
\newblock Cooperative evolution of intraband and interband excitations for
  high-harmonic generation in strained
  {${\mathrm{MoS}}_{2}$}\href{http://dx.doi.org/10.1103/PhysRevB.99.184306}{.
\newblock \emph{Phys. Rev. B} \textbf{99} no. 184306}.
\bibAnnoteFile{Guan2019}

\bibitem{Jia2020}
{L.~Jia} et~al.
\newblock Optical high-order harmonic generation as a structural
  characterization tool\href{http://dx.doi.org/10.1103/PhysRevB.101.144304}{.
\newblock \emph{Phys. Rev. B} \textbf{101} no.~14, p. 144\,304 (2020)}.
\bibAnnoteFile{Jia2020}

\bibitem{Avetissian2020bilayer}
{H.~K. Avetissian} et~al.
\newblock {High-harmonic generation at particle{\textendash}hole multiphoton
  excitation in gapped bilayer
  graphene}\href{http://dx.doi.org/10.1117/1.JNP.14.026004}{.
\newblock \emph{J. Nanophotonics} \textbf{14} no.~2, p. 026\,004 (2020)}.
\newblock \href{https://arxiv.org/abs/1905.08016}{arXiv:1905.08016}.
\bibAnnoteFile{Avetissian2020bilayer}

\bibitem{SilvaNatPhoton2018}
{R.~E.~F. Silva} et~al.
\newblock High-harmonic spectroscopy of ultrafast many-\allowbreak{}body
  dynamics in strongly correlated
  systems\href{http://dx.doi.org/10.1038/s41566-018-0129-0}{.
\newblock \emph{Nat. Photon.} \textbf{12} no.~2, pp. 266--270 (2018)}.
\newblock
  \href{https://arxiv.org/abs/1704.08471}{arXiv:\allowbreak{}1704.\allowbreak{}08471}.
\bibAnnoteFile{SilvaNatPhoton2018}

\bibitem{Takayoshi2019}
{S.~Takayoshi, Y.~Murakami and P.~Werner}.
\newblock High-harmonic generation in quantum spin
  systems\href{http://dx.doi.org/10.1103/PhysRevB.99.184303}{.
\newblock \emph{Phys. Rev. B} \textbf{99} no.~18, p. 184\,303 (2019)}.
\newblock \href{https://arxiv.org/abs/1901.07588}{arXiv:1901.07588}.
\bibAnnoteFile{Takayoshi2019}

\bibitem{SSH}
{W.~Su, J.~Schrieffer and A.~J. Heeger}.
\newblock Solitons in
  polyacetylene\href{http://dx.doi.org/10.1103/PhysRevLett.42.1698}{.
\newblock \emph{Phys. Rev. Lett.} \textbf{42}, p. 1698 (1979)}.
\bibAnnoteFile{SSH}

\bibitem{Bauer2018}
{D.~Bauer and K.~K. Hansen}.
\newblock High-harmonic generation in solids with and without topological edge
  states\href{http://dx.doi.org/10.1103/PhysRevLett.120.177401}{.
\newblock \emph{Phys. Rev. Lett.} \textbf{120} no.~6, p. 177\,401 (2018)}.
\newblock
  \href{https://arxiv.org/abs/1711.05783}{arXiv:\allowbreak{}1711.\allowbreak{}05783}.
\bibAnnoteFile{Bauer2018}

\bibitem{Drueke2019}
{H.~Dr\"ueke and D.~Bauer}.
\newblock Robustness of topologically sensitive harmonic generation in
  laser-driven linear
  chains\href{http://dx.doi.org/10.1103/PhysRevA.99.053402}{.
\newblock \emph{Phys. Rev. A} \textbf{99} no.~5, p. 053\,402 (2019)}.
\newblock
  \href{https://arxiv.org/abs/1901.01437}{arXiv:\allowbreak{}1901.\allowbreak{}01437}.
\bibAnnoteFile{Drueke2019}

\bibitem{Juerss2019}
{C.~J\"ur\ss{} and D.~Bauer}.
\newblock High-harmonic generation in {Su-Schrieffer-Heeger}
  chains\href{http://dx.doi.org/10.1103/PhysRevB.99.195428}{.
\newblock \emph{Phys. Rev. B} \textbf{99} no.~19, p. 195\,428 (2019)}.
\newblock \href{https://arxiv.org/abs/1902.04120}{arXiv:1902.04120}.
\bibAnnoteFile{Juerss2019}

\bibitem{Misha2018}
{R.~E.~F. Silva} et~al.
\newblock Topological strong-field physics on sub-laser-cycle
  timescale\href{http://dx.doi.org/10.1038/s41566-019-0516-1}{.
\newblock \emph{Nat. Photon.} \textbf{13} no.~12, pp. 849--854 (2019)}.
\newblock \href{https://arxiv.org/abs/1806.11232}{arXiv:1806.11232}.
\bibAnnoteFile{Misha2018}

\bibitem{DeVega2020}
{S.~de~Vega} et~al.
\newblock Strong-field-driven dynamics and high-harmonic generation in
  interacting one dimensional
  systems\href{http://dx.doi.org/10.1103/PhysRevResearch.2.013313}{.
\newblock \emph{Phys. Rev. Research} \textbf{2} no.~1, p. 013\,313 (2020)}.
\bibAnnoteFile{DeVega2020}

\bibitem{Jia2019}
{L.~Jia} et~al.
\newblock High harmonic generation in magnetically-doped topological
  insulators\href{http://dx.doi.org/10.1103/PhysRevB.100.125144}{.
\newblock \emph{Phys. Rev. B} \textbf{100} no.~12, p. 125\,144 (2019)}.
\bibAnnoteFile{Jia2019}

\bibitem{Avetissian2018}
{H.~K. Avetissian} et~al.
\newblock Multiphoton excitation and high-harmonics generation in topological
  insulator\href{http://dx.doi.org/10.1088/1361-648x/aab989}{.
\newblock \emph{J. Phys.: Condens. Matter} \textbf{30} no.~18, p. 185\,302
  (2018)}.
\newblock \href{https://arxiv.org/abs/1711.07840}{arXiv:1711.07840}.
\bibAnnoteFile{Avetissian2018}

\bibitem{Lee2019}
{C.~H. Lee} et~al.
\newblock Enhanced higher harmonic generation from nodal topology{.
\newblock \emph{arXiv}
  }\href{https://arxiv.org/abs/1906.11806}{arXiv:1906.11806}.
\bibAnnoteFile{Lee2019}

\bibitem{Asteria_2018}
{L.~Asteria} et~al.
\newblock Measuring quantized circular dichroism in ultracold topological
  matter\href{http://dx.doi.org/10.1038/s41567-019-0417-8}{.
\newblock \emph{Nat. Phys.} \textbf{15} no.~5, pp. 449--454 (2019)}.
\newblock \href{https://arxiv.org/abs/1805.11077}{arXiv:1805.11077}.
\bibAnnoteFile{Asteria_2018}

\bibitem{Zhang2018}
{G.~P. Zhang} et~al.
\newblock Generating high-order optical and spin harmonics from ferromagnetic
  monolayers\href{http://dx.doi.org/10.1038/s41467-018-05535-4}{.
\newblock \emph{Nat. Commun.} \textbf{9} no. 3031, pp. 1--7 (2018)}.
\bibAnnoteFile{Zhang2018}

\bibitem{Haldane1988}
{F.~D.~M. Haldane}.
\newblock Model for a quantum {Hall} effect without {Landau} levels:
  Condensed-matter realization of the ``parity
  anomaly''\href{http://dx.doi.org/10.1103/PhysRevLett.61.2015}{.
\newblock \emph{Phys. Rev. Lett.} \textbf{61} no.~18, pp. 2015--2018 (1988)}.
\bibAnnoteFile{Haldane1988}

\bibitem{YueGaarde2020}
{L.~Yue and M.~B. Gaarde}.
\newblock Structure gauges and laser gauges for the semiconductor bloch
  equations in high-order harmonic generation in
  solids\href{http://dx.doi.org/10.1103/PhysRevA.101.053411}{.
\newblock \emph{Phys. Rev. A} \textbf{101} no.~5, p. 053\,411 (2020)}.
\bibAnnoteFile{YueGaarde2020}

\bibitem{Kohmoto1985}
{M.~Kohmoto}.
\newblock Topological invariant and the quantization of the {Hall}
  conductance\href{http://dx.doi.org/10.1016/0003-4916(85)90148-4}{.
\newblock \emph{Ann. Phys.} \textbf{160} no.~2, pp. 343--354 (1985)}.
\bibAnnoteFile{Kohmoto1985}

\bibitem{Yang2013}
{F.~Yang and R.-B. Liu}.
\newblock Berry phases of quantum trajectories of optically excited
  electron-hole pairs in semiconductors under strong terahertz
  fields\href{http://dx.doi.org/10.1088/1367-2630/15/11/115005}{.
\newblock \emph{New J. Phys.} \textbf{15} no.~11, p. 115\,005 (2013)}.
\newblock \href{https://arxiv.org/abs/1211.3021}{arXiv:1211.3021}.
\bibAnnoteFile{Yang2013}

\bibitem{Saito2017}
{N.~Saito} et~al.
\newblock Observation of selection rules for circularly polarized fields in
  high-harmonic generation from a crystalline
  solid\href{http://dx.doi.org/10.1364/OPTICA.4.001333}{.
\newblock \emph{Optica} \textbf{4} no.~11, pp. 1333--1336 (2017)}.
\bibAnnoteFile{Saito2017}

\bibitem{Cooper2018}
{N.~R. Cooper, J.~Dalibard and I.~B. Spielman}.
\newblock Topological bands for ultracold
  atoms\href{http://dx.doi.org/10.1103/RevModPhys.91.015005}{.
\newblock \emph{Rev. Mod. Phys.} \textbf{91} no.~1, p. 015\,005 (2019)}.
\newblock \href{https://arxiv.org/abs/1803.00249}{arXiv:1803.00249}.
\bibAnnoteFile{Cooper2018}

\bibitem{bernevig2006}
{B.~A. Bernevig, T.~L. Hughes and S.-C. Zhang}.
\newblock Quantum spin {Hall} effect and topological phase transition in {HgTe}
  quantum wells\href{http://dx.doi.org/10.1126/science.1133734}{.
\newblock \emph{Science} \textbf{314} no. 5806, pp. 1757--1761 (2006)}.
\newblock
  \href{https://arxiv.org/abs/cond-mat/0611399}{arXiv:cond-mat/0611399}.
\bibAnnoteFile{bernevig2006}

\bibitem{konig2007}
{M.~K{\"o}nig} et~al.
\newblock Quantum spin {Hall} insulator state in {HgTe} quantum
  wells\href{http://dx.doi.org/10.1126/science.1148047}{.
\newblock \emph{Science} \textbf{318} no. 5851, pp. 766--770 (2007)}.
\newblock \href{https://arxiv.org/abs/0710.0582}{arXiv:0710.0582}.
\bibAnnoteFile{konig2007}

\bibitem{Kane2005}
{C.~L. Kane and E.~J. Mele}.
\newblock Quantum spin {Hall} effect in
  graphene\href{http://dx.doi.org/10.1103/PhysRevLett.95.226801}{.
\newblock \emph{Phys. Rev. Lett.} \textbf{95} no.~22, p. 226\,801 (2005)}.
\newblock
  \href{https://arxiv.org/abs/cond-mat/0411737}{arXiv:cond-mat/0411737}.
\bibAnnoteFile{Kane2005}

\bibitem{Jotzu2014}
{G.~Jotzu} et~al.
\newblock Experimental realization of the topological {Haldane} model with
  ultracold fermions\href{http://dx.doi.org/10.1038/nature13915}{.
\newblock \emph{Nature} \textbf{515} no. 7526, p. 237 (2014)}.
\newblock \href{https://arxiv.org/abs/1406.7874}{arXiv:1406.7874}.
\bibAnnoteFile{Jotzu2014}

\bibitem{Ching-Kai2016}
{C.-K. Chiu} et~al.
\newblock Classification of topological quantum matter with
  symmetries\href{http://dx.doi.org/10.1103/RevModPhys.88.035005}{.
\newblock \emph{Rev. Mod. Phys.} \textbf{88} no.~3, p. 035\,005 (2016)}.
\newblock \href{https://arxiv.org/abs/1505.03535}{arXiv:1505.03535}.
\bibAnnoteFile{Ching-Kai2016}

\bibitem{Blount1962}
{E.~Blount}.
\newblock Formalisms of band theory{.
\newblock In {F.~Seitz and D.~Turnbull} (eds.), \emph{Solid State Physics},
  vol.~13, pp. 305--373 (Academic Press, 1962)}.
\bibAnnoteFile{Blount1962}

\bibitem{VladPRBVG_LG0}
{S.~Y. Kruchinin, M.~Korbman and V.~S. Yakovlev}.
\newblock Theory of strong-field injection and control of photocurrent in
  dielectrics and wide band gap
  semiconductors\href{http://dx.doi.org/10.1103/PhysRevB.87.115201}{.
\newblock \emph{Phys. Rev. B} \textbf{87} no.~11, p. 115\,201 (2013)}.
\bibAnnoteFile{VladPRBVG_LG0}

\bibitem{Spivak1999}
{M.~Spivak}.
\newblock \emph{A comprehensive introduction to differential geometry} (Publish
  or Perish, Houston, 1999).
\bibAnnoteFile{Spivak1999}

\bibitem{SilvaPRB2019}
{R.~E.~F. Silva, F.~Mart\'{\i}n and M.~Ivanov}.
\newblock High harmonic generation in crystals using maximally localized
  wannier functions\href{http://dx.doi.org/10.1103/PhysRevB.100.195201}{.
\newblock \emph{Phys. Rev. B} \textbf{100} no.~19, p. 195\,201 (2019)}.
\bibAnnoteFile{SilvaPRB2019}

\bibitem{HaugKoch2004}
{H.~Haug and S.~W. Koch}.
\newblock \emph{Quantum Theory of the Optical and Electronic Properties of
  Semiconductors} (World Scientific, Singapore, 2004).
\bibAnnoteFile{HaugKoch2004}

\bibitem{VampaPRL2014}
{G.~Vampa} et~al.
\newblock Theoretical analysis of high-harmonic generation in
  solids\href{http://dx.doi.org/10.1103/PhysRevLett.113.073901}{.
\newblock \emph{Phys. Rev. Lett.} \textbf{113} no.~7, p. 073\,901 (2014)}.
\newblock \href{http://www.attoscience.ca/pdf/Vampa_PRL_2014.pdf}{JASL eprint}.
\bibAnnoteFile{VampaPRL2014}

\bibitem{Pechukas1976}
{P.~Pechukas} et~al.
\newblock Analytic structure of the eigenvalue problem as used in semiclassical
  theory of electronically inelastic
  collisions\href{http://dx.doi.org/10.1063/1.432297}{.
\newblock \emph{J. Chem. Phys.} \textbf{64} no.~3, pp. 1099--1105 (1976)}.
\bibAnnoteFile{Pechukas1976}

\bibitem{Hwang1977}
{J.-T. Hwang and P.~Pechukas}.
\newblock The adiabatic theorem in the complex plane and the semiclassical
  calculation of nonadiabatic transition
  amplitudes\href{http://dx.doi.org/10.1063/1.434630}{.
\newblock \emph{J. Chem. Phys.} \textbf{67} no.~10, pp. 4640--4653 (1977)}.
\bibAnnoteFile{Hwang1977}

\bibitem{Reuter2016}
{M.~G. Reuter}.
\newblock {A unified perspective of complex band structure: interpretations,
  formulations, and
  applications}\href{http://dx.doi.org/10.1088/1361-648x/29/5/053001}{.
\newblock \emph{J. Phys.: Condens. Matter} \textbf{29} no.~5, p. 053\,001
  (2016)}.
\newblock \href{https://arxiv.org/abs/1607.06724}{arXiv:1607.06724}.
\bibAnnoteFile{Reuter2016}

\bibitem{Hawkins2015}
{P.~G. Hawkins, M.~Y. Ivanov and V.~S. Yakovlev}.
\newblock Effect of multiple conduction bands on high-harmonic emission from
  dielectrics\href{http://dx.doi.org/10.1103/PhysRevA.91.013405}{.
\newblock \emph{Phys. Rev. A} \textbf{91} no.~1, p. 013\,405 (2015)}.
\newblock \href{https://arxiv.org/abs/1409.5707}{arXiv:1409.5707}.
\bibAnnoteFile{Hawkins2015}

\bibitem{Floss2018}
{I.~Floss} et~al.
\newblock \textit{Ab initio} multiscale simulation of high-order harmonic
  generation in solids\href{http://dx.doi.org/10.1103/PhysRevA.97.011401}{.
\newblock \emph{Phys. Rev. A} \textbf{97} no.~1, p. 011\,401 (2018)}.
\newblock \href{https://arxiv.org/abs/1705.10707}{arXiv:1705.10707}.
\bibAnnoteFile{Floss2018}

\bibitem{Fleischer2014}
{A.~Fleischer} et~al.
\newblock Spin angular momentum and tunable polarization in high-harmonic
  generation\href{http://dx.doi.org/10.1038/nphoton.2014.108}{.
\newblock \emph{Nat. Photon.} \textbf{8} no.~7, pp. 543--549 (2014)}.
\newblock \href{http://arxiv.org/abs/1310.1206}{arXiv:1310.1206}.
\bibAnnoteFile{Fleischer2014}

\bibitem{Alon1998}
{O.~E. Alon, V.~Averbukh and N.~Moiseyev}.
\newblock Selection rules for the high harmonic generation
  spectra\href{http://dx.doi.org/10.1103/PhysRevLett.80.3743}{.
\newblock \emph{Phys. Rev. Lett.} \textbf{80} no.~17, pp. 3743--3746 (1998)}.
\bibAnnoteFile{Alon1998}

\bibitem{Sorngard2013}
{S.~A. S\o{}rng\aa{}rd, S.~I. Simonsen and J.~P. Hansen}.
\newblock High-order harmonic generation from graphene: Strong attosecond
  pulses with arbitrary
  polarization\href{http://dx.doi.org/10.1103/PhysRevA.87.053803}{.
\newblock \emph{Phys. Rev. A} \textbf{87} no.~5, p. 053\,803 (2013)}.
\bibAnnoteFile{Sorngard2013}

\bibitem{Chen2019}
{Z.-Y. Chen and R.~Qin}.
\newblock Circularly polarized extreme ultraviolet high harmonic generation in
  graphene\href{http://dx.doi.org/10.1364/OE.27.003761}{.
\newblock \emph{Opt. Express} \textbf{27} no.~3, pp. 3761--3770 (2019)}.
\bibAnnoteFile{Chen2019}

\bibitem{Zurron2019}
{O.~Zurr\'{o}n-Cifuentes} et~al.
\newblock Optical anisotropy of non-perturbative high-order harmonic generation
  in gapless graphene\href{http://dx.doi.org/10.1364/OE.27.007776}{.
\newblock \emph{Opt. Express} \textbf{27} no.~5, pp. 7776--7786 (2019)}.
\bibAnnoteFile{Zurron2019}

\bibitem{JimenezGalan2019}
{A.~Jim\'enez-Gal\'an} et~al.
\newblock Ultrafast topology for strong-field valleytronics (2019).{
  }\href{http://arxiv.org/abs/1910.07398}{arXiv:1910.07398}.
\bibAnnoteFile{JimenezGalan2019}

\bibitem{Mauger2016}
{F.~Mauger, A.~D. Bandrauk and T.~Uzer}.
\newblock Circularly polarized molecular high harmonic generation using a
  bicircular laser\href{http://dx.doi.org/10.1088/0953-4075/49/10/10LT01}{.
\newblock \emph{J. Phys. B: At. Mol. Opt. Phys.} \textbf{49} no.~10, p. 10LT01
  (2016)}.
\newblock \href{https://hal.archives-ouvertes.fr/hal-01101871/}{HAL eprint}.
\bibAnnoteFile{Mauger2016}

\bibitem{Baykusheva2016}
{D.~Baykusheva} et~al.
\newblock Bicircular high-harmonic spectroscopy reveals dynamical symmetries of
  atoms and molecules\href{http://dx.doi.org/10.1103/PhysRevLett.116.123001}{.
\newblock \emph{Phys. Rev. Lett.} \textbf{116} no.~12, p. 123\,001 (2016)}.
\newblock \href{https://doi.org/10.3929/ethz-a-010625328}{ETH eprint}.
\bibAnnoteFile{Baykusheva2016}

\bibitem{Yuan2018}
{K.-J. Yuan and A.~D. Bandrauk}.
\newblock Symmetry in circularly polarized molecular high-order harmonic
  generation with intense bicircular laser
  pulses\href{http://dx.doi.org/10.1103/PhysRevA.97.023408}{.
\newblock \emph{Phys. Rev. A} \textbf{97} no.~2, p. 023\,408 (2018)}.
\bibAnnoteFile{Yuan2018}

\bibitem{Konishi2014}
{K.~Konishi} et~al.
\newblock Polarization-controlled circular second-harmonic generation from
  metal hole arrays with threefold rotational
  symmetry\href{http://dx.doi.org/10.1103/PhysRevLett.112.135502}{.
\newblock \emph{Phys. Rev. Lett.} \textbf{112} no.~13, p. 135\,502 (2014)}.
\bibAnnoteFile{Konishi2014}

\bibitem{Medisauskas2015}
{L.~Medi\v{s}auskas} et~al.
\newblock Generating isolated elliptically polarized attosecond pulses using
  bichromatic counterrotating circularly polarized laser
  fields\href{http://dx.doi.org/10.1103/PhysRevLett.115.153001}{.
\newblock \emph{Phys. Rev. Lett.} \textbf{115} no.~15, p. 153\,001 (2015)}.
\newblock \href{http://arxiv.org/abs/1504.06578}{arXiv:1504.06578}.
\bibAnnoteFile{Medisauskas2015}

\bibitem{Pisanty2017}
{E.~Pisanty and A.~Jim\'enez-Gal\'an}.
\newblock Strong-field approximation in a rotating frame: High-order harmonic
  emission from $p$ states in bicircular
  fields\href{http://dx.doi.org/10.1103/PhysRevA.96.063401}{.
\newblock \emph{Phys. Rev. A} \textbf{96} no.~6, p. 063\,401 (2017)}.
\newblock \href{http://arxiv.org/abs/1709.00397}{arXiv:1709.00397}.
\bibAnnoteFile{Pisanty2017}

\bibitem{Harada2018}
{Y.~Harada} et~al.
\newblock Circular dichroism in high-order harmonic generation from chiral
  molecules\href{http://dx.doi.org/10.1103/PhysRevA.98.021401}{.
\newblock \emph{Phys. Rev. A} \textbf{98} no.~2, p. 021\,401 (2018)}.
\newblock \href{http://hdl.handle.net/2115/71440}{HUSCAP eprint}.
\bibAnnoteFile{Harada2018}

\bibitem{Ayuso2018}
{D.~Ayuso} et~al.
\newblock Chiral dichroism in bi-elliptical high-order harmonic
  generation\href{http://dx.doi.org/10.1088/1361-6455/aaae5e}{.
\newblock \emph{J. Phys. B: At. Mol. Opt. Phys.} \textbf{51} no.~6, p. 06LT01
  (2018)}.
\bibAnnoteFile{Ayuso2018}

\bibitem{WilsonMassModel}
{K.~G. Wilson}.
\newblock {Quarks and Strings on a
  Lattice}\href{http://dx.doi.org/10.1007/978-1-4613-4208-3_6}{.
\newblock In {A.~Zichichi} (ed.), \emph{New Phenomena in Subnuclear Physics},
  pp. 69--142 (Springer US, Boston, MA, 1977)}.
\bibAnnoteFile{WilsonMassModel}

\bibitem{W1}
{J.~B. Kogut}.
\newblock The lattice gauge theory approach to quantum
  chromodynamics\href{http://dx.doi.org/10.1103/RevModPhys.55.775}{.
\newblock \emph{Rev. Mod. Phys.} \textbf{55} no.~3, pp. 775--836 (1983)}.
\bibAnnoteFile{W1}

\bibitem{W2a}
{L.~H. Karsten and J.~Smith}.
\newblock Lattice fermions: Species doubling, chiral invariance and the
  triangle anomaly\href{http://dx.doi.org/10.1016/0550-3213(81)90549-6}{.
\newblock \emph{Nucl. Phys. B} \textbf{183} no.~1, pp. 103--140 (1981)}.
\bibAnnoteFile{W2a}

\bibitem{W2b}
{H.~B. Nielsen and M.~Ninomiya}.
\newblock Absence of neutrinos on a lattice: {(I)}. proof by homotopy
  theory\href{http://dx.doi.org/10.1016/0550-3213(81)90361-8}{.
\newblock \emph{Nucl. Phys. B} \textbf{185} no.~1, pp. 20--40 (1981)}.
\bibAnnoteFile{W2b}

\bibitem{W5a}
{F.~Wilczek}.
\newblock Problem of strong $\mathrm{P}$ and $\mathrm{T}$ invariance in the
  presence of instantons\href{http://dx.doi.org/10.1103/PhysRevLett.40.279}{.
\newblock \emph{Phys. Rev. Lett.} \textbf{40} no.~5, pp. 279--282 (1978)}.
\bibAnnoteFile{W5a}

\bibitem{W5b}
{S.~Weinberg}.
\newblock A new light
  boson?\href{http://dx.doi.org/10.1103/PhysRevLett.40.223}{.
\newblock \emph{Phys. Rev. Lett.} \textbf{40} no.~4, pp. 223--226 (1978)}.
\bibAnnoteFile{W5b}

\bibitem{W5c}
{J.~Preskill, M.~B. Wise and F.~Wilczek}.
\newblock Cosmology of the invisible
  axion\href{http://dx.doi.org/10.1016/0370-2693(83)90637-8}{.
\newblock \emph{Phys. Lett. B} \textbf{120} no.~1, pp. 127--132 (1983)}.
\bibAnnoteFile{W5c}

\bibitem{W6a}
{X.-L. Qi, T.~L. Hughes and S.-C. Zhang}.
\newblock Topological field theory of time-reversal invariant
  insulators\href{http://dx.doi.org/10.1103/PhysRevB.78.195424}{.
\newblock \emph{Phys. Rev. B} \textbf{78} no.~19, p. 195\,424 (2008)}.
\newblock \href{https://arxiv.org/abs/0802.3537}{arXiv:0802.3537}.
\bibAnnoteFile{W6a}

\bibitem{W6b}
{A.~M. Essin, J.~E. Moore and D.~Vanderbilt}.
\newblock Magnetoelectric polarizability and axion electrodynamics in
  crystalline
  insulators\href{http://dx.doi.org/10.1103/PhysRevLett.102.146805}{.
\newblock \emph{Phys. Rev. Lett.} \textbf{102} no.~14, p. 146\,805 (2009)}.
\newblock \href{https://arxiv.org/abs/0810.2998}{arXiv:0810.2998}.
\bibAnnoteFile{W6b}

\bibitem{W7a}
{X.-L. Qi} et~al.
\newblock Inducing a magnetic monopole with topological surface
  states\href{http://dx.doi.org/10.1126/science.1167747}{.
\newblock \emph{Science} \textbf{323} no. 5918, pp. 1184--1187 (2009)}.
\newblock \href{https://arxiv.org/abs/0811.1303}{arXiv:0811.1303}.
\bibAnnoteFile{W7a}

\bibitem{W7b}
{G.~Rosenberg and M.~Franz}.
\newblock Witten effect in a crystalline topological
  insulator\href{http://dx.doi.org/10.1103/PhysRevB.82.035105}{.
\newblock \emph{Phys. Rev. B} \textbf{82} no.~3, p. 035\,105 (2010)}.
\newblock \href{https://arxiv.org/abs/1001.3179}{arXiv:1001.3179}.
\bibAnnoteFile{W7b}

\bibitem{PRLNathan}
{A.~Bermudez} et~al.
\newblock Wilson fermions and axion electrodynamics in optical
  lattices\href{http://dx.doi.org/10.1103/PhysRevLett.105.190404}{.
\newblock \emph{Phys. Rev. Lett.} \textbf{105} no.~19, p. 190\,404 (2010)}.
\newblock \href{https://arxiv.org/abs/1004.5101}{arXiv:1004.5101}.
\bibAnnoteFile{PRLNathan}

\bibitem{RodriguezL2018}
{P.~Rodriguez-Lopez} et~al.
\newblock Nonlocal optical response in topological phase transitions in the
  graphene family\href{http://dx.doi.org/10.1103/PhysRevMaterials.2.014003}{.
\newblock \emph{Phys. Rev. Materials} \textbf{2}, p. 014\,003 (2018)}.
\newblock
  \href{https://arxiv.org/abs/1711.06311}{arXiv:\allowbreak{}1711.\allowbreak{}06311}.
\bibAnnoteFile{RodriguezL2018}

\bibitem{Neupane2014}
{M.~Neupane} et~al.
\newblock Observation of quantum-tunnelling-modulated spin texture in ultrathin
  topological insulator {Bi}$_2${Se}$_3$
  films\href{http://dx.doi.org/10.1038/ncomms4841}{.
\newblock \emph{Nat. Commun.} \textbf{5} no. 3841, pp. 1--7 (2014)}.
\newblock \href{https://arxiv.org/abs/1404.2830}{arXiv:1404.2830}.
\bibAnnoteFile{Neupane2014}

\bibitem{Cao18-1}
{Y.~Cao} et~al.
\newblock Unconventional superconductivity in magic-angle graphene
  superlattices\href{http://dx.doi.org/10.1038/nature26160}{.
\newblock \emph{Nature} \textbf{556}, pp. 43--50 (2018)}.
\bibAnnoteFile{Cao18-1}

\bibitem{Cao18-2}
{Y.~Cao} et~al.
\newblock Correlated insulator behaviour at half-filling in magic-angle
  graphene superlattices\href{http://dx.doi.org/10.1038/nature26154}{.
\newblock \emph{Nature} \textbf{556}, pp. 80--84 (2018)}.
\newblock \href{https://arxiv.org/abs/1802.00553}{arXiv:1802.00553}.
\bibAnnoteFile{Cao18-2}

\bibitem{Lu19}
{X.~Lu} et~al.
\newblock Superconductors, orbital magnets and correlated states in magic-angle
  bilayer graphene\href{http://dx.doi.org/10.1038/s41586-019-1695-0}{.
\newblock \emph{Nature} \textbf{574}, pp. 653--657 (2019)}.
\newblock \href{https://arxiv.org/abs/1903.06513}{arXiv:1903.06513}.
\bibAnnoteFile{Lu19}

\bibitem{Balents2020}
{L.~Balents} et~al.
\newblock {Superconductivity and strong correlations in
  moir{\ifmmode\acute{e}\else\'{e}\fi} flat
  bands}\href{http://dx.doi.org/10.1038/s41567-020-0906-9}{.
\newblock \emph{Nat. Phys.} pp. 1--9}.
\newblock \href{https://arxiv.org/abs/1912.03375}{arXiv:1912.03375}.
\bibAnnoteFile{Balents2020}

\bibitem{AcinRoadmap}
{A.~Ac\'{\i}n} et~al.
\newblock The quantum technologies roadmap: a {European} community
  view\href{http://dx.doi.org/10.1088/1367-2630/aad1ea}{.
\newblock \emph{New J. Phys.} \textbf{20} no.~8, p. 080\,201 (2018)}.
\newblock \href{https://arxiv.org/abs/1712.03773}{arXiv:1712.03773}.
\bibAnnoteFile{AcinRoadmap}

\bibitem{LSA12}
{M.~Lewenstein, A.~Sanpera and V.~Ahufinger}.
\newblock \emph{Ultracold Atoms in Optical Lattices: Simulating quantum
  many-body systems} (Oxford University Press, 2012).
\bibAnnoteFile{LSA12}

\bibitem{Bloch08}
{I.~Bloch, J.~Dalibard and W.~Zwerger}.
\newblock Many-body physics with ultracold
  gases\href{http://dx.doi.org/10.1103/RevModPhys.80.885}{.
\newblock \emph{Rev. Mod. Phys.} \textbf{80} no.~3, pp. 885--964 (2008)}.
\newblock \href{https://arxiv.org/abs/0704.3011}{arXiv:0704.3011}.
\bibAnnoteFile{Bloch08}

\bibitem{specialissueNaturePhys}
{J.~I. Cirac and P.~Zoller}.
\newblock Goals and opportunities in quantum
  simulation\href{http://dx.doi.org/10.1038/nphys2275}{.
\newblock \emph{Nat. Phys.} \textbf{8} no.~4, pp. 264--266 (2012)}.
\bibAnnoteFile{specialissueNaturePhys}

\bibitem{specialissueNaturePhys1}
{I.~Bloch, J.~Dalibard and S.~Nascimb\`{e}ne}.
\newblock Quantum simulations with ultracold quantum
  gases\href{http://dx.doi.org/10.1038/nphys2259}{.
\newblock \emph{Nat. Phys.} \textbf{8} no.~4, pp. 267--276 (2012)}.
\bibAnnoteFile{specialissueNaturePhys1}

\bibitem{specialissueNaturePhys2}
{R.~Blatt and C.~F. Roos}.
\newblock Quantum simulations with trapped
  ions\href{http://dx.doi.org/10.1038/nphys2252}{.
\newblock \emph{Nat. Phys.} \textbf{8} no.~4, pp. 277--284 (2012)}.
\bibAnnoteFile{specialissueNaturePhys2}

\bibitem{specialissueNaturePhys3}
{A.~Aspuru-Guzik and P.~Walther}.
\newblock Photonic quantum
  simulators\href{http://dx.doi.org/10.1038/nphys2253}{.
\newblock \emph{Nat. Phys.} \textbf{8} no.~4, pp. 285--291 (2012)}.
\newblock
  \href{https://walther.quantum.at/fileadmin/user_upload/a_walther/News/Papers/2012.Apr.02_Paper_Photonic_quantum_simulators.pdf}{U
  Wien eprint}.
\bibAnnoteFile{specialissueNaturePhys3}

\bibitem{specialissueNaturePhys4}
{A.~A. Houck, H.~E. T\"{u}reci and J.~Koch}.
\newblock On-chip quantum simulation with superconducting
  circuits\href{http://dx.doi.org/10.1038/nphys2251}{.
\newblock \emph{Nat. Phys.} \textbf{8} no.~4, pp. 292--299 (2012)}.
\newblock
  \href{http://faculty.ee.princeton.edu/tureci/sites/default/files/nphys2251_0.pdf}{U
  Princeton eprint}.
\bibAnnoteFile{specialissueNaturePhys4}

\bibitem{Sala2017}
{S.~Sala, J.~F\"orster and A.~Saenz}.
\newblock Ultracold-atom quantum simulator for attosecond
  science\href{http://dx.doi.org/10.1103/PhysRevA.95.011403}{.
\newblock \emph{Phys. Rev. A} \textbf{95} no.~1, p. 011\,403 (2017)}.
\newblock \href{https://arxiv.org/abs/1311.2304}{arXiv:1311.2304}.
\bibAnnoteFile{Sala2017}

\bibitem{Senaratne2018}
{R.~Senaratne} et~al.
\newblock Quantum simulation of ultrafast dynamics using trapped ultracold
  atoms\href{http://dx.doi.org/10.1038/s41467-018-04556-3}{.
\newblock \emph{Nature Commun.} \textbf{9} no.~1, p. 2065 (2018)}.
\bibAnnoteFile{Senaratne2018}

\bibitem{Ramos2019}
{R.~Ramos} et~al.
\newblock Measuring the time a tunnelling atom spends in the barrier{
  }\href{https://arxiv.org/abs/1907.13523}{arXiv:1907.13523}.
\bibAnnoteFile{Ramos2019}

\end{thebibliography}

\end{document}